\documentclass[a4paper,11pt]{article}
\pdfoutput=1 

\usepackage{jheppub} 
    
\usepackage{amsmath}
\usepackage[T1]{fontenc} 
\usepackage{ulem}
\usepackage[dvipsnames]{xcolor}
\usepackage{booktabs}
\usepackage{bbold}
\usepackage{multicol}
\usepackage{multirow}
\usepackage{stmaryrd} 
\usepackage{cancel}
\usepackage{pifont}
\usepackage{simplewick}
\usepackage{tikz}
\usepackage{comment}
\usepackage{subcaption}
\usepackage{hyperref}
\usepackage{tikz}
\usepackage{makecell}

\usepackage[export]{adjustbox}

\title{Systematic Improvement of \\
Hamiltonian Truncation Effective Theory}

\author[a]{E. Demiray,}
\author[b]{K. Farnsworth,}
\author[a]{R. Houtz}

\affiliation[a]{Department of Physics, University of Florida, Gainesville, FL 32611, USA}
\affiliation[b]{D\'epartment de Physique Th\'eorique, Universit\'e de Gen\`eve, CH-1211 Gen\`eve, Switzerland}

\abstract{
Hamiltonian Truncation Effective Theory is a framework that aims to improve the results of Hamiltonian truncation in a systematic, order-by-order fashion using Effective Field Theory methodology. The result is a truncated effective Hamiltonian with corrections that result from a matching procedure. We establish the rigor of this method by calculating nontrival next-to-leading order corrections in a $1/E_{\rm max}$ expansion, where $E_{\rm max}$ is our effective theory cutoff. We illustrate this explicitly using 1+1D $\lambda \phi^4$ theory, calculating corrections up to order $1/E_{\rm max}^3$. At this order, novel nonlocal contributions to the matching conditions must be incorporated. We show that by including these nonlocal terms, the error scales as $1/E_{\rm max}^4$, as expected from the Effective Field Theory power counting, providing a nontrivial check that this method is consistent and robust. We also estimate the critical coupling at which this theory flows to the 2D Ising conformal field theory and confirm that separation of scales, an essential feature of Effective Field Theory, persists at this order. These results establish Hamiltonian Truncation Effective Theory as a generic, systematic framework for improving convergence in Hamiltonian truncation and lay the groundwork to apply this method to more complex systems in higher dimensions.
}

\begin{document}

\maketitle

\flushbottom

\clearpage


\parskip=5pt
\normalsize

\setcounter{page}{2}
\setcounter{footnote}{0}

\section{Introduction\label{intro}}

Strongly coupled systems in quantum field theory (QFT), such as QCD at low energies or strongly correlated condensed matter systems, are challenging to study precisely as they cannot rely on perturbative techniques.  
Currently, lattice field theory is the state of the art method used to probe systems at strong coupling with errors that are quantifiable and under control. Lattice methods, however, struggle to model chiral fermions
and theories with supersymmetry. Due to the sign problem, performing time evolution is also computationally expensive on the lattice, meaning dynamical variables are difficult to model \cite{Troyer:2004ge, Alexandru:2016gsd}. Other numerical tools have emerged to compliment lattice methods and extract observables cheaply in the absence of available lattice results, including the density matrix renormalization group used in condensed matter physics and Hamiltonian truncation, which is the focus of this work.

Hamiltonian truncation is based on the Rayleigh-Ritz variational method in quantum mechanics. To use it, the Hamiltonian of a system is split into two parts, one that is solvable, $H_0$, and an interaction term $V$ which need not be small:
\begin{align}
H = H_0 + V \,.
\end{align}
The system is discretized and then truncated to a finite size so that the Hamiltonian can be diagonalized numerically. This results in approximate eigenvalues and eigenvectors of the full system which can be used to construct any observable of interest, including information about real-time dynamics. It was first used in the context of QFT in \cite{Brooks:1983sb}, and further developed in \cite{Yurov:1989yu, Yurov:1991my}. There has been recent renewed interest in  applying these methods to strongly coupled QFTs after the works of \cite{Katz:2013qua, Hogervorst:2014rta}. Since then various versions of Hamiltonian truncation have been constructed, each with its own relative advantages and disadvantages, which use different methods to discretize and truncate the system, see \cite{Katz:2014uoa, Rychkov:2014eea, Rychkov:2015vap, Elias-Miro:2015bqk, Bajnok:2015bgw, Katz:2016hxp, Rakovszky:2016ugs, Anand:2017yij, Elias-Miro:2017tup, Rutter:2018aog, Fitzpatrick:2018ttk, Hogervorst:2018otc, Delacretaz:2018xbn,  Fitzpatrick:2019cif, Elias-Miro:2020qwz, Anand:2020gnn, Anand:2020qnp,Tilloy:2021yre,  Hogervorst:2021spa, Chen:2021bmm, Anand:2021qnd, Cohen:2021erm, EliasMiro:2021aof, Emonts:2022vim, Chen:2022zms, Delacretaz:2022ojg, Henning:2022xlj, EliasMiro:2022pua, Chen:2023glf, Lajer:2023unt, Fitzpatrick:2023mbt, Fitzpatrick:2023aqm, Delouche:2023wsl, Schmoll:2023eez, Ingoldby:2024fcy, Delouche:2024tjf, Fitzpatrick:2024rks, Ingoldby:2025bdb} for developments and applications as well as the reviews~\cite{James:2017cpc, Fitzpatrick:2022dwq}. 

Although a promising alternative to lattice methods, one major limitation of Hamiltonian truncation is that the number of basis states grows exponentially with the truncation parameter, resulting in the fact that most of the progress so far has been restricted to low (two and three) dimensional systems. To confront this limitation,\footnote{One other solution to this problem could be the use of quantum computers once they are viable, see e.g. \cite{Ingoldby:2024fcy, Ingoldby:2025bdb} for work on developing quantum computational methods for Hamiltonian truncation.} significant effort has been put into constructing effective Hamiltonians \cite{Lee:2000gm, Feverati:2006ni, Giokas:2011ix, Hogervorst:2014rta, Rychkov:2014eea, Elias-Miro:2015bqk, Elias-Miro:2017tup, Rutter:2018aog, Fitzpatrick:2018ttk, Elias-Miro:2020qwz, Anand:2020gnn,  Cohen:2021erm, EliasMiro:2022pua, Chen:2023glf, Lajer:2023unt, Delouche:2023wsl}, with the goal of incorporating the effects of the neglected states above the truncation cutoff without increasing the size of the truncated Hilbert space. These types of techniques are crucial to be able to apply Hamiltonian truncation to systems with a large number of degrees of freedom and/or in higher dimensions.

Hamiltonian Truncation Effective Theory (HTET) \cite{Cohen:2021erm} 
seeks to achieve this by using Effective Field Theory (EFT) techniques to systematically improve convergence order-by-order in inverse powers of the truncation parameter. Taking advantage of EFT intuition, we utilize the massive Fock space approach of \cite{Rychkov:2014eea}, which is formulated in finite volume using equal-time quantization. Our truncated Hilbert space consists of eigenstates of the free Hamiltonian $H_0$,
\begin{align}
H_0|i\rangle = E_i |i\rangle \,,
\end{align}
with energy $E_i$ below some total energy cutoff $E_{\rm max}$. We then construct the interaction term in this truncated Hilbert space to get the 
full truncated Hamiltonian
\begin{align}
\langle f |H_{\rm trunc} |i\rangle = E_i \delta_{fi} + \langle f|V|i\rangle \,,
\end{align}
which can be diagonalized numerically. The goal of HTET is to construct further corrections
\begin{align}
H_{\rm eff} = H_{\rm trunc}+ H_{\rm corr}
\end{align}
to this truncated Hamiltonian which act on the truncated Hilbert space and systematically improve the convergence order-by-order in a $1/E_{\rm max}$ expansion. Within the EFT framework, this is accomplished by matching a particular observable in the full and truncated theories at the ``scale of new physics'' $E_{\rm max}$. We expect these matching corrections to display separation of scales, that is their coefficients should be insensitive to changes in the IR physics as long as the IR scales (e.g. mass or inverse volume) are much lower than $E_{\rm max}$. Separation of scales indicates that the matching corrections are correctly incorporating the effects of states above the truncation cutoff, and without this we expect our EFT expansion to break down. We choose the matching observable such that this procedure:
\begin{itemize}
\item uniquely defines our effective Hamiltonian $H_{\rm eff}$,
\item  yields a systematic expansion in powers of IR scales over $E_{\rm max}$,
\item and ensures a manifest separation of scales at each order in this expansion. 
\end{itemize}
Similar improvement methods have been employed in previous work \cite{Lee:2000gm, Feverati:2006ni, Giokas:2011ix, Hogervorst:2014rta, Rychkov:2014eea, Elias-Miro:2015bqk, Elias-Miro:2017tup, Rutter:2018aog, Fitzpatrick:2018ttk, Elias-Miro:2020qwz, Anand:2020gnn,  Cohen:2021erm, EliasMiro:2022pua, Chen:2023glf, Lajer:2023unt, Delouche:2023wsl}, but often violate these assumptions or require the inclusion of additional states to the truncated basis. In HTET we {\it require} these features of EFT to be realized, and study the numerical convergence of observables of the system under these assumptions. If the convergence matches our expectation from EFT, this is strong evidence that this method is reliable and can be applied straightforwardly to other systems. The leading order correction was computed for 1+1D $\lambda \phi^4$ theory in \cite{Cohen:2021erm} and the convergence was found to agree with EFT expectations. In that case, the included correction was completely local, but at higher orders in the EFT expansion we generically expect nonlocal and non-Hermitian terms to appear as the effective Hamiltonian compensates for the nonlocal regulator $E_{\rm max}$ \cite{Hogervorst:2014rta, Elias-Miro:2017tup, Rutter:2018aog}. These nonlocal terms must be carefully constructed, as any inaccuracy corresponds to our EFT matching corrections reproducing a \textit{nonlocal} full theory instead of the local theory we are interested in. This is a qualitatively new effect that appears at NLO and was not necessary to address in the leading order, local, corrections computed in \cite{Cohen:2021erm}.

Here we calculate the matching corrections to the next order in the $1/E_{\rm max}$ expansion for this theory, providing a nontrivial test of the power of HTET by including nonlocal and non-Hermitian terms. We show that the power counting continues to be consistent after the inclusion of these terms and that separation of scales is still manifest. This is conclusive evidence that the HTET methodology is robust and can be extended to arbitrarily high order, which is critical to apply this technique to systems in higher dimensions.

This paper is organized as follows. In Section \ref{sec:HTET} we present HTET for 1+1D $\lambda \phi^4$ theory, including the EFT power counting in Section \ref{pc},  the matching calculations in Section \ref{sec:match} and the final expressions for the effective Hamiltonian in Section \ref{sec:finalexpress} at this order. In Section \ref{sec:numerics} we present the numerical results and show the agreement with EFT predictions as well as a prediction of the critical coupling for which this theory flows to the 2D Ising model in the IR. We conclude in Section \ref{sec:conclusions}. In Appendix \ref{sec:rules}, for completeness we include the diagrammatic rules used to calculate the matching corrections, first presented in \cite{Cohen:2021erm}.  In Appendix \ref{sos} we demonstrate separation of scales for the higher order calculations done in the main text. 

\section{Hamiltonian Truncation Effective Theory for 1+1D $\lambda \phi^4$ theory\label{sec:HTET}}

We are considering the Lagrangian 
\begin{align}
\mathcal{L} = \frac{1}{2}(\partial \phi)^2 - \frac{1}{2} m^2 \phi^2 - \frac{\lambda}{4!} \phi^4 \,. 
\label{eq:Lagrangian}
\end{align}
This choice of theory has several advantages. In 1+1D, $\phi$ is dimensionless and both $m^2$ and $\lambda$ have mass dimension 2, i.e. they are strongly relevant and we expect Hamiltonian truncation methods to converge well. 
This allows us to apply perturbation theory in the UV, where this theory is weakly coupled, while still accessing nonperturbative physics in the IR. Moreover this theory is ``super-renormalizable,'' and all UV divergences can be removed by simply normal-ordering the operators. A useful feature of this setup is that, for a particular nonperturbative choice of the dimensionless ratio, $\lambda/m^2$, this theory flows to the 2D Ising model in the IR, where we can compare with known analytic results. This system has been previously studied in great detail using truncated methods \cite{Rychkov:2014eea, Elias-Miro:2015bqk, Bajnok:2015bgw, Anand:2017yij, Elias-Miro:2017tup, Lajer:2023unt}, providing a clean arena to explore the errors and subsequent improvement associated with truncation systematically.

To apply Hamiltonian truncation to this system, we first split the Hamiltonian into a solvable part $H_0$, and an interaction term $V$. We will choose $H_0$ to describe the free massive scalar, and $V$ to parameterize the $\lambda \phi^4$ interaction. 
We then discretize our system by placing the theory on a spatial circle and imposing periodic boundary conditions
\begin{align}
\phi (t,x)  = \phi (t, x + 2\pi R) \,,
\end{align}
where $R$ is the radius of the circle.  The solution to the free massive equations of motion on this spacetime for $\phi$ gives the mode expansion
\begin{align}
\phi(t,x) = \phi(t,R\,\theta) =  \frac{1}{\sqrt{2\pi R}} \sum_k \frac{1}{\sqrt{2\omega_k}} \left[ a^\dagger_k e^{i (\omega_k t -  k \theta)} + \text{h.c.}\right] \,,
\end{align}
with $k\in \mathbb{Z}$ and $\omega_k \equiv \sqrt{m^2 + k^2/R^2}$. The raising and lowering operators obey the usual commutation relation $[a_{k_1}, a_{k_2}^\dagger] = \delta_{k_1, k_2}$. Our expressions for $H_0$ and $V$ can be written
\begin{align}
H_0 	=&\ \sum_k \omega_k\, a^\dagger_k a_k\,,	\\
V 	=&\ \frac{\lambda}{4!}  \int R\,d\theta :\!\phi^4\!: \\
	=&\ \frac{\lambda}{4!} \frac{1}{8\pi R} \! \sum_{k_1,k_2,k_3,k_4} \!\! \frac{\delta_{k_1+k_2+k_3+k_4}}{\sqrt{ \omega_{k_1}\omega_{k_2} \omega_{k_3} \omega_{k_4}}} \bigg[a_{k_1}^\dagger a_{k_2}^\dagger a_{k_3}^\dagger  a_{k_4}^\dagger +4  a_{k_1}^\dagger  a_{k_2}^\dagger a_{k_3}^\dagger  a_{-k_4} 
		+6 a_{k_1}^\dagger  a_{k_2}^\dagger  a_{-k_3}  a_{-k_4}
	\nonumber	\\
	&\qquad{} \qquad{} \qquad{} \qquad{} \qquad{}  
		\qquad \qquad \quad
		+ 4 a_{-k_1}^\dagger a_{k_3}  a_{k_4} a_{k_2}
		+   a_{k_1} a_{k_2} a_{k_3} a_{k_4} \bigg]\,.
\end{align}
 Note here our interaction $V$ corresponds to the \it normal-ordered \rm operator $:\!\phi^4\!:$\,. We will also have contributions to our effective Hamiltonian of the form $:\!\phi^2\!:$, which can be written in terms of raising and lowering operators as
\begin{align}
\int R\,d\theta :\!\phi^2\!:\ =&\ \sum_k \frac{1}{2\omega_k} \left[ a^\dagger_k a^\dagger_{-k} + 2 a^\dagger_k a_{-k} + a_k a_{-k}\right] \,.
\end{align}
These are related to the non-normal-ordered operators $\phi^4$ and $\phi^2$ via the equations
\begin{align}
\begin{split}
\phi^4 &=\ :\!\phi^4\!: + 6Z:\!\phi^2\!: + 3Z^2 \,,\\
\phi^2 &=\ :\!\phi^2\!:  + Z \,,
\end{split}
\label{NOops}
\end{align}
where 
\begin{align}
Z = \frac{1}{4\pi R} \sum_k \frac{1}{\omega_k}\,.
\end{align}
This is the only source of UV divergences in this theory and can be regulated by normal-ordering the operators.\footnote{The choice to use only normal-ordered operators corresponds to a particular choice of renormalization scheme. However that we expect separation of scales, a hallmark of EFT, to only be manifest for ``non-normal-ordered'' operators using a slightly different renormalization scheme, see Appendix \ref{sos} and \cite{Cohen:2021erm} for a further discussion of this point.}

The basis we use is the Fock space basis which is spanned by the eigenstates of $H_0$ 
\begin{align}
H_0|i\rangle = E_i |i\rangle\,,\quad{} i =0, 1, 2, \dots
\end{align}
where $|i\rangle$ are the harmonic oscillator eigenstates on which $a_k$ and $a_k^\dagger$ act in the usual way. We then truncate our system to a finite size by only keeping eigenstates with energy below a cutoff $E_{\rm max}$, as measured with respect to the free Hamiltonian $H_0$, i.e. states with $E_i \leq E_{\rm max}$. Note this is a generically \it nonlocal \rm cutoff, since it depends on the total energy of the state, taking into account spectator states that are not necessarily localized together. 
We expect these effects to be suppressed by inverse powers of $E_{\rm max}$, but they will make an appearance in our effective Hamiltonian corrections as we try to approximate a local theory in the UV.

After performing this truncation, our Hamiltonian is a finite-dimensional matrix and can be diagonalized numerically. Based on  known power counting shown in more detail in the next section, we expect the error in the calculation of this ``raw'' truncation to scale like $1/E_{\rm max}^2$. 
 However, the number of states in our truncated basis scales exponentially as we increase $E_{\rm max}$. 
Rather than increasing the basis size, we will instead use  EFT techniques to compute an effective Hamiltonian 
\begin{align}
H_{\rm eff} = H_0 + H_1 + H_2 + H_3 + \cdots \,,
\end{align}
where $H_n\sim \mathcal{O}(V^n)$ are corrections to the truncated Hamiltonian. These corrections act on the truncated Hilbert space but take into account effects from states above the cutoff and therefore improve convergence. We solve for these terms in our effective Hamiltonian by matching the overlap between the free and interacting eigenstates 
of our system at energy $E_{\rm max}$. 
Adding in these corrections means we can improve the numerical convergence of our calculation by inverse powers of $E_{\rm max}$ without increasing the basis size.

\subsection{Power counting\label{pc}}

We organize our effective Hamiltonian in powers of the interaction coupling $\lambda$, which has mass dimension 2. 
We also 
assume 
\begin{align}
\label{eq:localapprox}
\omega_i, \omega_f \ll E_i, E_f \ll E_{\rm max} \,,
\end{align}
where $\omega_i, \omega_f$ correspond to the energies of the individual incoming and outgoing particles in the initial and final states, and $E_i,\  E_f$ correspond to the total energy of the low-lying initial and final states we are considering. By performing our matching calculation at the scale where $E_{\rm max}$ is much larger than these and any other IR scales in the problem (e.g. $m$ or $R^{-1}$), we can systematically expand our effective Hamiltonian in powers of IR scales over $E_{\rm max}$.

At each order in $\lambda$, the effective Hamiltonian can be parameterized as
\begin{align}
H_n \sim \frac{\lambda^n}{E_{\rm max}^{2n-2}}\sum_a \int\, R\, d\theta\, C_{na}\left( \frac{H_0}{E_{\rm max}}, \frac{R^{-1}}{E_{\rm max}}, \frac{m}{E_{\rm max}}\right)\mathcal{O}_{na} \left( \frac{\partial_x}{E_{\rm max}}, \frac{\partial_t}{E_{\rm max}}, \phi\right) \,.
\label{eq:powerct}
\end{align}
Here $\mathcal{O}_{na}$ are local, Hermitian, dimensionless operators, and $C_{na}$ are dimensionless coefficients involving ratios of the IR scales over $E_{\rm max}$. 
 The cutoff $E_{\rm max}$ is the only source of nonlocality in this theory, so we expect nonlocal terms to appear in our effective Hamiltonian only  as powers of $H_0$ times a local operator. This also means they are generically suppressed by powers of $1/E_{\rm max}$.
The explicit form of these operators can be seen in the diagrammatic expansion shown in the next section.

At order $\lambda^2$ we have, schematically,
\begin{align}
\begin{split}
H_2 \sim& \frac{\lambda^2}{E_{\rm max}^2} \int\, R\,  d\theta\, \bigg[ \phi^2 + \phi^4 + \frac{R^{-1} + H_0}{E_{\rm max}} \left( 1 + \phi^2 + \phi^4\right) \\
&\ \ \ \ \  \ \ \ \ \ \ \ \ \ \ + \frac{1}{E_{\rm max}}\left(\left[ H_0, \phi^2\right]+\left[ H_0, \phi^4\right]\right) + \mathcal{O}(E_{\rm max}^{-2})\bigg] \,,
\end{split}
\end{align}
where here we have replaced the terms $\partial_t \phi$ with the commutator $\left[H_0, \phi\right] = -i \partial_t \phi$. At leading order, the first two terms scale as $1/E_{\rm max}^2$ and are local and Hermitian. However at the next order, the terms with $1/E_{\rm max}^3$ scaling start to contain nonlocal, non-Hermitian terms like $H_0\, \int\, R\, d\theta\,  \phi^2$.

Based on the power counting in \eqref{eq:powerct}, we also see $H_3 \sim \lambda^3/E_{\rm max}^4$. Therefore in order to improve convergence systematically, we should first include the terms in $H_2$ that scale like $1/E_{\rm max}^2$ followed by the terms in $H_2$ that scale like $1/E_{\rm max}^3$ before including any terms coming from $H_3$. The first of these corrections was studied in \cite{Cohen:2021erm}, where the error in the eigenvalues of the raw Hamiltonian (without corrections) was shown to scale like $1/E_{\rm max}^2$ and the error after including the leading order corrections was shown to scale like $1/E_{\rm max}^3$. Here we extend this analysis to the next order by including the terms in $H_2$ that scale like $1/E_{\rm max}^3$, resulting in errors in the eigenvalues that scale like $1/E_{\rm max}^4$, exactly as predicted by the power counting in~\eqref{eq:powerct}. 

\subsection{Matching \label{sec:match}}

To define a unique effective Hamiltonian that has separation of scales at all orders in our $1/E_{\rm max}$ expansion, we match the low energy interacting eigenstates of our full and truncated systems at the truncation scale $E_{\rm max}$. 
This also automatically matches the energy eigenvalues of the full and truncated theories at this scale. 

To find the interacting eigenstates we first modify our Hamiltonian to be
\begin{align}
H \rightarrow H_\epsilon = H_0 + e^{-\epsilon|t|}V \,.
\end{align}
In the limit $\epsilon \rightarrow 0$, this Hamiltonian adiabatically turns off the interaction for $t\rightarrow \pm \infty$. We can then use the time evolution operator corresponding to this Hamiltonian in the interaction picture as a map between our free and interacting eigenstates. More concretely, for $t\rightarrow \pm \infty$, $H_\epsilon$ becomes our free Hamiltonian $H_0$, with the corresponding eigenstates $|i\rangle$ and eigenvalues $E_i$:
\begin{align}
H_\epsilon \xrightarrow[t\rightarrow \pm \infty]{} H_0 \quad{} \textnormal{with}\quad{} H_0|i \rangle = E_i |i \rangle \,.
\end{align}
For $\epsilon\rightarrow 0$, the interaction is adiabatically turned on, and at $t = 0$, $H_\epsilon$ becomes the full interacting Hamiltonian $H = H_0 + V$:
\begin{align}
H_\epsilon \xrightarrow[t\rightarrow 0]{} H \quad{} \textnormal{with}\quad{}  H|\Psi_i \rangle = \mathcal{E}_i |\Psi_i \rangle \,.
\end{align}
Here $|\Psi_i\rangle$ are the eigenstates of the full Hamiltonian $H$ with eigenvalues $\mathcal{E}_i$.

The time evolution operator in the interaction picture of this Hamiltonian $H_\epsilon$, 
\begin{align}
U_{\rm IP, \epsilon}(t_f, t_i) = T \exp\left\{ -i \int_{t_i}^{t_f} dt\, V_{\rm IP, \epsilon} (t) \right\}\,,\ \quad{} \textnormal{with}\quad{} V_{\rm IP, \epsilon} (t)  = e^{i H_0 t} V e^{-\epsilon |t|} e^{-i H_0 t} \,,
\end{align}
then maps eigenstates of $H_0$ at $t \rightarrow \pm \infty$ to eigenstates of $H$ at $t = 0$, i.e. 
\begin{align}
|\Psi_i \rangle &= \lim_{t_i \rightarrow -\infty} U_{IP, \epsilon}(0, t_i) |i\rangle \,.
\label{eq:intstate}
\end{align}
In the limit $\epsilon\rightarrow 0$, this is precisely the M\o\,\!ller operator $\Omega^{\pm} = \lim\limits_{\epsilon\rightarrow 0} U_{IP, \epsilon}(0,\pm \infty)$ that relates free and interacting eigenstates of a system.\footnote{See e.g. Chapter 7 of \cite{kleinert2016particles} for a pedagogical discussion.}

With this definition, we can look at the overlap of our low energy interacting eigenstates with our free eigenstates, an observable we call $\Sigma_{fi}$,\footnote{This is the same matching operator that was used in \cite{Cohen:2021erm}, but with a slightly different physical interpretation. We thank Joan Elias Mir\'{o} and James Ingoldby for discussions regarding this.}
\begin{align}
\Sigma_{fi} \equiv \langle \Psi_f | i \rangle =  \lim_{t_f \rightarrow \infty} \langle f |U_{IP, \epsilon}(t_f, 0 ) |i\rangle\,.
\label{eq:Sigma}
\end{align}
Here $|i\rangle, |f\rangle$ are low-lying energy eigenstates of the free Hamiltonian $H_0$ and $|\Psi_i\rangle, |\Psi_f\rangle$ are eigenstates of the full interacting Hamiltonian $H$ defined by \eqref{eq:intstate}. We then define our effective Hamiltonian by matching, order-by-order, the observable $\Sigma_{fi}$ in both the full and truncated theories at the scale $E_{\rm max}$. This also ensures that the energy eigenvalues of the full and truncated Hamiltonians agree order-by-order.

Expanding $\Sigma_{fi}$ in powers of the interaction gives
\begin{align}
\Sigma_{fi} = \delta_{fi} + \frac{\langle f|V|i\rangle}{E_{fi}  +i\epsilon} + \sum_\alpha \frac{\langle f|V|\alpha \rangle \langle \alpha |V|i \rangle }{(E_{fi} + i \epsilon)(E_{f\alpha} + i \epsilon)} + \mathcal{O}(V^3) \,,
\end{align}
with $E_{fi} \equiv E_f - E_i$. This calculation can be done explicitly for the full, untruncated theory at the scale $E_{\rm max}$ where the theory is weakly coupled and this expansion can be done perturbatively. An analogous calculation can be done in the effective theory if we assume that our effective Hamiltonian can be written 
\begin{align}
H_{\rm eff} = H_0 + H_1 + H_2 + H_3 + \cdots \,,
\end{align}
with $H_n \sim \mathcal{O}(V^n)$. In this case, the expansion of $\Sigma_{fi}$ for the effective theory gives
\begin{align}
\Sigma^{(\rm eff)}_{fi} = \delta_{fi} + \frac{\langle f|H_1 |i\rangle}{E_{fi}  +i\epsilon}+\frac{\langle f|H_2|i\rangle}{E_{fi}  +i\epsilon} + \sum^<_\alpha \frac{\langle f|H_1|\alpha \rangle \langle \alpha |H_1|i \rangle }{(E_{fi} + i \epsilon)(E_{f\alpha} + i \epsilon)} + \mathcal{O}(V^3) \,,
\end{align}
with $\sum^<_\alpha$ indicating that we are only summing over states with energy $E_\alpha \leq E_{\rm max}$ in the truncated theory.

Matching $\Sigma_{fi}$ in the full and effective theories at the scale $E_{\rm max}$, we can calculate the form of $H_n$ that accomplishes the required matching order-by-order. The resulting matching conditions are:

\begin{subequations}
\begin{align}
\Sigma_{fi} 
	&=\left.\Sigma^{(\rm eff)}_{fi} \right|_{\mathcal{O}(V)}
	\ \Rightarrow \langle f|H_1|i \rangle =\langle f| V|i\rangle	\,,
	\\
\Sigma_{fi} 
	&= \left.\Sigma^{(\rm eff)}_{fi} \right|_{\mathcal{O}(V^2)}
	 \Rightarrow\langle f| H_2|i\rangle =\sum^>_\alpha \frac{\langle f|V|\alpha\rangle \langle \alpha|V|i \rangle}{E_f - E_\alpha} \,,
	\\
\Sigma_{fi} 
	&=\left.\Sigma^{(\rm eff)}_{fi} \right|_{\mathcal{O}(V^3)} 
	 \Rightarrow \langle f|H_3 |i \rangle 
		= \sum_{\alpha, \beta}^> \!
					\frac{\langle f|V|\alpha\rangle \langle \alpha|V|\beta\rangle \langle \beta|V|i\rangle }
						{(E_f - E_\alpha)(E_f - E_\beta)} 
		\nonumber\\
		& \qquad\qquad\qquad\qquad
			- \sum_\alpha^< 
				\! \sum_\beta ^>  \!
					\frac{\langle f|V|\alpha\rangle \langle \alpha|V|\beta\rangle \langle \beta|V|i\rangle }
						{(E_{\alpha} - E_\beta)(E_f - E_\beta)} \,,
\end{align}
\end{subequations}
etc. Here $\sum_\alpha^>$ indicates the sum is only taken over states with energy $E_\alpha > E_{\rm max}$.
We have also taken $\epsilon \rightarrow 0$ in these expressions since the energy denominators will never vanish for these sums. As dictated by the power counting in Section \ref{pc}, we will only need to calculate $\langle f|H_1|i \rangle$  and $\langle f|H_2|i \rangle$ for the particular order in $1/E_{\rm max}$ corrections we are interested in. 

\subsubsection{Matching calculations for $H_1$}
These matching calculations can be done by explicit calculation 
or using a diagrammatic expansion as in \cite{Cohen:2021erm} using the diagrammatic rules reproduced in Appendix \ref{sec:rules}. The matching condition at $\mathcal{O}(V)$ is simply
\begin{align}
\langle f |H_1|i \rangle = \langle f |V|i \rangle \,,
\end{align}
with $|i\rangle,\ |f\rangle$ in the truncated basis. Diagrammatically, we can find this contribution by looking at all possible diagrams containing one power of the interaction $V$.  These are (without drawing the spectator particles that do not participate in the interaction)
\begin{align}
\label{eq:phi4tree}
\includegraphics[valign=c,scale=0.65]{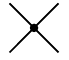}\quad{} \includegraphics[valign=c,scale=0.65]{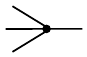}\quad{} \includegraphics[valign=c,scale=0.65]{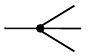}\quad{} \reflectbox{\includegraphics[valign=c,scale=0.65]{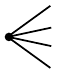}} \quad{}\includegraphics[valign=c,scale=0.65]{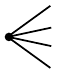} \,.
\end{align}
Calculating one of these using our diagrammatic rules gives
\begin{align}
\includegraphics[valign=c,scale=0.35]{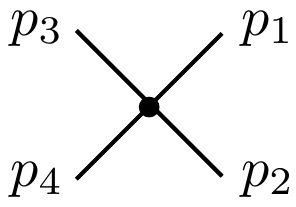} &= \frac{\lambda}{2\pi R} \delta_{1+2+3+4} \langle f| \phi_4^{(-)}\phi_3^{(-)} \phi_2^{(+)} \phi_1^{(+)} |i\rangle \,,
\end{align}
where here we have introduced the shorthand $\phi_1 \equiv \phi_{p_1}$ and $\delta_{1+2} \equiv \delta_{p_1+p_2}$. The other diagrams from \eqref{eq:phi4tree} have the same coefficient, with $\phi_n^{(\pm)}$ depending on the external states. These diagrams combine to give 
\begin{align}
\includegraphics[valign=c,scale=0.65]{figs/phi4_1_f}
+ \includegraphics[valign=c,scale=0.65]{figs/phi4_2_f}
+ \includegraphics[valign=c,scale=0.65]{figs/phi4_3_f}
+ \reflectbox{\includegraphics[valign=c,scale=0.65]{figs/phi4_4_f}}
+ \includegraphics[valign=c,scale=0.65]{figs/phi4_5_f} = \frac{\lambda}{4!} \int R\, d\theta\, \langle f|\! :\!\phi^4\!:\! |i\rangle \,.
\end{align}
The fact that all the diagrams in \eqref{eq:phi4tree} have the same coefficient is an artifact of the interaction being local at this order. As we will see, at higher orders this will no longer be the case.

This calculation corresponds to what we call our raw truncation, in which we use 
\begin{align}
H_{\rm eff}^{(\rm raw)} = H_0 + V
\label{eq:raw}
\end{align}
for our effective Hamiltonian. According to our power counting, at this order we expect the error in our calculation to scale as IR scales multiplied by $1/E_{\rm max}^2$.

\subsubsection{Matching calculations for  $H_2$\label{sec:H2}}
The matching condition at $\mathcal{O}(V^2)$ is
\begin{align}
\langle f| H_2|i\rangle =\sum^>_\alpha \frac{\langle f|V|\alpha\rangle \langle \alpha|V|i \rangle}{E_f - E_\alpha} \,,
\label{eq:H2match}
\end{align}
where again the sum indicates we are summing only over states with energy $E_\alpha > E_{\rm max}$. 

To calculate this diagrammatically, we first draw all diagrams with two vertices (again not explicitly drawing the spectator particles):
\begin{subequations}
\label{eq:v2diagrams1}
\begin{gather}
\includegraphics[valign=c,scale=0.65]{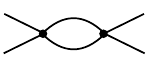}\quad{} \includegraphics[valign=c,scale=0.65]{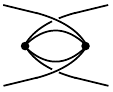}
\quad{} \includegraphics[valign=c,scale=0.85]{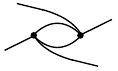}
\quad{} \includegraphics[valign=c,scale=0.65]{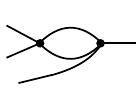}
\quad{} \includegraphics[valign=c,scale=0.65]{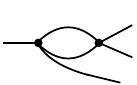}
\quad{} \includegraphics[valign=c,scale=0.65]{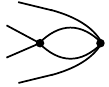}
\quad{}\includegraphics[valign=c,scale=0.65]{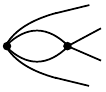} \label{eq:4extlegs}\\
\includegraphics[valign=c,scale=0.65]{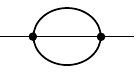} 
\quad{} \includegraphics[valign=c,scale=0.65]{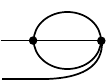}
\quad{} \includegraphics[valign=c,scale=0.65]{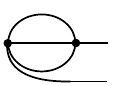}
\quad{} \includegraphics[valign=c,scale=0.65]{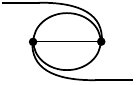}\label{eq:2extlegs}\\
\includegraphics[valign=c,scale=0.65]{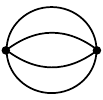}\, \label{eq:0extlegs} \,.
\end{gather}
\end{subequations}
Here we have not included diagrams like 
\begin{gather}
\includegraphics[valign=c,scale=0.30]{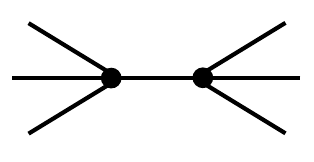}\quad{} \includegraphics[valign=c,scale=0.30]{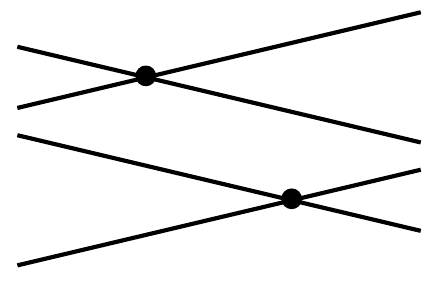} \,.
\end{gather}
These and other tree level diagrams give the same contribution in both the full and truncated theory, so they cancel in the matching calculation with our assumption \eqref{eq:localapprox}. 

The matching condition \eqref{eq:H2match} can be rewritten as
\begin{align}
\langle f| H_2|i\rangle =\sum_\alpha \frac{\langle f|V|\alpha\rangle \langle \alpha|V|i \rangle}{E_f - E_\alpha}\bigg|_{\rm full} - \sum^<_\alpha \frac{\langle f|V|\alpha\rangle \langle \alpha|V|i \rangle}{E_f - E_\alpha}\bigg|_{\rm eff} \,,
\label{eq:H2exp}
\end{align}
which we can calculate diagrammatically separately in the full theory and the effective theory before taking the difference. Using our diagrammatic rules from Appendix \ref{sec:rules}, the explicit expressions for a subset of the $2\rightarrow 2$ diagrams are:
\begin{subequations}
\begin{align}
\label{eq:diag1}
\hspace{-.25cm}
\includegraphics[valign=c,scale=0.65]{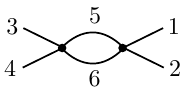} \!
	- \! \biggl[ \! \includegraphics[valign=c,scale=0.65]{figs/V221_num_f} \! \biggr]_\text{eff} \!
	&= \frac 18 \left( \frac{\lambda}{2\pi R} \right)^{\!\! 2} \sum_{1, \dots, 6} \delta_{12,56}\, \delta_{34,56}
		\langle f| \phi_4^{(-)} \phi_3^{(-)} \phi_2^{(+)} \phi_1^{(+)}|i\rangle\nonumber\\
	&\qquad \times
		\frac{1}{2\omega_5} \frac{1}{2\omega_6}
		\frac{\Theta(E_f-\omega_3 - \omega_4 + \omega_5 + \omega_6 - E_{\rm max})}{\omega_3 + \omega_4 - \omega_5 - \omega_6 + i \epsilon}\,,
\\[10pt]
\label{eq:diag2}
\raisebox{.05cm}{\includegraphics[valign=c,scale=0.65]{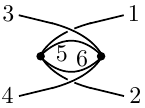}}
	- \!\left[ \raisebox{.05cm}{\includegraphics[valign=c,scale=0.65]{figs/V222_num_f}} \right]_\text{eff}
	&= \frac 18 \left( \frac{\lambda}{2\pi R} \right)^{\!\! 2}
		\sum_{1, \ldots, 6} \delta_{12,56} \delta_{34,56}
		\langle f| \phi_4^{(-)} \phi_3^{(-)} \phi_2^{(+)} \phi_1^{(+)} |i\rangle
	\nonumber\\
	&\qquad \times
		\frac{1}{2\omega_5} \frac{1}{2\omega_6}
		\frac{\Theta(E_f+\omega_1 + \omega_2 + \omega_5 + \omega_6 - E_{\rm max})}{ - \omega_1 - \omega_2 - \omega_5 - \omega_6 + i \epsilon}\,,
\end{align}
\end{subequations}

These expressions have several interesting and perhaps unfamiliar features.
First, the step function in these expressions is ensuring that we are only summing over intermediate states with $E_\alpha > E_{\rm max}$, as is required from the matching calculation \eqref{eq:H2match}. We rewrote $E_\alpha = E_f - E_{f\alpha}$ in \eqref{eq:H2exp} to make the $E_f$ dependence clear, since $E_f$ is the \it total \rm energy of our final state and the source of the nonlocality in the effective Hamiltonian. 
Second, these two expressions are not the same, particularly the arguments of the step functions and the $1/E_{f\alpha}$ energy denominators. Without the step functions, if we looked at external states with the same energy so that $\omega_1 + \omega_2 = \omega_3 + \omega_4$, these two terms would sum together as expected in old fashioned perturbation theory, giving a local contribution to the operator $\phi^4$. However in this case we have enforced a nonlocal cutoff on the theory in the form of the step function, and now these terms do not generically combine into a local operator. 

Expanding in $1/E_{\rm max}$, we will see locality restored in the leading order term and that the nonlocality is suppressed by higher powers in $1/E_{\rm max}$. For the sum of the two previous diagrams, this expansion gives:
\begin{align}
\begin{split}
\includegraphics[valign=c,scale=0.65]{figs/V221_num_f}&+\includegraphics[valign=c,scale=0.65]{figs/V222_num_f}
- \biggl[ \includegraphics[valign=c,scale=0.65]{figs/V221_num_f}+ \includegraphics[valign=c,scale=0.65]{figs/V222_num_f} \biggr]_\text{eff}\\
&= \frac 18 \left( \frac{\lambda}{2\pi R} \right)^{\!\! 2}\delta_{12,34} \sum_{5, 6} \delta_{12,56}
\langle f| \phi_4^{(-)} \phi_3^{(-)} \phi_2^{(+)} \phi_1^{(+)}|i\rangle \frac{1}{2\omega_5} \frac{1}{2\omega_6}\\
&\qquad{} \times\bigg[2\frac{\Theta(\omega_5 + \omega_6 - E_{\rm max})}{ - \omega_5 - \omega_6}+ 2\frac{E_f \delta( \omega_5 + \omega_6 - E_{\rm max})}{ - \omega_5 - \omega_6 }\\
&\qquad{}\qquad{} +E_{if}\frac{\Theta(\omega_5 + \omega_6 - E_{\rm max})}{( \omega_5 +\omega_6)^2}+E_{if}\frac{\delta( \omega_5 + \omega_6 - E_{\rm max})}{ - \omega_5 - \omega_6}+\cdots\bigg] \,,
\end{split}
\end{align}
with $E_{if} \equiv E_i-E_f = \omega_1 + \omega_2 - \omega_3 - \omega_4$. We also used the Taylor series approximation $\Theta(E_f + X) \approx \Theta(X) + E_f\, \delta(X)$, treating $\Theta$ as a distribution. 
The first term in this expansion, without any explicit dependence on $E_i$ or $E_f$, is local and will contribute as a constant coefficient multiplied by $\int\,R\, d\theta :\!\phi^4\!:$ in the effective Hamiltonian. This was included in the calculations done in \cite{Cohen:2021erm}. The other three terms contribute to the new corrections in this work; they are nonlocal and will appear in our effective Hamiltonian as $H_0 \int\, R\, d\theta :\!\phi^4\!:$ ($E_f$ contributions) or $\int\, R\, d\theta :\!\phi^4\!: H_0$ ($E_i$ contributions). 

We can perform the same
expansion for the other diagrams at $\mathcal{O}(V^2)$ \eqref{eq:v2diagrams1}, which all have a similar structure. The total contribution with 4 external legs can be calculated using the diagams \eqref{eq:4extlegs} and will contribute a term to the effective Hamiltonian of the form
\begin{align}
\langle f |H_{2}|i\rangle \supset&\ \frac{1}{4!}  \mathbb{C}_4\int R\,d\theta\, \langle f |\!:\!\phi^4\!:\!|i\rangle =  \bigg[\includegraphics[valign=c,scale=0.65]{figs/V221_f}+\cdots\bigg]- \biggl[ \includegraphics[valign=c,scale=0.65]{figs/V221_f}+\cdots \biggr]_\text{eff} \,,
\end{align}
with 
\begin{align}
\mathbb{C}_4 &=  \frac{3\lambda^2}{ 8\pi R} \sum_{k}   \bigg[\frac{\Theta(2\omega_k-E_{\rm max})}{\omega_k^2(-2\omega_k)}+\frac{1}{2}(E_i+E_f)\frac{\delta(2\omega_k-E_{\rm max})}{\omega_k^2(-2\omega_k)}-\frac{1}{2}E_{fi}\frac{\Theta( 2\omega_k-E_{\rm max})}{\omega_k^2(-2\omega_k)^2}\bigg] \,. 
\label{eq:c4main}
\end{align}
Similarly, the total contribution from the diagrams \eqref{eq:2extlegs} 
contributes a term to the effective Hamiltonian
\begin{align}
\langle f |H_{2} |i\rangle \supset  \frac{1}{2}  \mathbb{C}_2\int R\,d\theta\, \langle f |\!:\!\phi^2\!:\!|i\rangle = \bigg[\includegraphics[valign=c,scale=0.65]{figs/sunset_f} &+\cdots\bigg]
- \biggl[ \includegraphics[valign=c,scale=0.65]{figs/sunset_f} +\cdots \biggr]_\text{eff} \,,
\end{align}
with 
\begin{align}
\begin{split}
\mathbb{C}_2 &= \frac{1}{24} \frac{\lambda^2}{(2\pi R)^2}   \sum_{k, k'} \bigg[\frac{\Theta(  \omega_{k}+ \omega_{k'}+ \omega_{k+k'} -E_{\rm max})}{\omega_{k}\omega_{k'}\omega_{k+k'}( - \omega_{k}- \omega_{k'}- \omega_{k+k'})} \\
&\quad{}+  \frac{1}{2}(E_i+E_f) \frac{\delta(\omega_{k}+ \omega_{k'}+ \omega_{k+k'} -E_{\rm max})}{\omega_{k}\omega_{k'}\omega_{k+k'}(- \omega_{k}- \omega_{k'}- \omega_{k+k'})} \\
&\quad{} -   \frac{1}{2}E_{fi}  \frac{\Theta( \omega_{k}+ \omega_{k'}+ \omega_{k+k'} -E_{\rm max})}{\omega_{k}\omega_{k'}\omega_{k+k'}(- \omega_{k}- \omega_{k'}- \omega_{k+k'})^2}\bigg] \,.
\label{eq:c2main}
\end{split}
\end{align}
And finally using the diagram \eqref{eq:0extlegs}, we can calculate the contribution to the identity operator in our effective Hamiltonian:
\begin{align}
\langle f |H_{2} |i\rangle \supset  \mathbb{C}_0 \int R\,d\theta\, \langle f |\mathbb{1} |i\rangle = \includegraphics[valign=c,scale=0.65]{figs/bubbled_f}- \biggl[ \includegraphics[valign=c,scale=0.65]{figs/bubbled_f} \biggr]_{\text{eff}} \,,
\label{eq:c0-diagram}
\end{align}
with 
\begin{align}
\begin{split}
\mathbb{C}_0&=\frac{1}{384} \frac{\lambda^2}{(2\pi R)^3} \sum_{1,\ldots,4}\delta_{1234,0} \bigg[\frac{\Theta( \omega_1+\omega_2+\omega_3+\omega_4 - E_\text{max})}{\omega_1\omega_2\omega_3\omega_4(-\omega_1-\omega_2-\omega_3-\omega_4)}\\
&\qquad{}\qquad{}+E_f \frac{\delta(\omega_1+\omega_2+\omega_3+\omega_4 - E_\text{max})}{\omega_1\omega_2\omega_3\omega_4(-\omega_1-\omega_2-\omega_3-\omega_4)}+\cdots \bigg] \,.
\label{eq:c0main}
\end{split}
\end{align}
Note there are two different deltas here. One is a Kronecker delta imposing momentum conservation $\delta_{1+2,0}$, and one is a Dirac delta function $\delta(\omega_1+\omega_2 - E_{\rm max})$ from the Taylor expansion of the step function distributions $\Theta(E_f +\omega_1+\omega_2 - E_{\rm max}) \approx \Theta(\omega_1+\omega_2 - E_{\rm max})+E_f\, \delta(\omega_1+\omega_2 - E_{\rm max})$.

For each of these calculations we can separate out the leading and subleading behavior of $H_2$:
\begin{align}
H_2 = H_{2}^{( \rm LO)} + H_{2}^{( \rm NLO)} +\cdots \,.
\end{align}
The leading order term $H_{2}^{( \rm LO)}$ is local and was calculated in \cite{Cohen:2021erm}. There it was shown that adding this term to the raw Hamiltonian \eqref{eq:raw} increases convergence from $1/E_{\rm max}^2$ to $1/E_{\rm max}^3$. The next-to-leading order term $H_{2}^{( \rm NLO)}$ is the new correction included in this work, and we expect this NLO correction to improve the convergence to $1/E_{\rm max}^4$ based on the power counting argument of Section \ref{pc}.

\subsection{Final expressions\label{sec:finalexpress}}
The final expressions for our effective Hamiltonian are
\begin{subequations}
 \label{eq:heff}
\begin{align}
H_{\rm eff}^{(\rm raw)} =&\ H_0 + V \,,
\label{eq:hraw}
\\
H_{\rm eff}^{(\rm LO)} =&\ H_0 + V + H_2^{(\rm LO)}\,,
\label{eq:hLO}
\\
H_{\rm eff}^{(\rm NLO)} =&\ H_0 + V+ H_2^{(\rm LO)}+ H_2^{(\rm NLO)} \,.
\label{eq:hNLO}
\end{align}
\end{subequations}
The leading order piece is
\begin{align}
H_{2}^{( \rm LO)} =  \frac{1}{2} m_{2}^2 \int R\, d\theta :\!\phi^2\!:+ \frac{1}{4!} \lambda_2 \int R\, d\theta :\!\phi^4\!: \,,
\end{align}
with 
\begin{subequations}
\label{eq:LOcoeffs}
\begin{align}
 m_2^2 
 =&\ \frac{-\lambda^2}{24 (2 \pi R)^2}\sum_{123}\delta_{1+2+3,0} \frac{\Theta(\omega_1 + \omega_{2} + \omega_{3} - E_{\rm max})}{\omega_1 \omega_{2}\omega_{3}(\omega_1 + \omega_{2} + \omega_{3})}\,,\\
\lambda_2 =&\  \frac{-3\lambda^2}{16 \pi R} \sum_{k} \frac{\Theta(2\omega_k-E_{\rm max})}{\omega_k^3} \,. 
\end{align}
\end{subequations}
Here we see that at leading order the corrections are constant coefficients multiplying local operators, and the effective Hamiltonian is local. At the next order the explicit expression for the correction is
\begin{align}
\begin{split}
H_{2}^{(\rm NLO)} 
=&\ \zeta H_0\label{H2NLO}\\
& +\frac{1}{2}\alpha_1^{\rm (1)}  \int R\, d\theta\, \{ H_0 :\!\phi^2\!:\}+ \frac{1}{2} \alpha_2^{\rm (2)} \int  R\, d\theta\left[ H_0, :\!\phi^2\!:\right] \\
& +\frac{1}{4!}\beta_1^{\rm (1)}\int  R\, d\theta\,  \{H_0,  :\!\phi^4\!:\}+\frac{1}{4!} \beta_2^{\rm (2)} \int  R\, d\theta\left[ H_0, :\!\phi^4\!:\right] \,,
\end{split}
\end{align}
with
 \begin{subequations}
 \label{eq:NLOcoeffs}
\begin{align}
\zeta =&\ \frac{-\lambda^2}{384(2\pi R)^2}
\sum_{1234}\delta_{1+2+3+4,0} \frac{\delta(\omega_1 + \omega_2 + \omega_3 +\omega_{4} - E_{\rm max})}{\omega_1\omega_2\omega_3\omega_{4}(\omega_1+\omega_2+\omega_3 + \omega_{4})}\,,\\
\alpha_1^{\rm (1)}=&\ \frac{-\lambda^2}{48 (2 \pi R)^2}\sum_{123}\delta_{1+2+3,0} \frac{\delta(\omega_1 + \omega_{2} + \omega_{3} - E_{\rm max})}{\omega_1 \omega_{2}\omega_{3}(\omega_1 + \omega_{2} + \omega_{3})}\,,\\
\alpha_2^{\rm (2)}=&\ \frac{-\lambda^2}{48 (2 \pi R)^2}\sum_{123}\delta_{1+2+3,0} \frac{\Theta(\omega_1 + \omega_{2} + \omega_{3} - E_{\rm max})}{\omega_1 \omega_{2}\omega_{3}(\omega_1 + \omega_{2} + \omega_{3})^2}\,,\\
\beta_1^{\rm (1)}=&\ \frac{-3\lambda^2}{4\pi }\frac{1}{E_{\rm max}^2 \sqrt{E_{\rm max}^2 - 4m^2}}\,,\\
\beta_2^{\rm (2)}=&\ \frac{-3\lambda^2}{64\pi R}\sum_{k} \frac{\Theta(2\omega_k-E_{\rm max})}{\omega_k^4} \,.
\end{align}
\end{subequations}
 At this order we see explicitly that our effective Hamiltonian has nonlocal contributions. We can now check numerically using these expressions if our EFT power counting is correct.

\section{Numerical results for 1+1D $\lambda \phi^4$ theory\label{sec:numerics}}
In this section, we present the numerical improvements for 1+1D $\lambda \phi^4$ theory using the improved effective Hamiltonian described above. 
Our results are organized according to the different orders of truncation improvement, labeled $H_{\rm eff}^{\rm (raw)}$, $H_{\rm eff}^{\rm (LO)}$, and $H_{\rm eff}^{\rm (NLO)}$ as defined in \eqref{eq:heff}. 

The Hamiltonian is represented numerically as a matrix. In the basis of energy eigenstates of the free theory $H_0$, we directly construct the matrices for the following operators 
	\begin{align}
	H_0 \,,
		\quad\int  R\, d\theta\, \! : \! \phi^2 \! : \,,
		\quad\int R\, d\theta\, \! : \! \phi^4 \! : \,,
		\quad\int R\, d\theta\, \llbracket H_0, : \! \phi^2 \! : \rrbracket\,,
		\quad\int R\, d\theta\, \llbracket H_0, : \! \phi^4 \! : \rrbracket \,,
	\end{align}
where $\llbracket  H_0, :\!\phi^n\!: \rrbracket$ indicates either a commutator or anticommutator. The full $H_{\rm eff}$ is built as a linear combination of these matrices, where the coefficients are given by our final expressions above in Section~\ref{sec:finalexpress}.\footnote{The code to construct $H_{\rm eff}$ can be found at \url{https://github.com/EkremDemiray/HTET_NLO} and \url{github.com/rahoutz/hamiltonian-truncation}. All computations were performed on a laptop computer.} Numerical evaluation of these coefficients is straightforward. The sums in  \eqref{eq:LOcoeffs} and \eqref{eq:NLOcoeffs} are computed directly by summing over $k$, from $k=0$ up to $k_{\rm UV}=1000$. This $k_{\rm UV}$ can be thought of as a Wilsonian momentum cutoff far above the energy scales we are modeling with $H_{\rm eff}$. 
Any state containing a particle with momentum $k_{\rm UV}$ has energy much larger than $E_{\rm max}$.

For our numerical results, we construct $H_{\rm eff}$ using the same Fock space defined in~\cite{Cohen:2021erm}. Namely, we use the eigenstates of $H_0$ such that their energy eigenvalue is smaller than $E_{\rm max}$, and we select a subset of states such that the total spatial momentum is zero. We also note that our theory is symmetric under a $\mathbb{Z}_2$ transformation which takes $\phi \to - \phi$, and so we can separate our basis into $\mathbb Z_2$-even and -odd sectors. The $\mathbb Z_2$ even (odd) sector has an even (odd) number of particles in each state. Finally, we quote all our results in units of $m=1$.

Our previous work~\cite{Cohen:2021erm} demonstrated the error for the $H_{\rm eff}^{\rm (raw)}$ scales as ${\mathcal{O}}(1/E_{\rm max}^2)$, and $H_{\rm eff}^{\rm (LO)}$ as $\mathcal{O}(1/E_{\rm max}^3)$. With the inclusion of our NLO corrections, we find that  $H_{\rm eff}^{\rm (NLO)}$ errors are reduced to scale as ${\mathcal{O}}(1/E_{\rm max}^4)$ as predicted by the EFT power counting in Section \ref{pc}.

We first present the convergence of the energy eigenstates in Sections~\ref{sec:lowV} and~\ref{sec:highV}, focusing on two benchmark radii: $R=10/2\pi$ and $R=20/2\pi$. The lower volume case, presented in Section~\ref{sec:lowV}, exhibits significant noise,\footnote{Since this noise decreases for higher states and for the higher volume benchmark, it may be due, at least in part, to finite volume effects \cite{Luscher:1985dn, Luscher:1986pf} which are comparable to $1/E_{\rm max}^4$ in the range of $E_{\rm max}$ we explore.} which partially obscures the expected scaling for some low-lying energy excitations. In contrast, increasing the volume reduces this noise, 
allowing more reliable extraction of the scaling behavior, as shown in Section~\ref{sec:highV}. We examine $H_{\rm eff}^{\rm (NLO)}$ near the critical coupling in Section~\ref{sec:cc} and extract a prediction for the value of critical coupling using the finite volume theory.

\subsection{Lower volume benchmark: $R=10/2\pi$}\label{sec:lowV}

\begin{figure}[htp!]
	\centering
	\begin{minipage}{.9\textwidth}
	\centering
	\begin{subfigure}[t]{0.5\textwidth}
		\centering
		\includegraphics[trim= 0cm	.05cm	0cm	0cm, clip=true, width=1.15\textwidth]{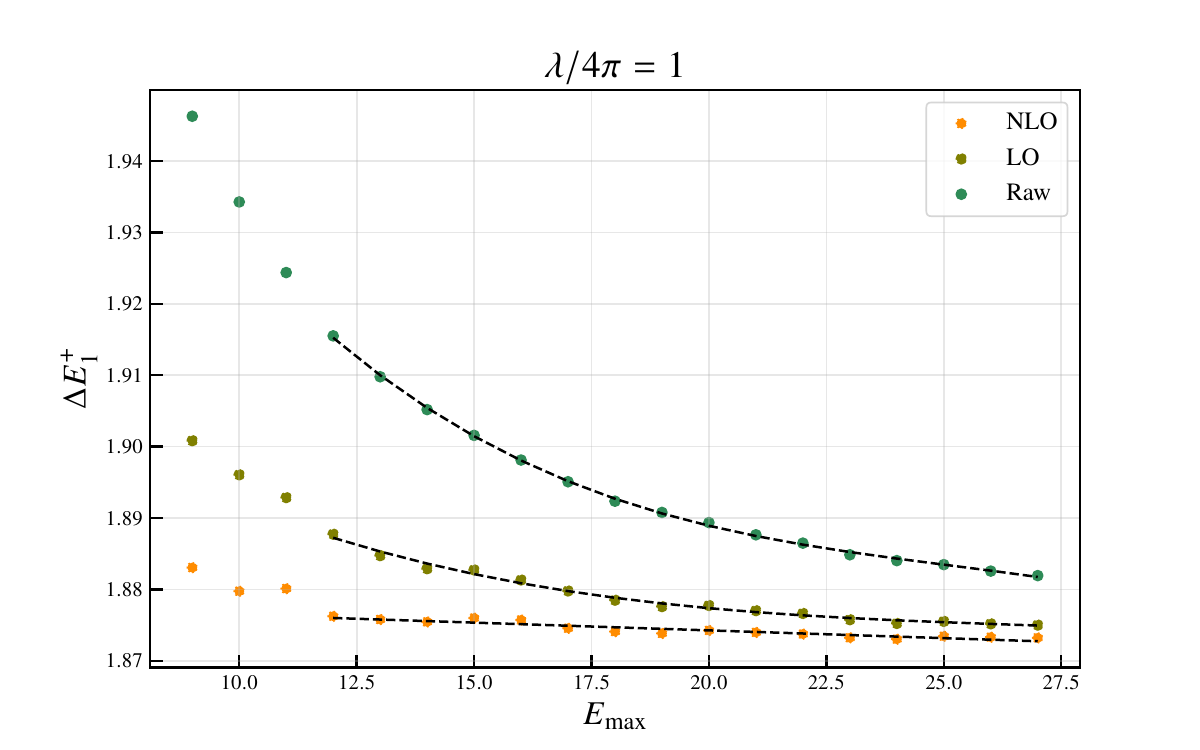}
		\caption{}
		\label{fig:E10_r10_vsEmax}
	\end{subfigure}%
	\begin{subfigure}[t]{0.5\textwidth}
		\centering
		\hspace{-1.2cm}
		\includegraphics[trim= 0cm	.05cm	0cm	0cm, clip=true, width=0.68\textwidth]{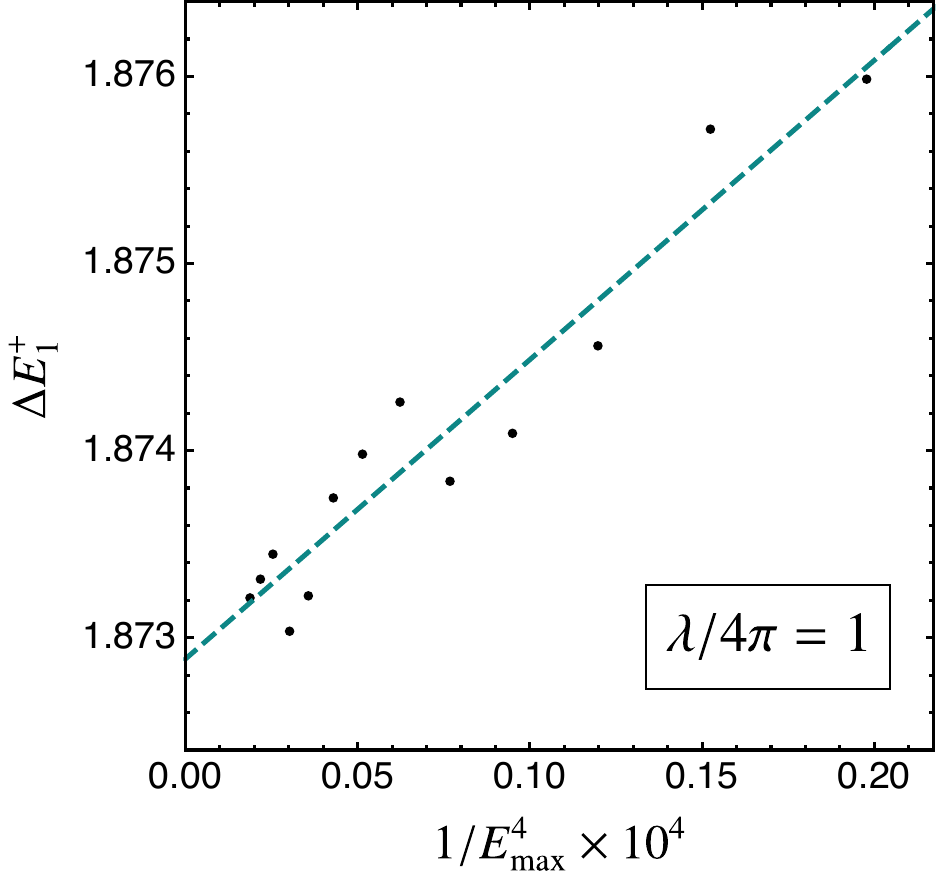}
		\caption{ }
		\label{fig:E10_r10_vsEmax4}
	\end{subfigure}
	\caption{\bf{Energy scaling of the first excited $\mathbb{Z}_2$-even state:} \rm The plot on the left shows the first $\mathbb{Z}_2$-even excitation above the ground state energy, $\Delta E_1^+$, at various values of the cutoff energy $E_{\rm max}$ for $H_{\rm eff}^{\rm (raw)}$ (green), $H_{\rm eff}^{\rm (LO)}$ (light green), and $H_{\rm eff}^{\rm (NLO)}$ (yellow). The black dashed lines indicate the corresponding power-law fits. The plot on the right shows $\Delta E_1^+$ versus $1/E_{\rm max}^4$ for $H_{\rm eff}^{\rm (NLO)}$, with the light blue dashed line indicating the fit.}
	\label{fig:E10_r10_fit}
 \end{minipage}
\end{figure}

We first present the results for the lower volume benchmark, $R = 10 /(2\pi)$. Fig.~\ref{fig:E10_r10_fit} shows the $E_{\rm max}$ scaling of the first excited $\mathbb{Z}_2$-even energy level above the ground state, $\Delta E_1^+ = E_1^+ - E_0$ for all three effective Hamiltonians: $H_{\rm eff}^{\rm (raw)}$, $H_{\rm eff}^{\rm (LO)}$ and $H_{\rm eff}^{\rm (NLO)}$. We fit $\Delta E_1^+$ to a power-law function of $E_{\rm max}$ that will be used for each energy gap $\Delta E_n$:
	\begin{align}
	\Delta E_n
		&= \Delta E_n^\infty + c_p \frac1{ (E_{\rm max})^p} \,,
	\label{eq:fitp}
	\end{align}
using $p= 2, 3, 4$ for $H_{\rm eff}^{\rm (raw)}$, $H_{\rm eff}^{\rm (LO)}$ and $H_{\rm eff}^{\rm (NLO)}$, respectively. Fig.~\ref{fig:E10_r10_vsEmax} shows $\Delta E_1^+$ versus $E_{\rm max}$ for $H_{\rm eff}^{\rm (raw)}$, $H_{\rm eff}^{\rm (LO)}$, and $H_{\rm eff}^{\rm (NLO)}$ as well as their respective power law fits, while Fig.~\ref{fig:E10_r10_vsEmax4} shows $\Delta E_1^+$ for $H_{\rm eff}^{\rm (NLO)}$ and its fit plotted versus $1/E_{\rm max}^4$. 

\begin{figure}[t]
	\centering
	\begin{minipage}{.9\textwidth}
	\centering
	\raisebox{-.0cm}{\includegraphics[trim= 0cm	0cm	0cm	0cm, clip=true,  width=.24\textwidth]{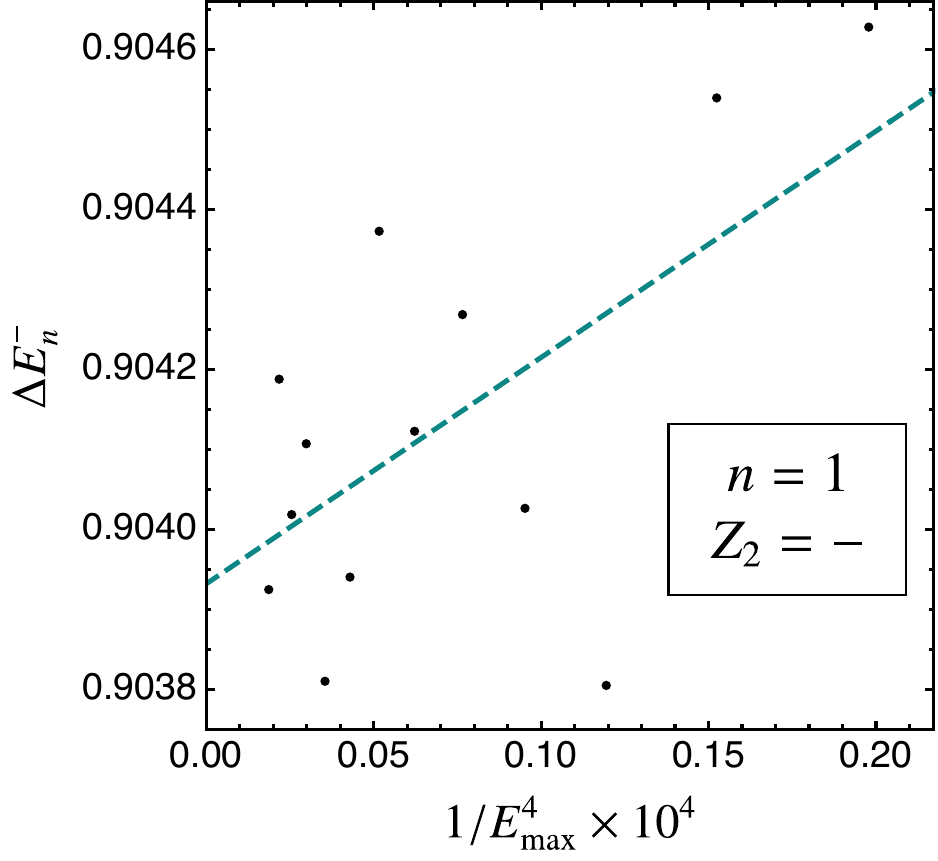}}
	\raisebox{-.0cm}{\includegraphics[trim= 0cm	0cm	0cm	0cm, clip=true,  width=.24\textwidth]{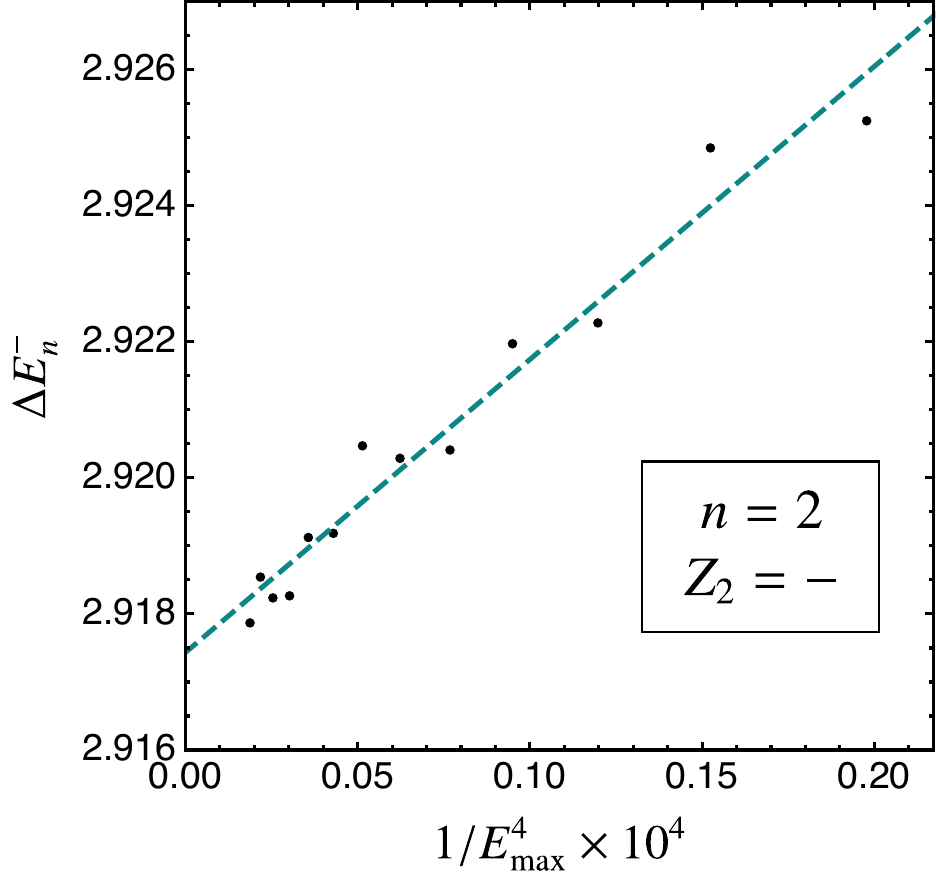}}
	\raisebox{-.0cm}{\includegraphics[trim= 0cm	0cm	0cm	0cm, clip=true,  width=.24\textwidth]{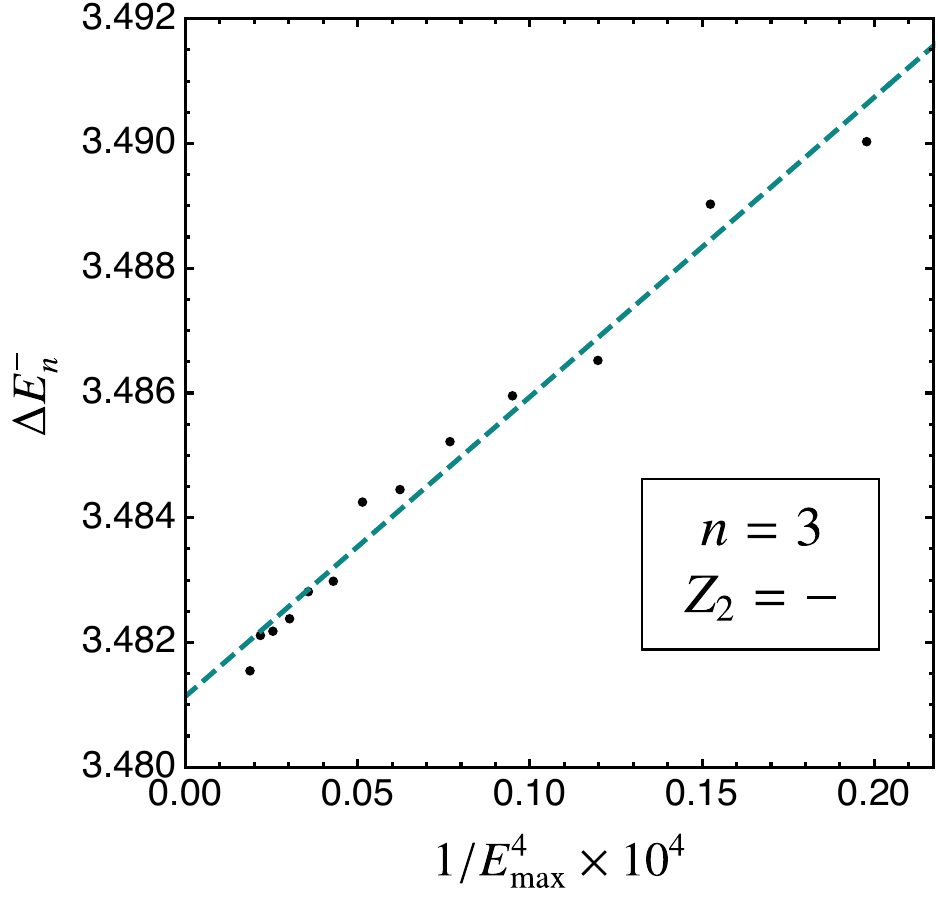}}
	\raisebox{-.0cm}{\includegraphics[trim= 0cm	0cm	0cm	0cm, clip=true,  width=.24\textwidth]{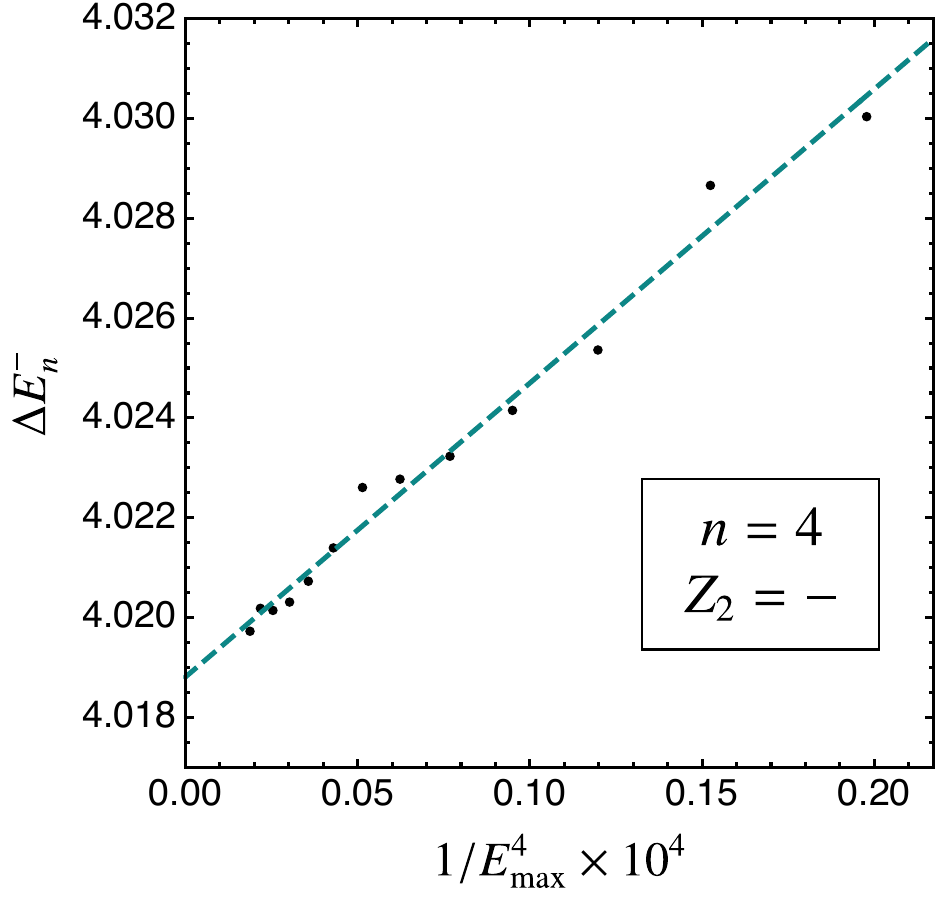}}
	\hspace*{0.5mm}
	\raisebox{-.0cm}{\includegraphics[trim= 0cm	0cm	0cm	0cm, clip=true,  width=.24\textwidth]{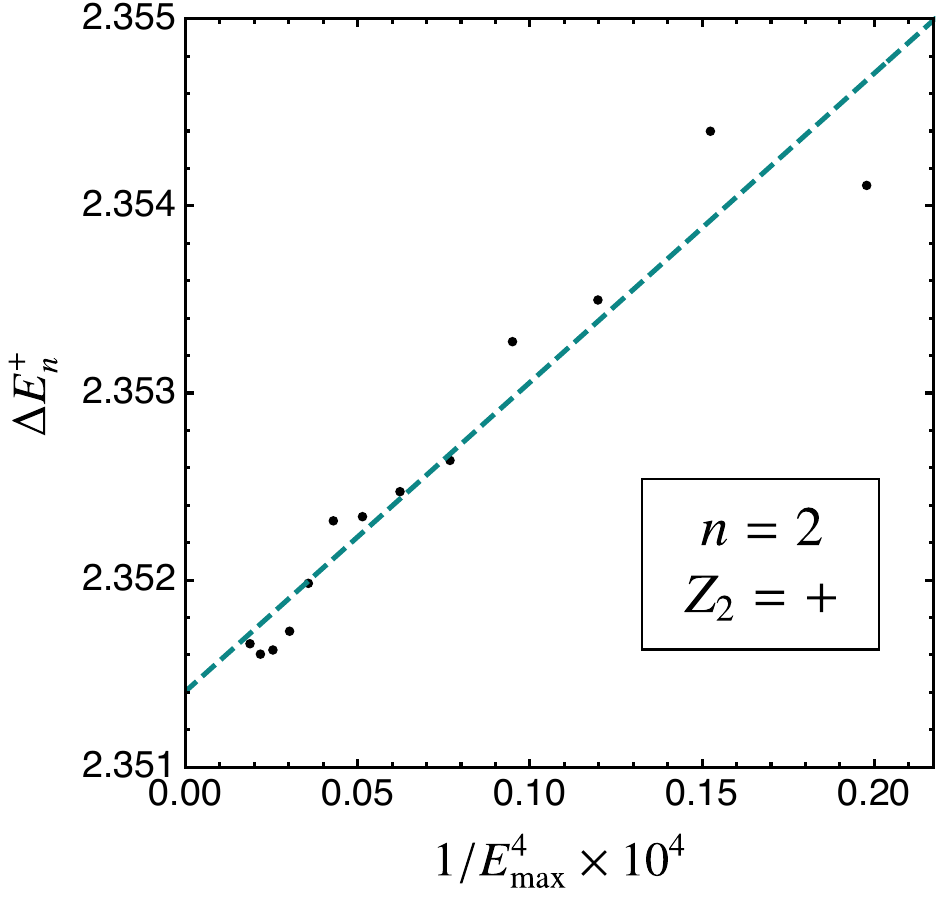}}
	\raisebox{-.0cm}{\includegraphics[trim= 0cm	0cm	0cm	0cm, clip=true,  width=.24\textwidth]{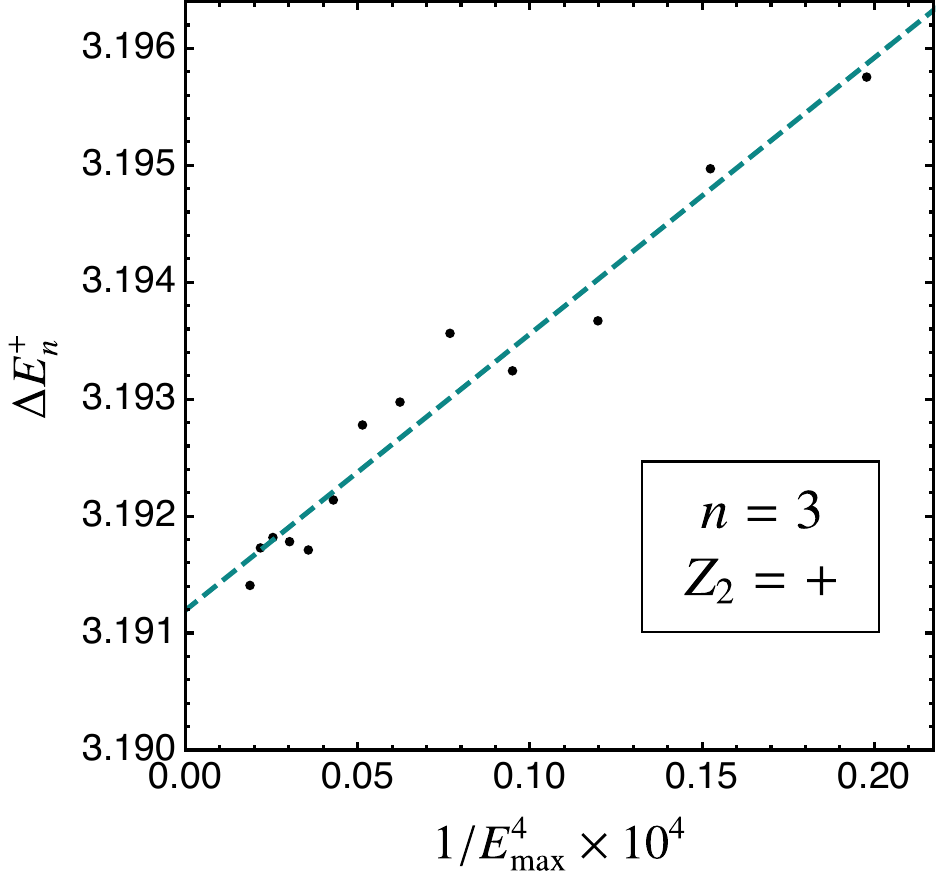}}
	\raisebox{-.0cm}{\includegraphics[trim= 0cm	0cm	0cm	0cm, clip=true,  width=.24\textwidth]{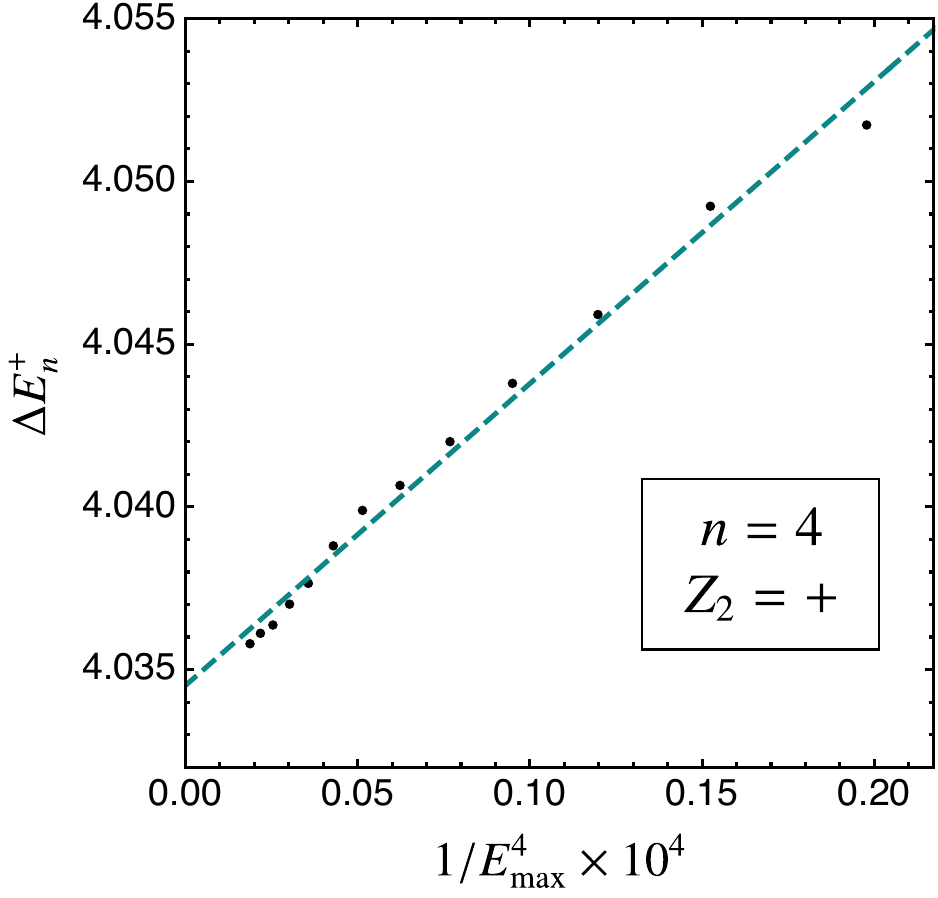}}
	\raisebox{-.0cm}{\includegraphics[trim= 0cm	0cm	0cm	0cm, clip=true,  width=.24\textwidth]{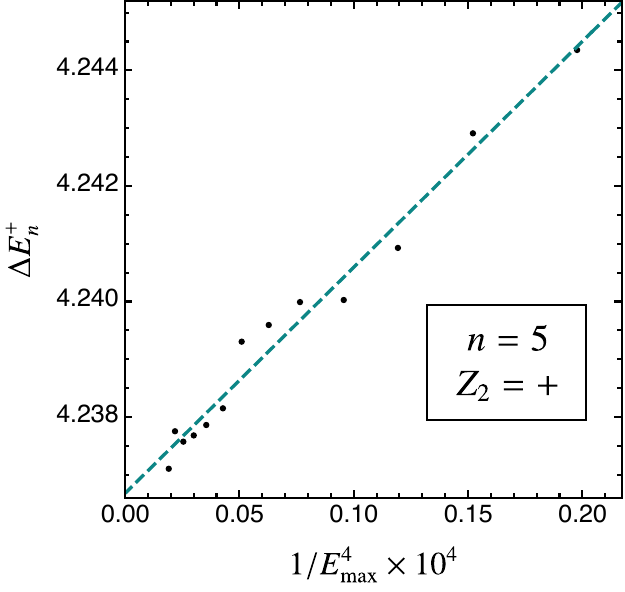}}

	\caption{\bf Energy scaling of higher excitation levels: \rm Energy gaps $\Delta E_n$ are shown versus $1/E_{\rm max}^4$ for the NLO theory. The first few excited energy levels are plotted for both $\mathbb{Z}_2$-even (top row) and $\mathbb{Z}_2$-odd (bottom row) eigenstates. In all panels, $R=10/2\pi,\ \lambda/4\pi = 1$. Numerical noise obscures the scaling behavior of $\Delta E_1^-$ in the top left panel. 
	}

	\label{fig:En-r10}
	\end{minipage}
\end{figure}

To extract the scaling behavior more clearly, we examine other excited states, as shown in Fig.~\ref{fig:En-r10}. We plot the excitation energies, $\Delta E_n = E_n - E_0$, where $E_n$ is the $n^{\rm th}$ excited energy level above the ground state. States belonging to the $\mathbb{Z}_2$-even and $\mathbb{Z}_2$-odd sectors are labelled by $\Delta E_n^+$ and $\Delta E_n^-$, respectively.  
For the excited states above $\Delta E_1^+$, the expected $1/E_{\rm max}^4$ scaling is evident and persists for several excitation levels. 

The upper left panel of Fig.~\ref{fig:En-r10} displays the scaling of the first excited energy gap above the ground state, $\Delta E_1^-$. For this eigenstate, numerical noise dominates over the $1/E_{\rm max}^4$ scaling. In order to study the behavior of the first excited state above the ground state to this level of precision, one must use the higher volume benchmark, see Sec.~\ref{sec:highV}. 
We suggest that the noise reduction for high excited states arises because they converge more slowly than $\Delta E_1^-$, which remains nearly flat over the accessible range of $E_{\rm max}$. The scaling of $\Delta E_1^-$ with $E_{\rm max}$ is therefore obscured by numerical noise. In contrast, the slower convergence of $\Delta E_n$ at higher $n$ produces a visible downward trend, allowing the scaling behavior to emerge despite a comparable level of noise.\footnote{Note the difference in range of the vertical axis between Fig.~\ref{fig:E10_r10_fit} and the upper left panel of Fig.~\ref{fig:En-r10}. }

\begin{figure}[h!]
	\centering
	\begin{minipage}{.9\textwidth}
	\centering
	\raisebox{-.0cm}{\includegraphics[trim= 0cm	0cm	0cm	0cm, clip=true,  width=0.32\textwidth]{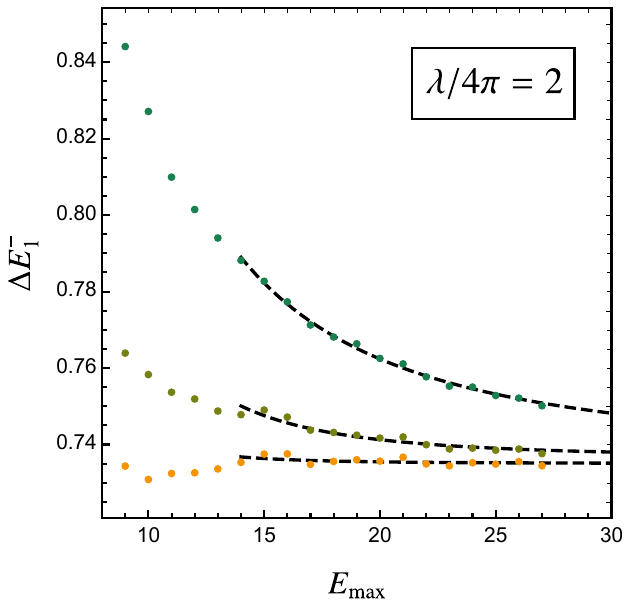}}
	\raisebox{-.0cm}{\includegraphics[trim= 0cm	0cm	0cm	0cm, clip=true,  width=0.32\textwidth]{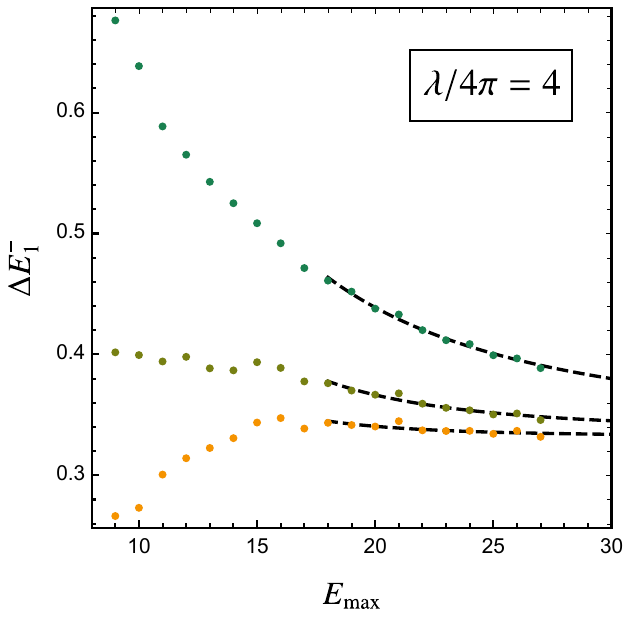}}
	\raisebox{-.0cm}{\includegraphics[trim= 0cm	0cm	0cm	0cm, clip=true,  width=0.32\textwidth]{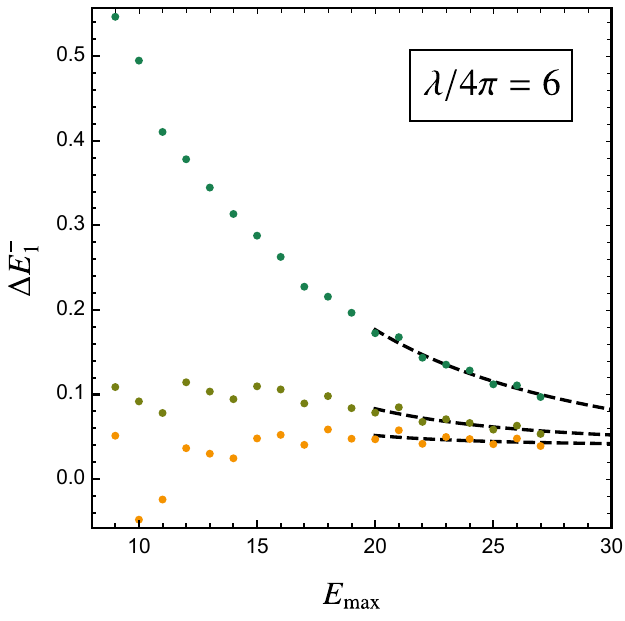}}
	\hspace*{-4mm}
	\raisebox{-.0cm}{\includegraphics[trim= 0cm	0cm	0cm	0cm, clip=true,  width=0.32\textwidth]{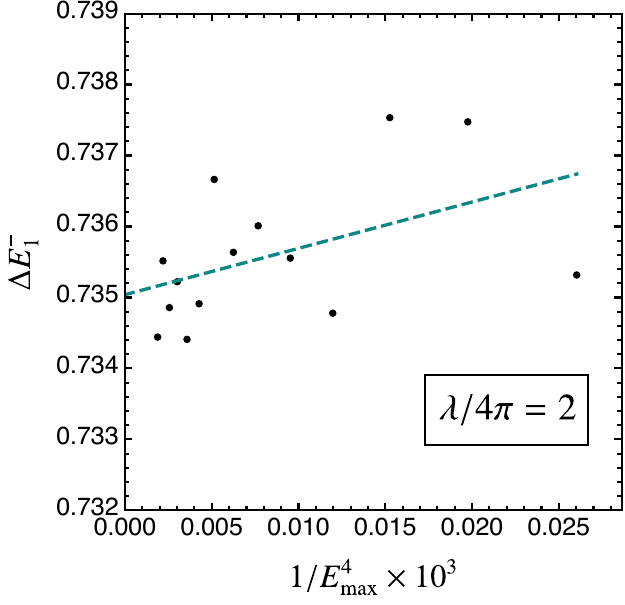}}
	\raisebox{-.0cm}{\includegraphics[trim= 0cm	0cm	0cm	0cm, clip=true,  width=0.32\textwidth]{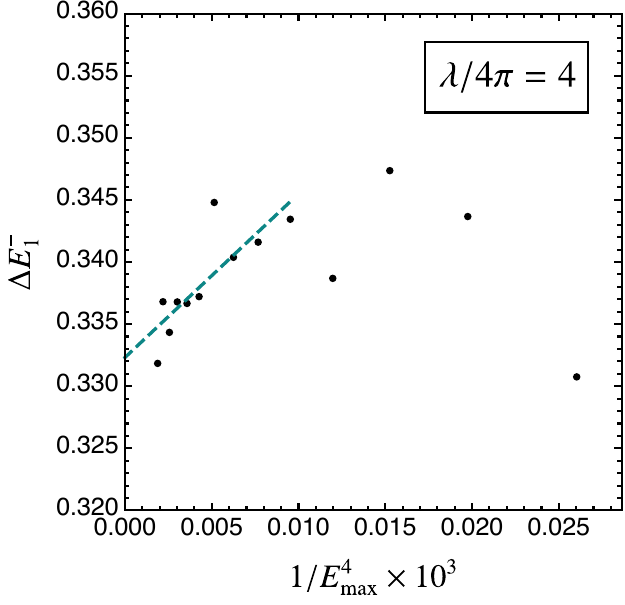}}
	\raisebox{-.0cm}{\includegraphics[trim= 0cm	0cm	0cm	0cm, clip=true,  width=0.32\textwidth]{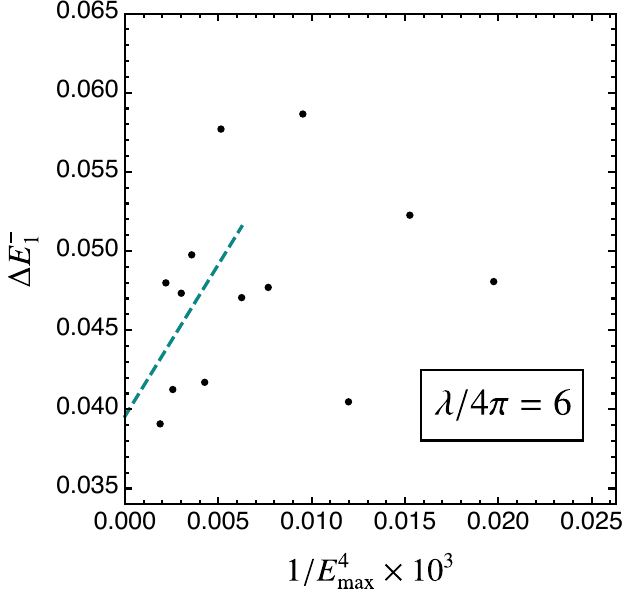}}
	\caption{\bf Scaling at stronger coupling (odd sector): \rm The first $\mathbb{Z}_2$-odd energy gap is shown versus $E_{\rm max}$ for increasing values of the coupling. Fits are applied only in the regime where the scaling behavior emerges in the high-energy tails. For $\lambda/4\pi=6$, numerical noise dominates, obscuring the fit. 
	} 
	\label{fig:E1_r10_vary_g_odd}
	\end{minipage}
\end{figure}

\begin{figure}[h!]
	\centering
	\begin{minipage}{.9\textwidth}
	\centering
	\raisebox{-.0cm}{\includegraphics[trim= 0cm	0cm	0cm	0cm, clip=true,  width=0.32\textwidth]{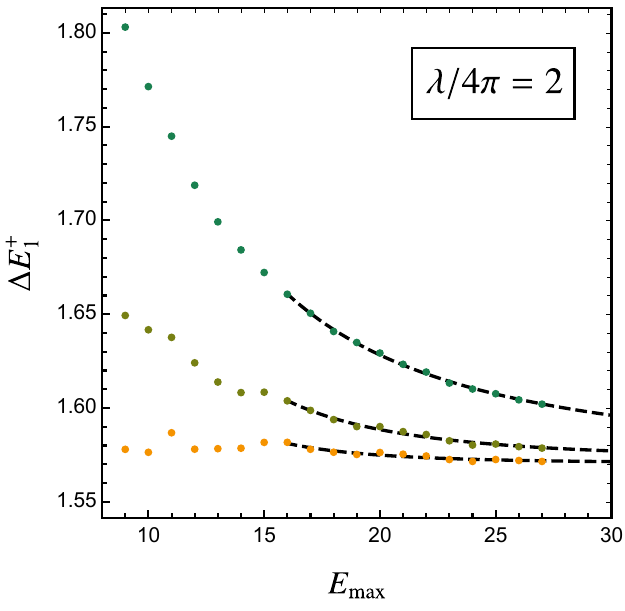}}
	\raisebox{-.0cm}{\includegraphics[trim= 0cm	0cm	0cm	0cm, clip=true,  width=0.32\textwidth]{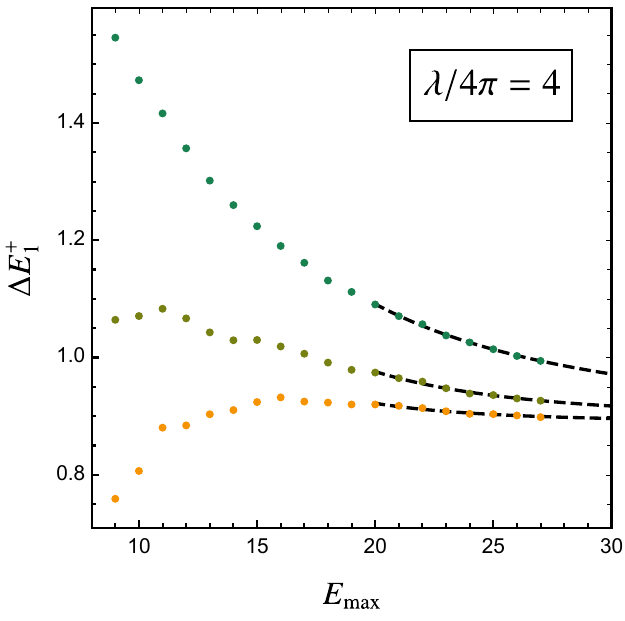}}
	\raisebox{-.0cm}{\includegraphics[trim= 0cm	0cm	0cm	0cm, clip=true,  width=0.32\textwidth]{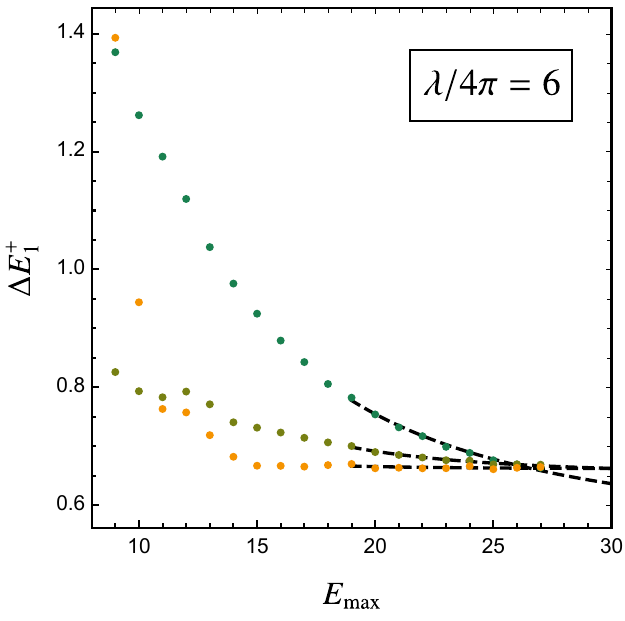}}
	\hspace*{-4mm}
	\raisebox{-.0cm}{\includegraphics[trim= 0cm	0cm	0cm	0cm, clip=true,  width=0.32\textwidth]{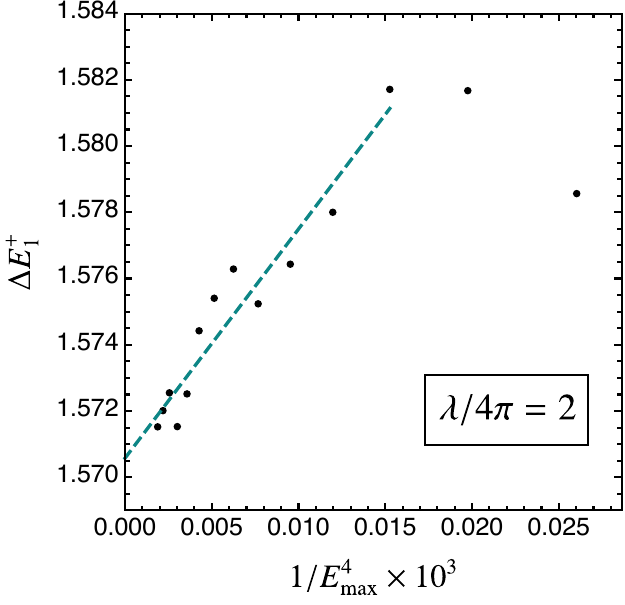}}
	\raisebox{-.0cm}{\includegraphics[trim= 0cm	0cm	0cm	0cm, clip=true,  width=0.32\textwidth]{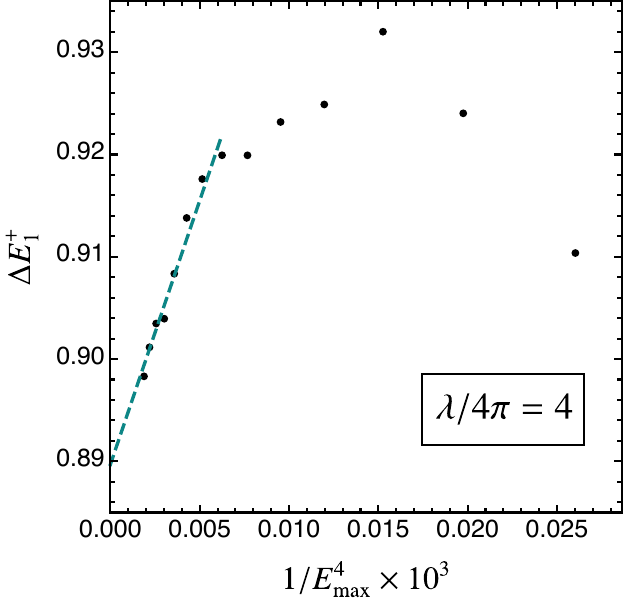}}
	\hspace*{-1mm}
	\raisebox{-.0cm}{\includegraphics[trim= 0cm	0cm	0cm	0cm, clip=true,  width=0.32\textwidth]{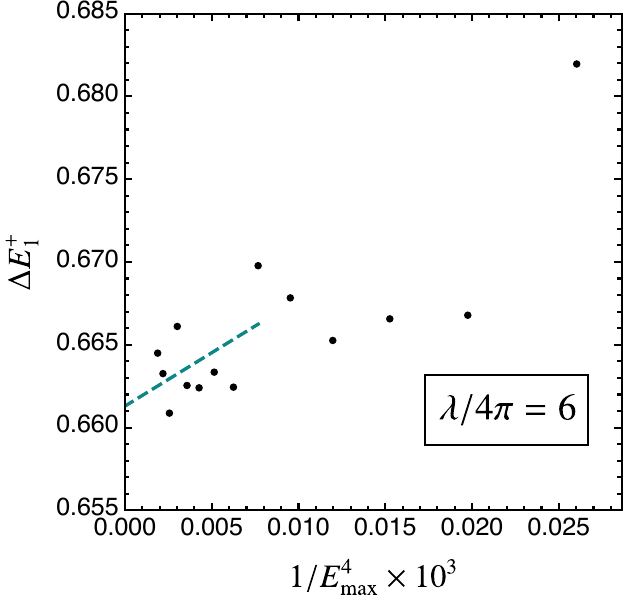}}
	\caption{\bf Scaling at stronger coupling (even sector): \rm The first $\mathbb{Z}_2$-even energy gap is shown vs. $E_{\rm max}$ for increasing coupling. Fits are applied only in the regime where scaling emerges in the high-energy tails. For $\lambda/4\pi=6$, numerical noise dominates, obscuring the fit.} 
	\label{fig:E1_r10_vary_g_even}
	\end{minipage}
\end{figure}
Next, we examine the convergence as we increase the coupling. Fig.~\ref{fig:E1_r10_vary_g_odd} shows $\Delta E_1^-$ for increasing values of $\lambda/4\pi$. Numerical noise stays high, though we do see some rough $1/E_{\rm max}^4$ scaling in the high energy tails. Convergence deteriorates near the critical coupling, where $E_1^-$ and the ground state become nearly degenerate. In contrast, Fig.~\ref{fig:E1_r10_vary_g_even} shows that $\Delta E_1^+$ retains a clear $1/E_{\rm max}^4$ scaling even at stronger coupling, up to $\lambda/4\pi =4$. For $\lambda/4\pi=6$, the scaling becomes difficult to disentangle from the noise.

\subsection{Higher volume benchmark: $R=20/2\pi$}
\label{sec:highV}

\begin{figure}[h!]
	\centering
	\begin{minipage}{.9\textwidth}
	\centering
	\begin{subfigure}[t]{0.5\textwidth}
		\includegraphics[trim=0cm .05cm 0cm 0cm, clip=true, width=1.15
		\textwidth]{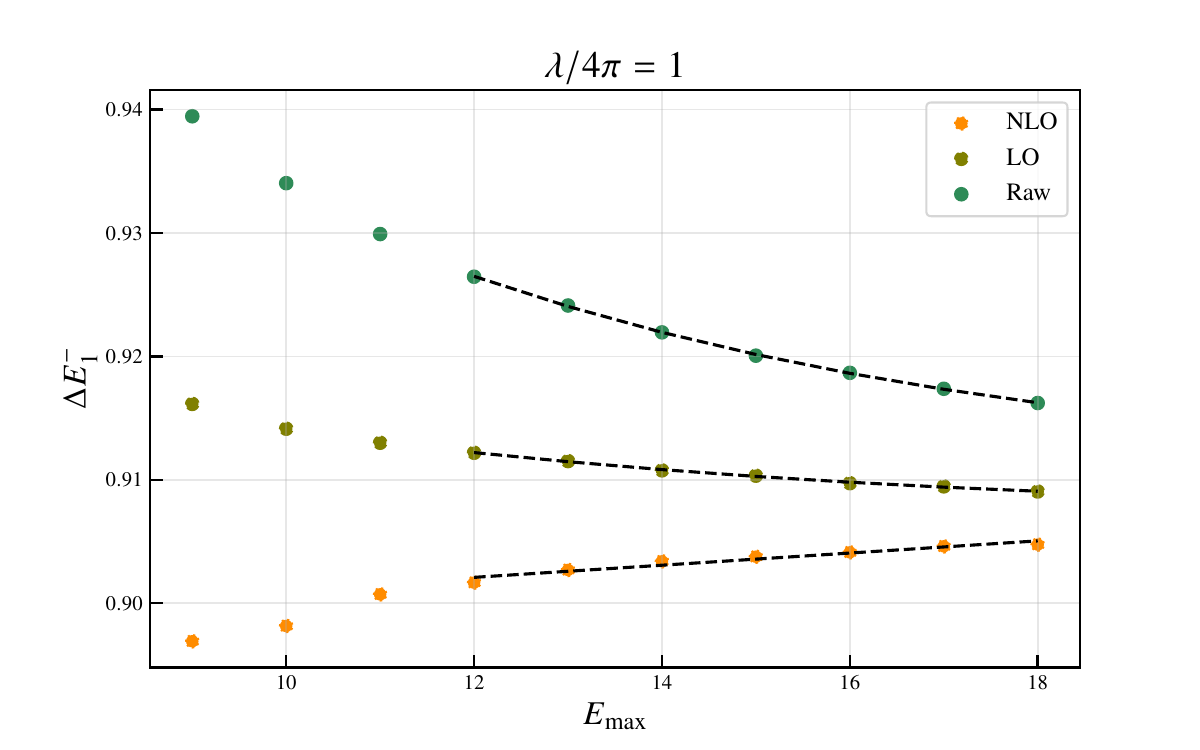}
		\caption{}
		\label{fig:e10_r20_left}
	\end{subfigure}%
	\begin{subfigure}[t]{0.5\textwidth}
		\centering
		\hspace{-1.2cm}
		\includegraphics[trim=0cm .05cm 0cm 0cm, clip=true, width=0.68\textwidth]{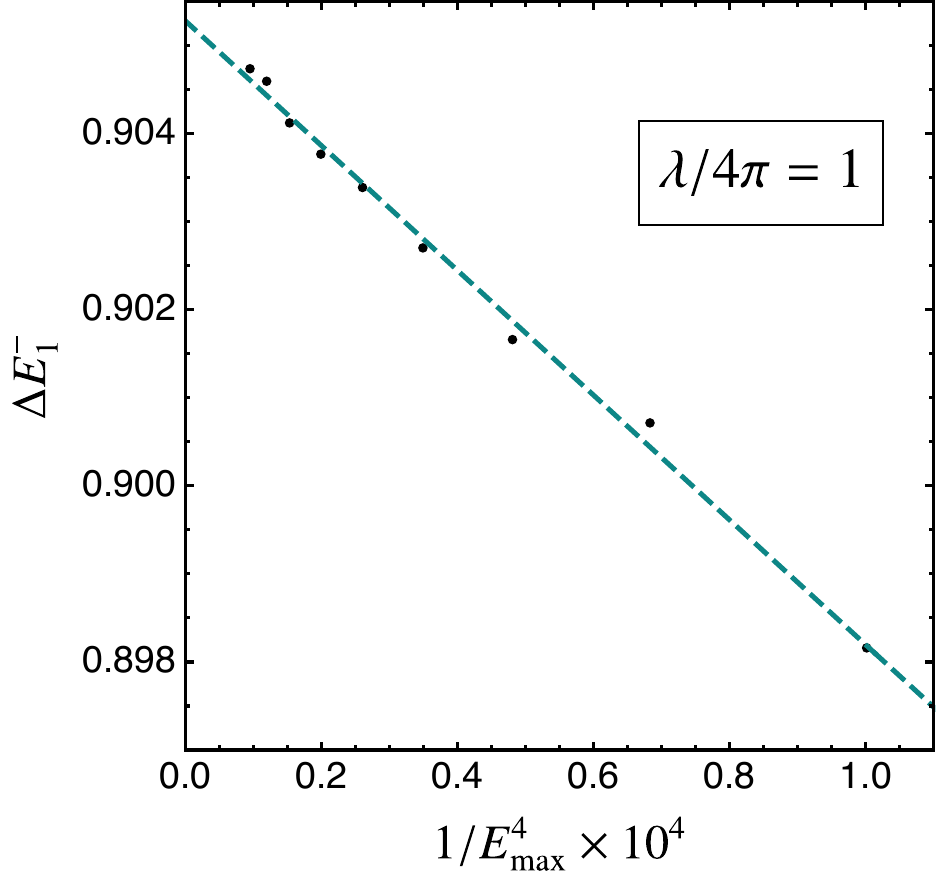}
		\caption{ }
		\label{fig:e10_r20_right}
	\end{subfigure}
	\caption{\bf {Energy scaling of the first excited state: }\rm The plot on the left shows the first energy excitation above the ground state energy, $\Delta E_1^-$, at various values of the cutoff energy $E_{\rm max}$ for $H_{\rm eff}^{\rm (raw)}$ (green), $H_{\rm eff}^{\rm (LO)}$ (light green), and $H_{\rm eff}^{\rm (NLO)}$ (yellow). Black dashed lines indicate the corresponding power-law fits. The plot on the right shows $\Delta E_1^-$ versus $1/E_{\rm max}^4$ for $H_{\rm eff}^{\rm (NLO)}$, with the light blue dashed line indicating the fit. }
	\label{fig:E10_r20_fit}
	\end{minipage}
\end{figure}

We now turn to the results at higher volume, putting the theory on a circle of radius $ R = 20/2\pi$. As shown in Fig.~\ref{fig:E10_r20_fit}, the increased volume leads to a clearer scaling behavior, with significantly reduced noise compared to the low-volume case. This improvement enables more robust extraction of the expected $ 1/E_{\rm max}^4$ scaling, particularly for the first excited energy gap $\Delta E_1^-$, using the same power-law fitting function in~\eqref{eq:fitp}. The number of states for each $E_{\rm max}$ grows with $R$, and so using the same computational resources as in Sec.~\ref{sec:lowV} only allows for a maximum energy cutoff of $E_{\rm max} = 18$. 

\begin{figure}[h!]
	\centering
	\begin{minipage}{.9\textwidth}
	\centering
	\raisebox{-.0cm}{\includegraphics[trim= 0cm	0cm	0cm	0cm, clip=true,  width=.24\textwidth]{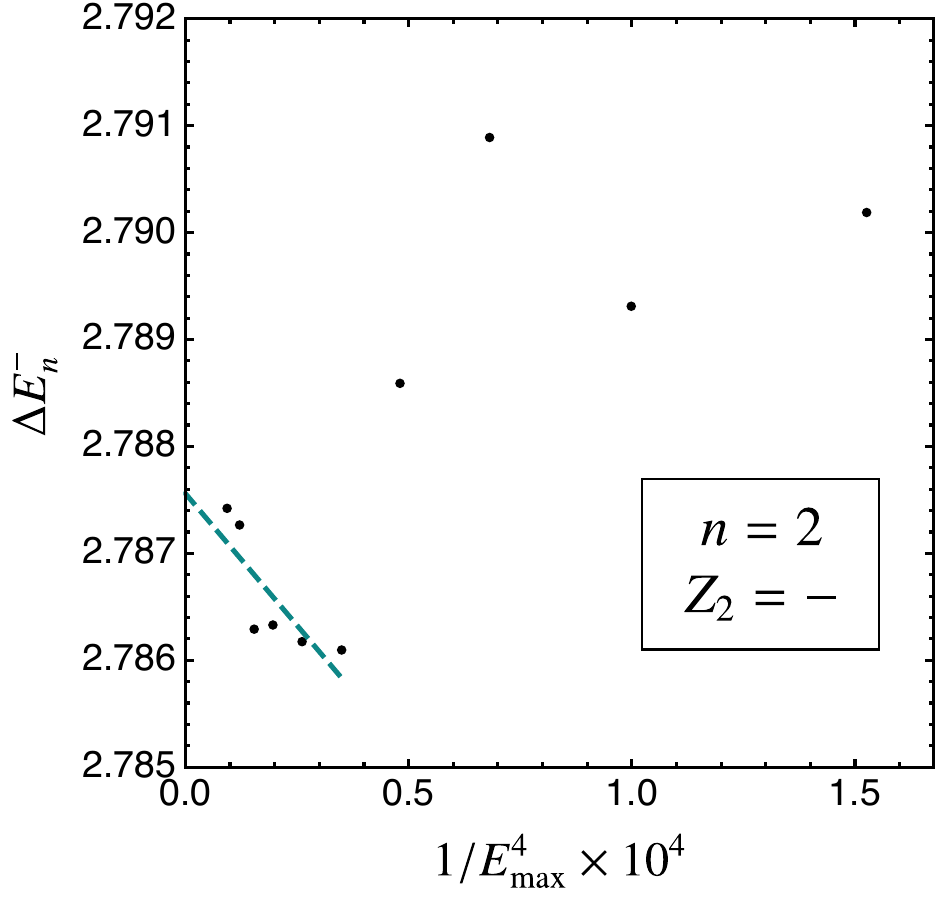}}
	\raisebox{-.0cm}{\includegraphics[trim= 0cm	0cm	0cm	0cm, clip=true,  width=.24\textwidth]{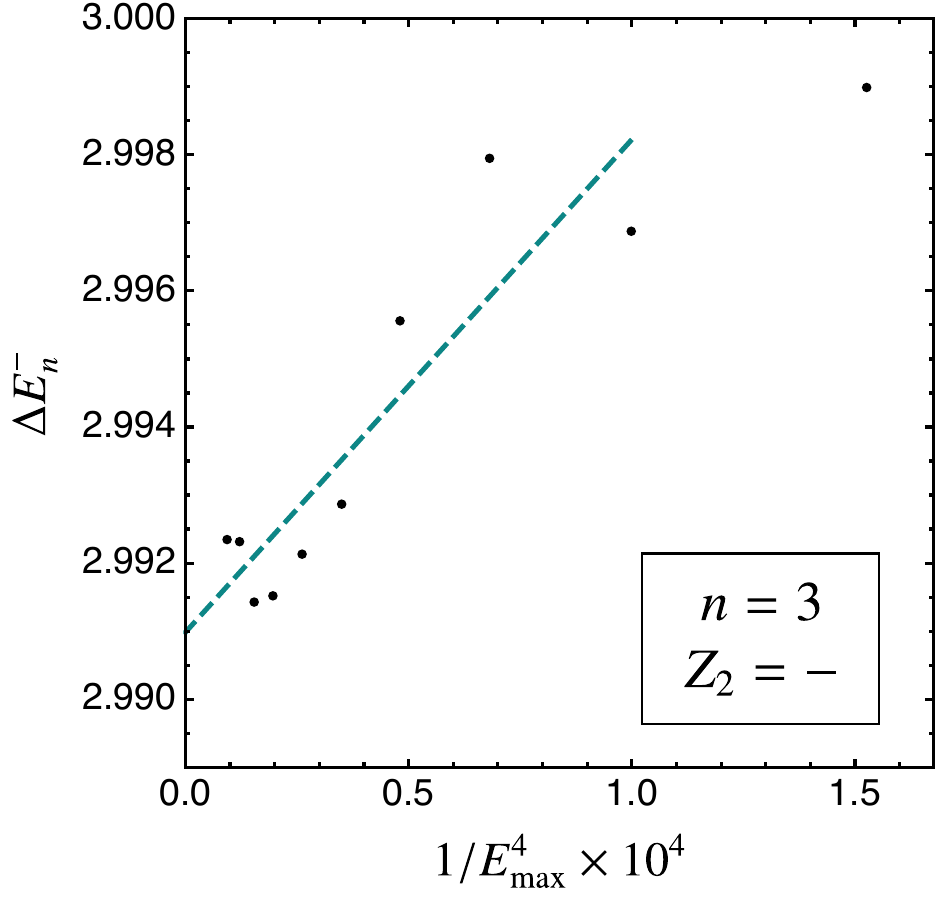}}
	\raisebox{-.0cm}{\includegraphics[trim= 0cm	0cm	0cm	0cm, clip=true,  width=.24\textwidth]{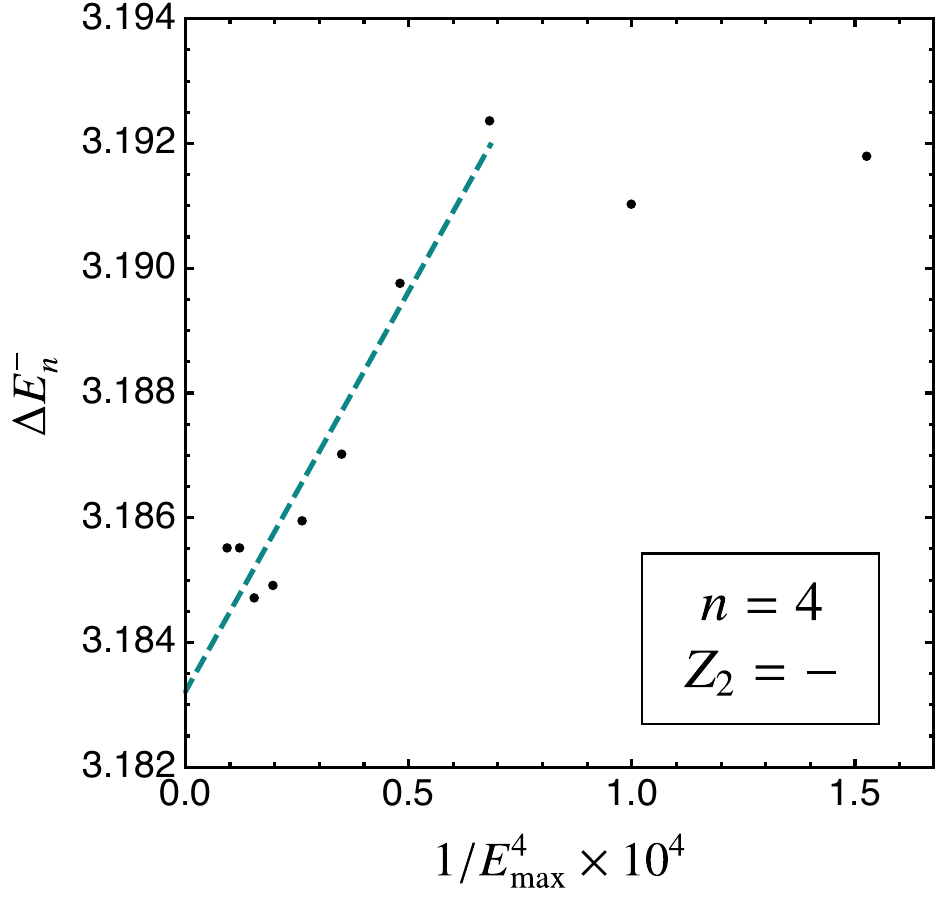}}
	\raisebox{-.0cm}{\includegraphics[trim= 0cm	0cm	0cm	0cm, clip=true,  width=.24\textwidth]{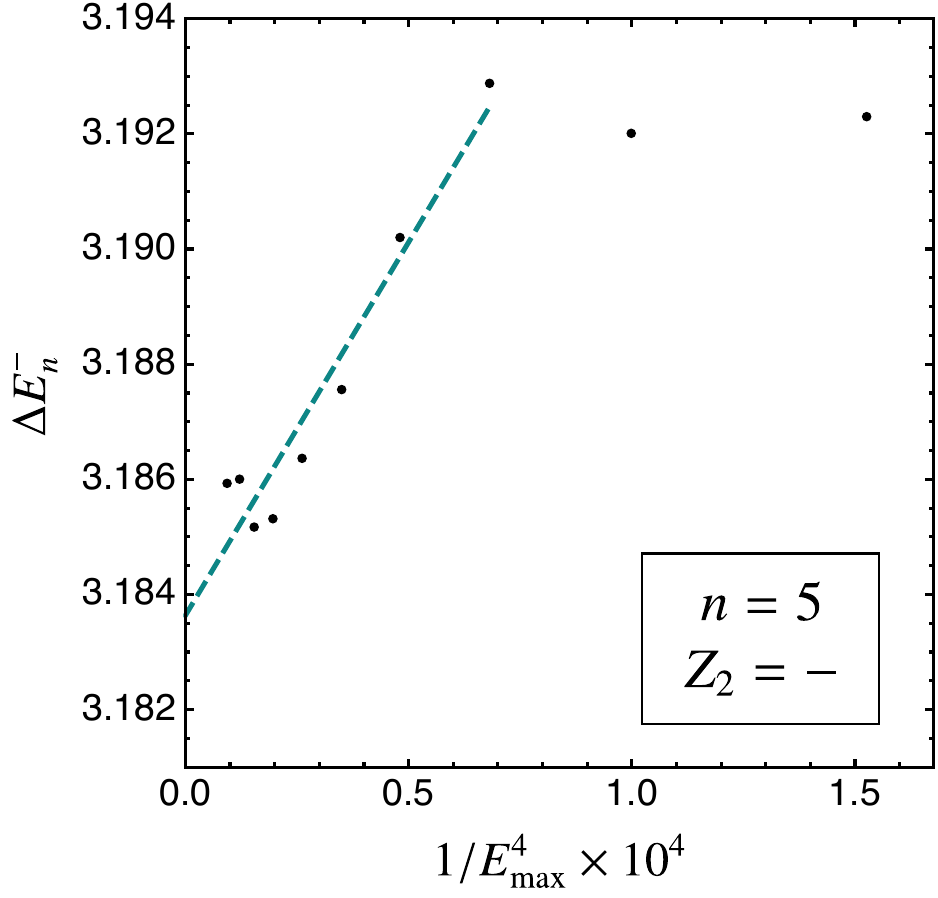}}
	\raisebox{-.0cm}{\includegraphics[trim= 0cm	0cm	0cm	0cm, clip=true,  width=.24\textwidth]{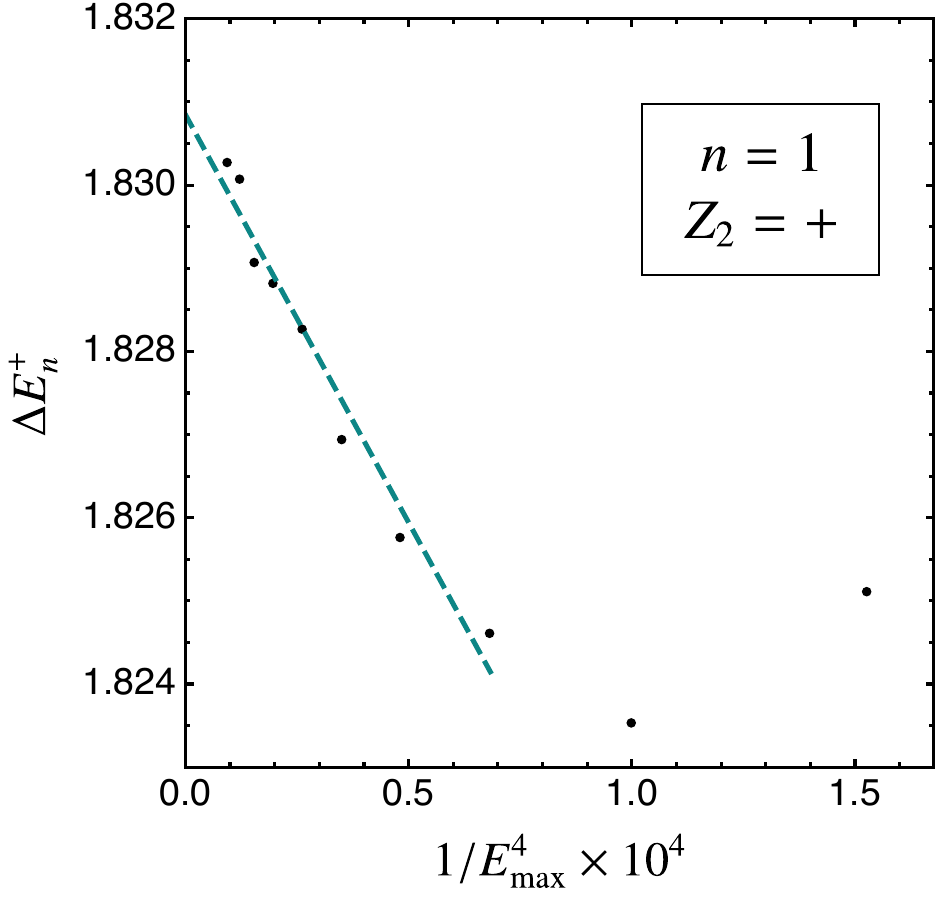}}
	\raisebox{-.0cm}{\includegraphics[trim= 0cm	0cm	0cm	0cm, clip=true,  width=.24\textwidth]{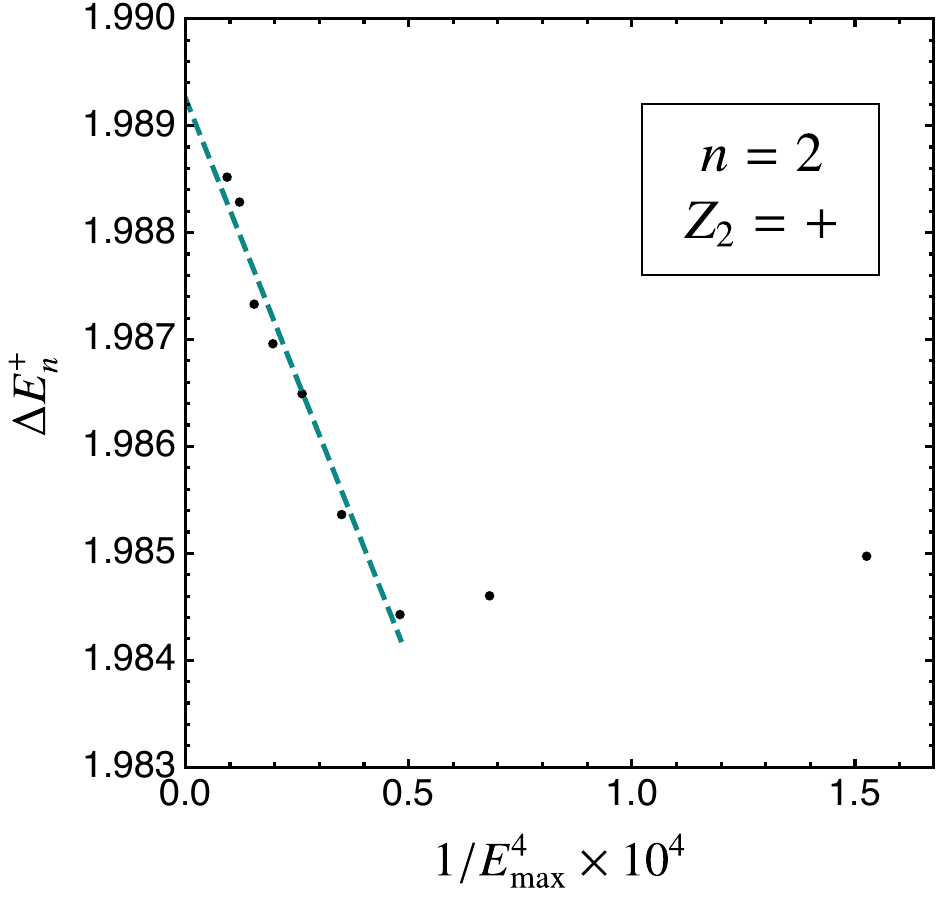}}
	\raisebox{-.0cm}{\includegraphics[trim= 0cm	0cm	0cm	0cm, clip=true,  width=.24\textwidth]{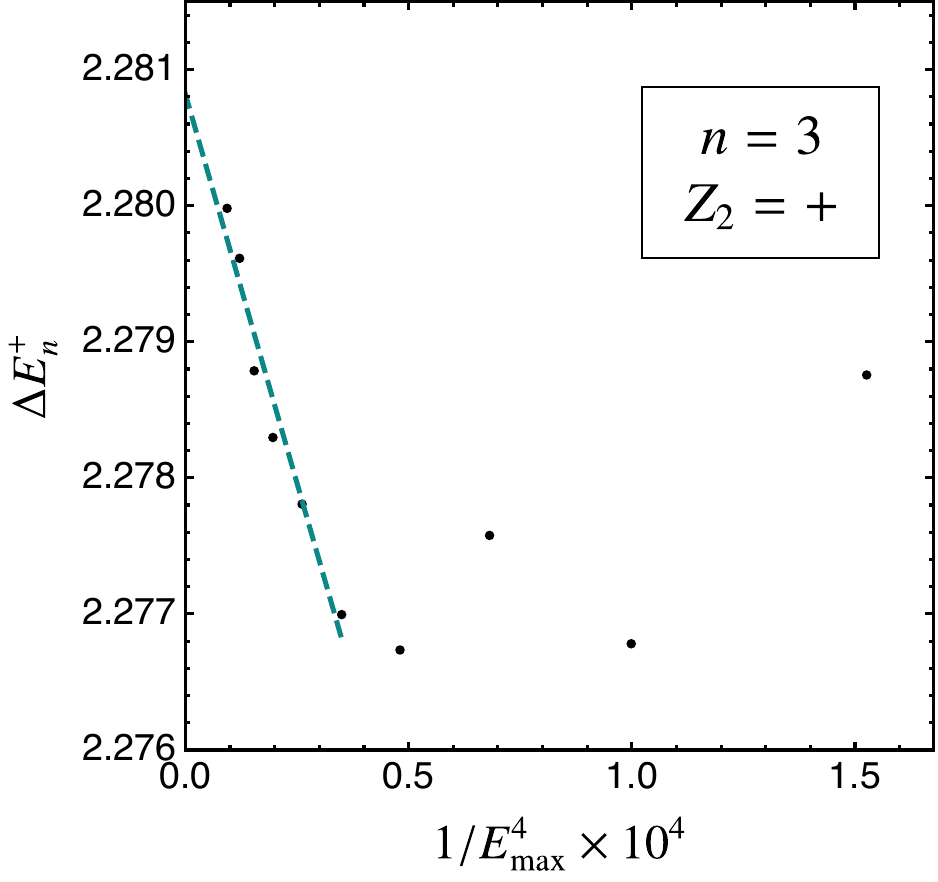}}
	\raisebox{-.0cm}{\includegraphics[trim= 0cm	0cm	0cm	0cm, clip=true,  width=.24\textwidth]{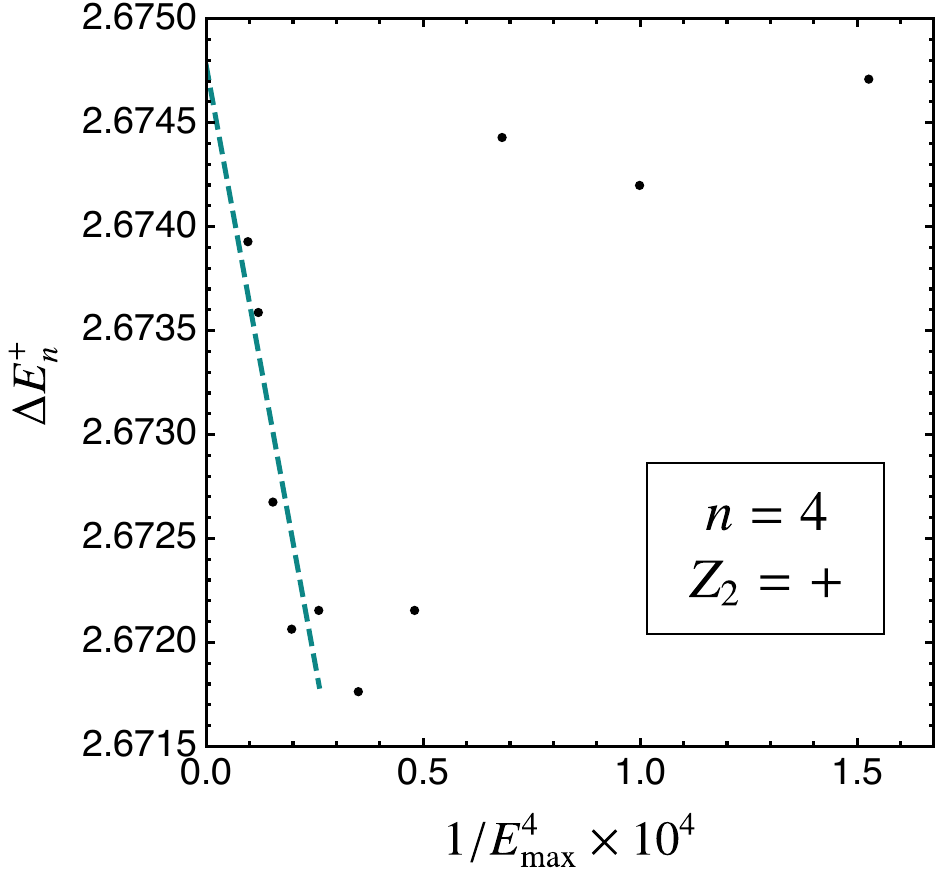}}
	\caption{\bf Energy scaling of higher excitation levels: \rm Higher excited energy levels relative to the ground state energy, $\Delta E_n$, are shown versus $1/E_{\rm max}^4$ for the NLO theory. The  first few exciations are plotted for both $\mathbb{Z}_2$-even (top row) and $\mathbb{Z}_2$-odd (bottom row) eigenstates. In all panels $R=20/2\pi,$ and $\lambda/4\pi = 1$.
	}
	\label{fig:En-r20}
	\end{minipage}
\end{figure}

\begin{figure}[h!]
	\centering
	\begin{minipage}{.9\textwidth}
	\centering
	\raisebox{-.0cm}{\includegraphics[trim= 0cm	0cm	0cm	0cm, clip=true,  width=0.32\textwidth]{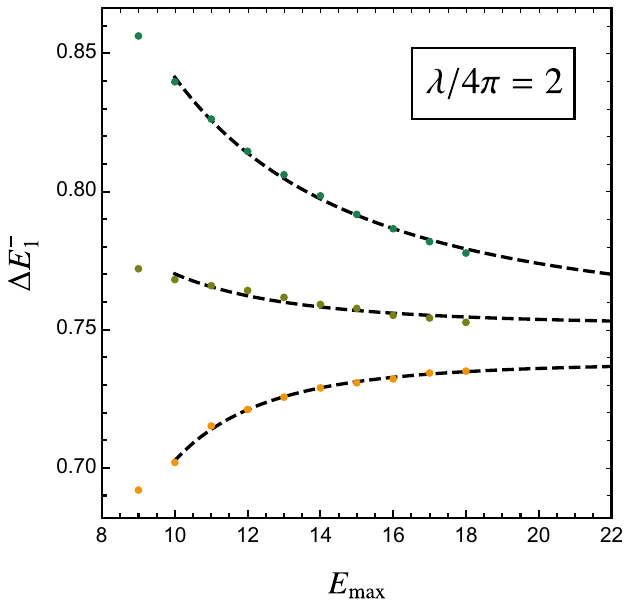}}
	\raisebox{-.0cm}{\includegraphics[trim= 0cm	0cm	0cm	0cm, clip=true,  width=0.32\textwidth]{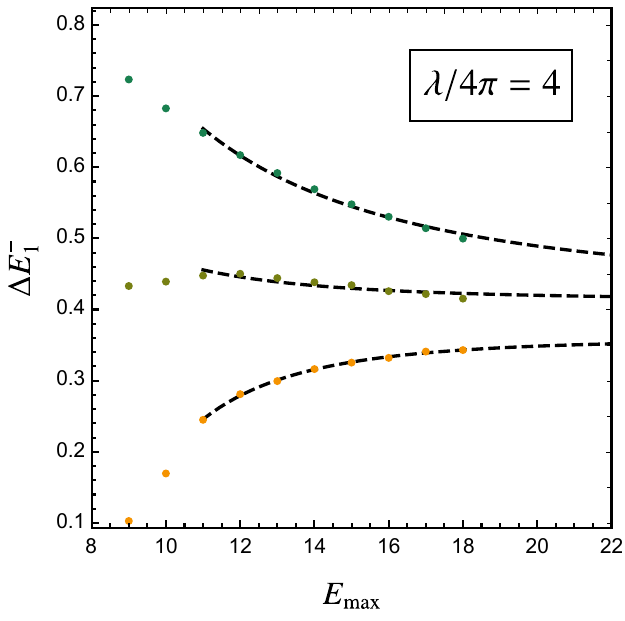}}
	\raisebox{-.0cm}{\includegraphics[trim= 0cm	0cm	0cm	0cm, clip=true,  width=0.32\textwidth]{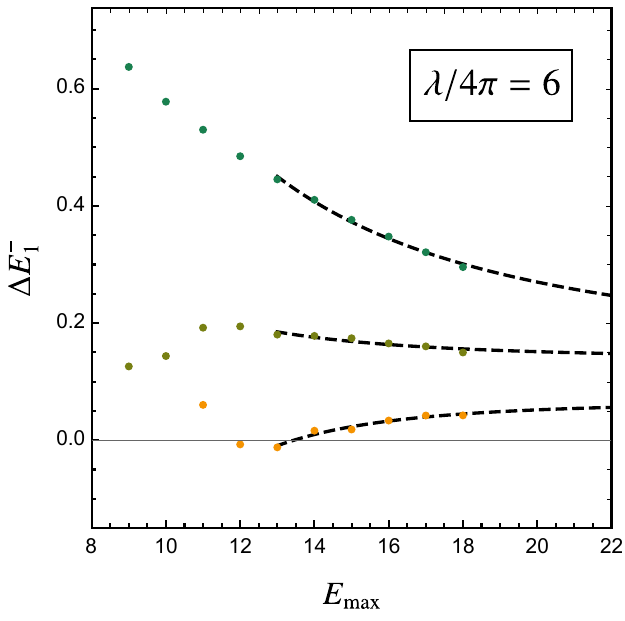}}
	\raisebox{-.0cm}{\includegraphics[trim= 0cm	0cm	0cm	0cm, clip=true,  width=0.32\textwidth]{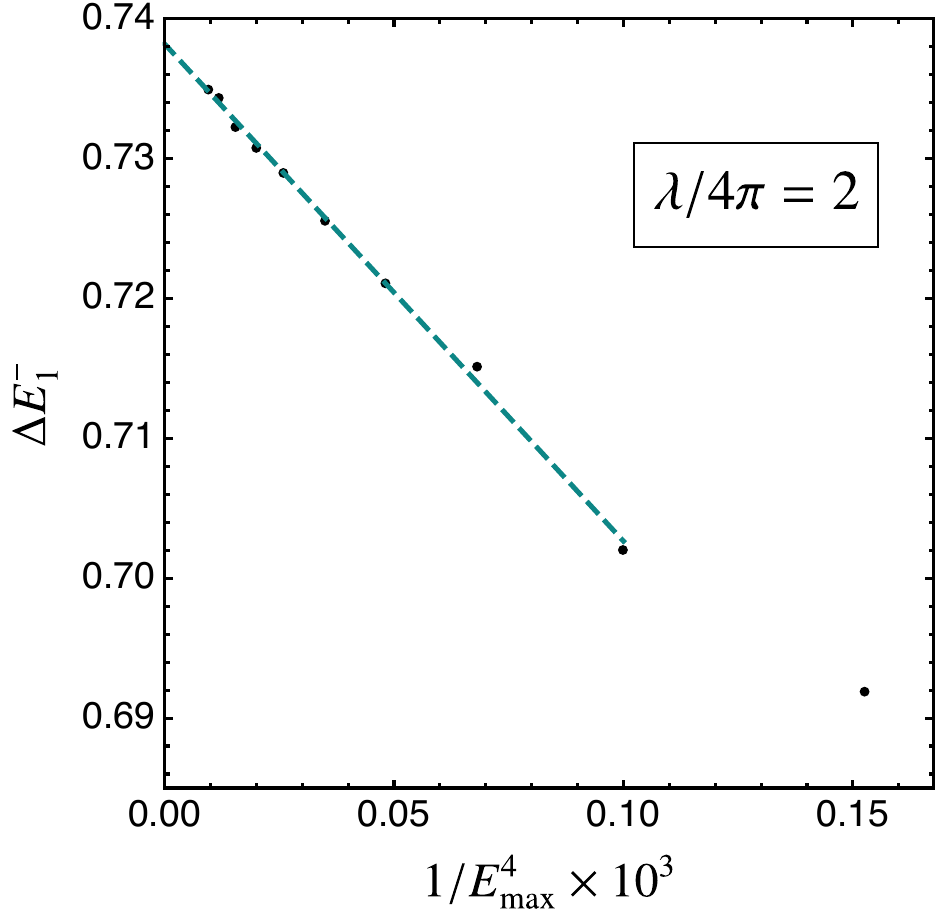}}
	\raisebox{-.0cm}{\includegraphics[trim= 0cm	0cm	0cm	0cm, clip=true,  width=0.32\textwidth]{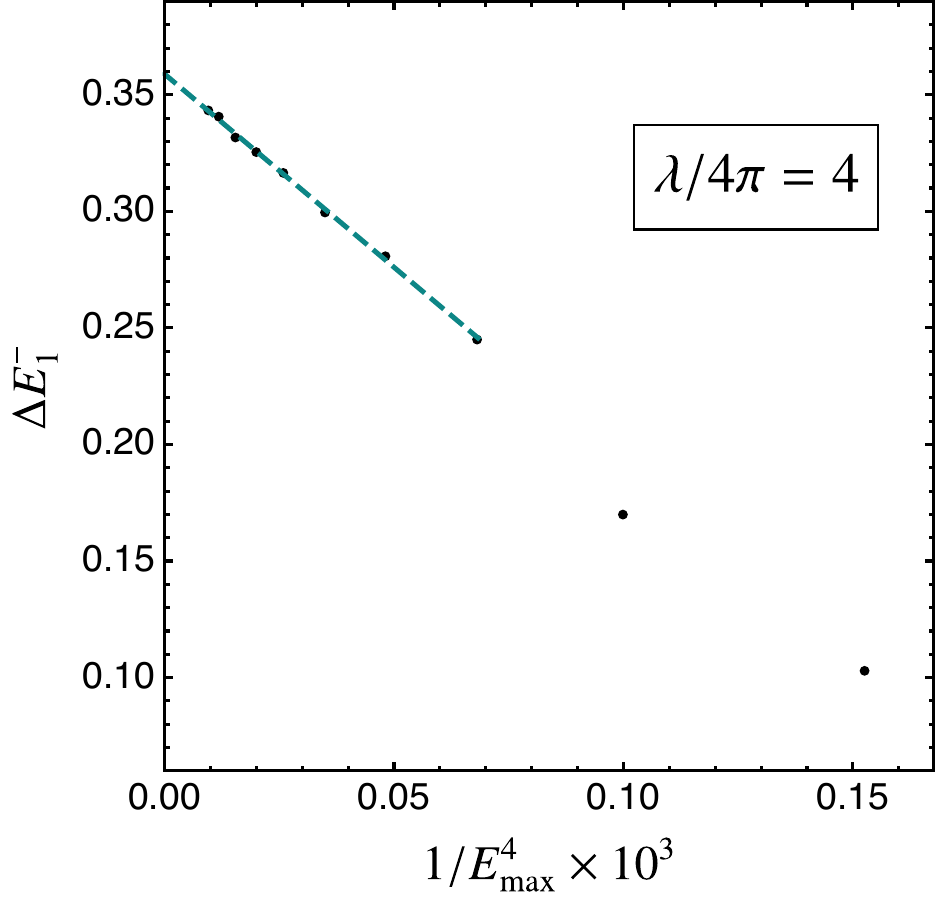}}
	\raisebox{-.0cm}{\includegraphics[trim= 0cm	0cm	0cm	0cm, clip=true,  width=0.32\textwidth]{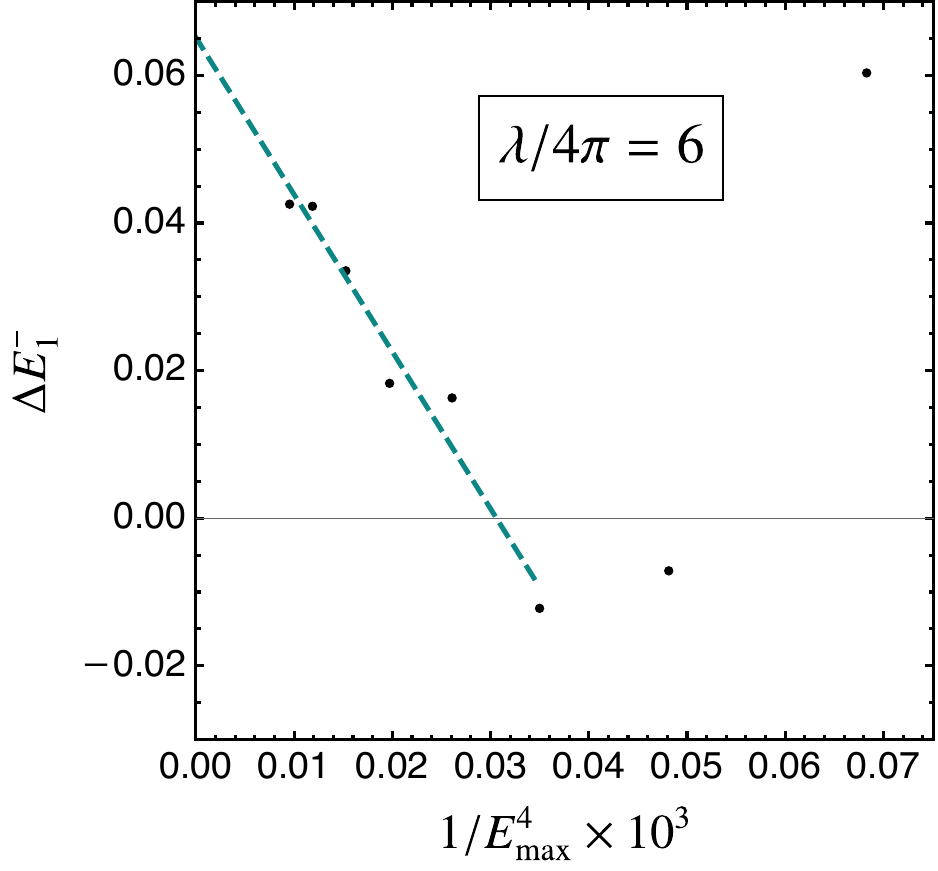}}
	\caption{\bf Scaling at stronger coupling (odd sector): \rm The first $\mathbb{Z}_2$-odd energy gap is shown versus $E_{\rm max}$ for increasing values of the coupling. The fits are applied only in the regime where scaling behavior emerges in the high-energy tails.}
	\label{fig:E1_r20_vary_g_odd}
	\end{minipage}
\vspace{11pt}
\end{figure}

\begin{figure}[h!]
	\centering
	\begin{minipage}{.9\textwidth}
	\centering
	\raisebox{-.0cm}{\includegraphics[trim= 0cm	0cm	0cm	0cm, clip=true,  width=0.32\textwidth]{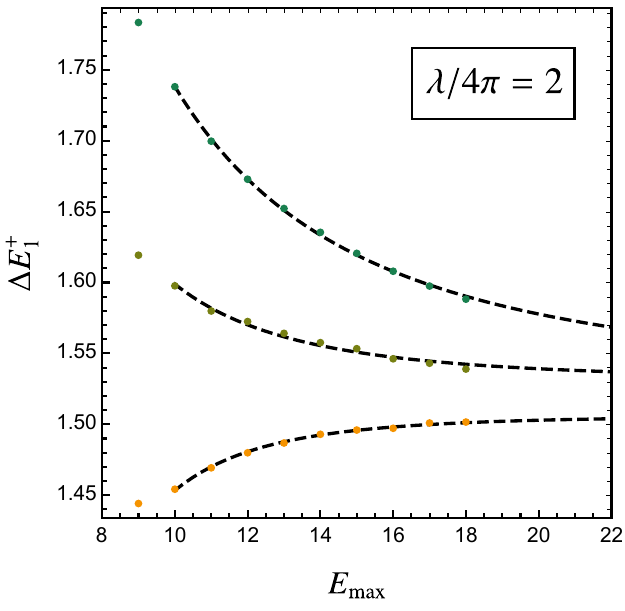}}
	\raisebox{-.0cm}{\includegraphics[trim= 0cm	0cm	0cm	0cm, clip=true,  width=0.32\textwidth]{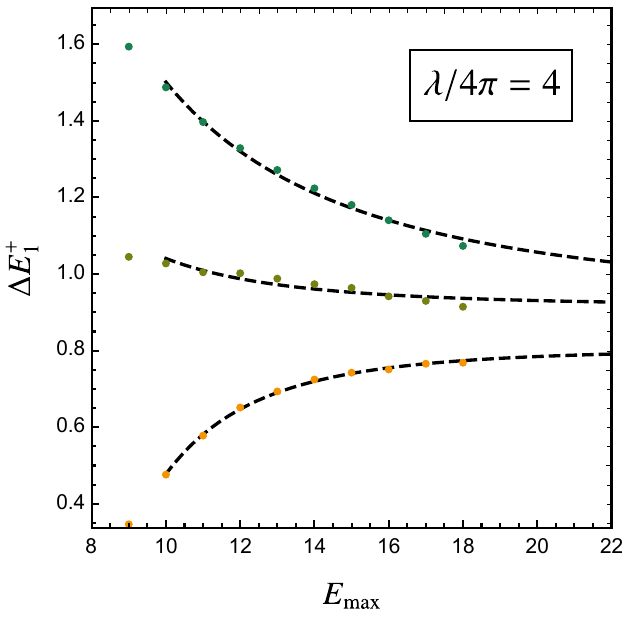}}
	\raisebox{-.0cm}{\includegraphics[trim= 0cm	0cm	0cm	0cm, clip=true,  width=0.32\textwidth]{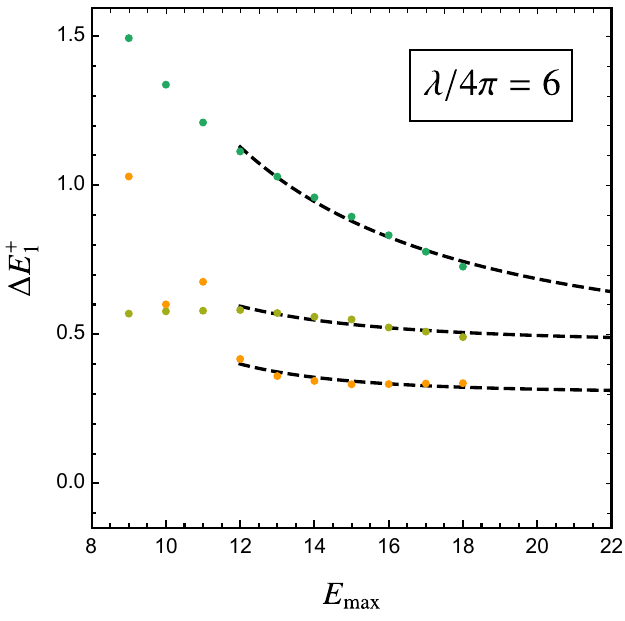}}
	\raisebox{-.0cm}{\includegraphics[trim= 0cm	0cm	0cm	0cm, clip=true,  width=0.32\textwidth]{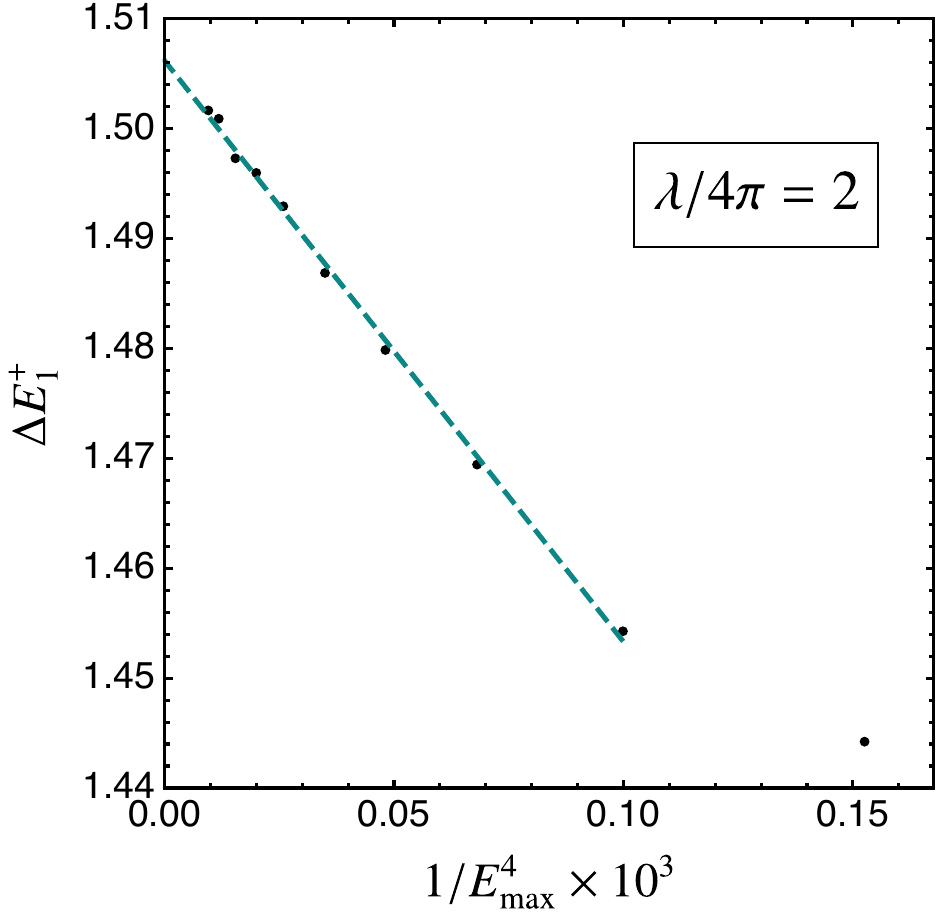}}
	\raisebox{-.0cm}{\includegraphics[trim= 0cm	0cm	0cm	0cm, clip=true,  width=0.32\textwidth]{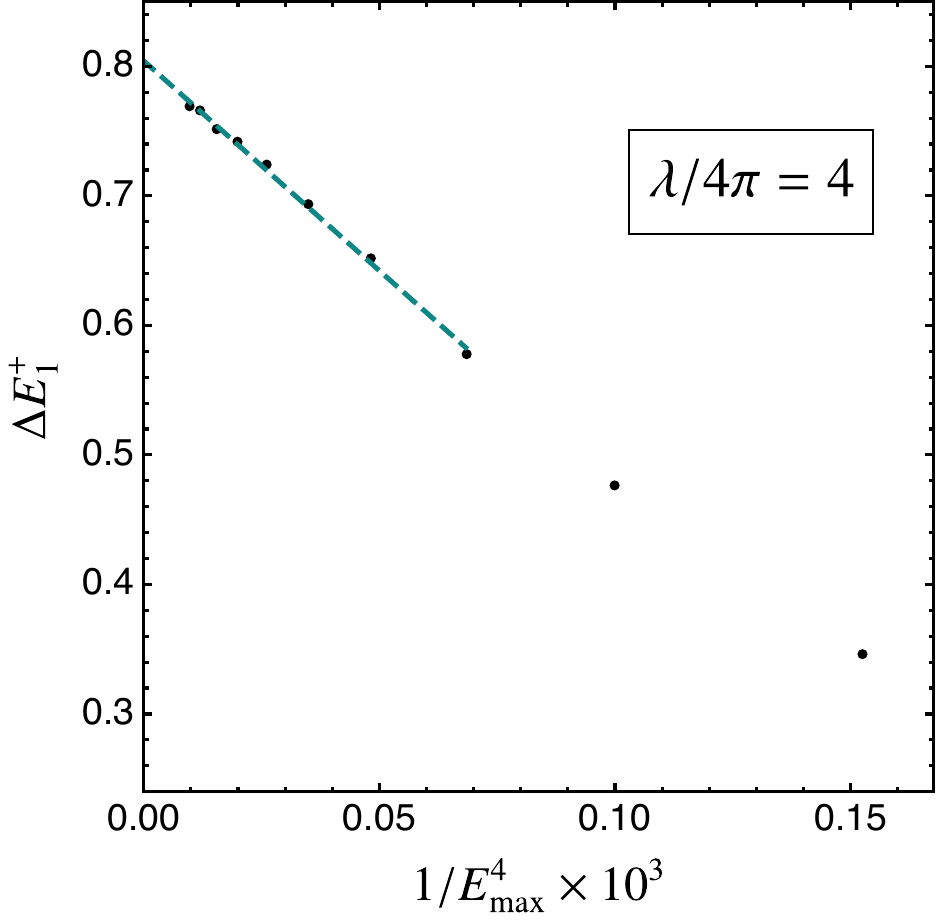}}
	\raisebox{-.0cm}{\includegraphics[trim= 0cm	0cm	0cm	0cm, clip=true,  width=0.32\textwidth]{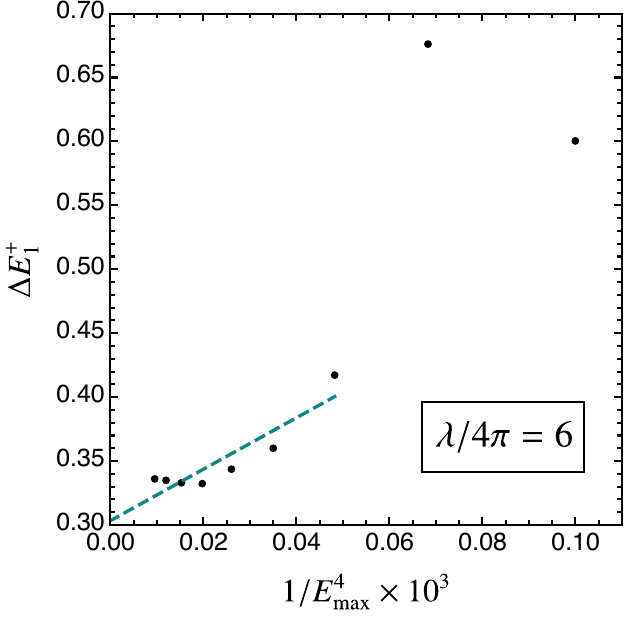}}
	\caption{\bf Scaling at stronger coupling (even sector): \rm The first $\mathbb{Z}_2$-even energy gap is shown versus $E_{\rm max}$ for increasing values of the coupling.  Fits are applied only in the regime where scaling behavior emerges in the high energy tails. }
	\label{fig:E1_r20_vary_g_even}
	\end{minipage}
\vspace{11pt}
\end{figure}

We now examine the first excited state energy $ \Delta E^-_1 = E^-_1 - E_0$ for the raw, LO and NLO effective Hamiltonians. The fit quality is noticeably better, with the scaling trend for $ H_{\rm eff}^{\rm (NLO)}$ now clearly visible for $\lambda/4\pi=1$. In Fig.~\ref{fig:En-r20}, we extend the analysis to higher excited states, again labeling excitations from the $ \mathbb{Z}_2$-even and $\mathbb{Z}_2$-odd sectors as $\Delta E_n^+$ and $\Delta E_n^-$, respectively. For higher $n$, however, the limited energy range at this volume ($E_{\rm max} = 18$) restricts the window over which clean convergence can be observed: reliable scaling behavior only shows up in the high-$E_{\rm max}$ tails. This is shown in the domain of the fit line in each panel of Fig.~\ref{fig:En-r20}. Still, the scaling behavior is evident for multiple levels, with a consistent $1/E_{\rm max}^4$ scaling across both $\mathbb{Z}_2$ sectors. 
Compared to the lower volume case, the enhancement in convergence visibility here is due to the suppression of numerical noise. The increased volume yields a finer energy resolution and a denser set of states, which in turn allows for smoother extrapolation in $E_{\rm max}$. 

We now examine the behavior of convergence as the coupling increases. Fig.~\ref{fig:E1_r20_vary_g_odd} shows $\Delta E_1^-$ for increasing values of $\lambda/4\pi$. In contrast to the lower volume case, the reduced numerical noise at this volume allows the expected $1/E_{\rm max}^4$ scaling to remain visible even at stronger couplings. We also do not observe a clear deterioration in convergence near the critical coupling. This is likely due to the lower maximum cutoff which prevents the near-degeneracy between $E_1^-$ and the ground state from being fully resolved. Fig.~\ref{fig:E1_r20_vary_g_even} shows the corresponding behavior for $\Delta E_1^+$, which maintains clear scaling up to $\lambda/4\pi = 6$, at which point the scaling may emerge only at the very end of the high $E_{\rm max}$ tail. Higher $E_{\rm max}$ would be required to definitively establish the $\Delta E_1^+$ scaling for $\lambda/4\pi =6$. Overall, the improved signal quality at high volume allows the $1/E_{\rm max}^4$ scaling to be clearer in the strong coupling regime.

\subsection{Critical coupling}
\label{sec:cc}

\begin{figure}[h!]
	\centering
	\begin{minipage}{0.9\textwidth}
	\centering
	\raisebox{-.02cm}{\includegraphics[trim= 2.5cm	1cm	 2.5cm	0.05cm, clip=true, width=0.96\textwidth]{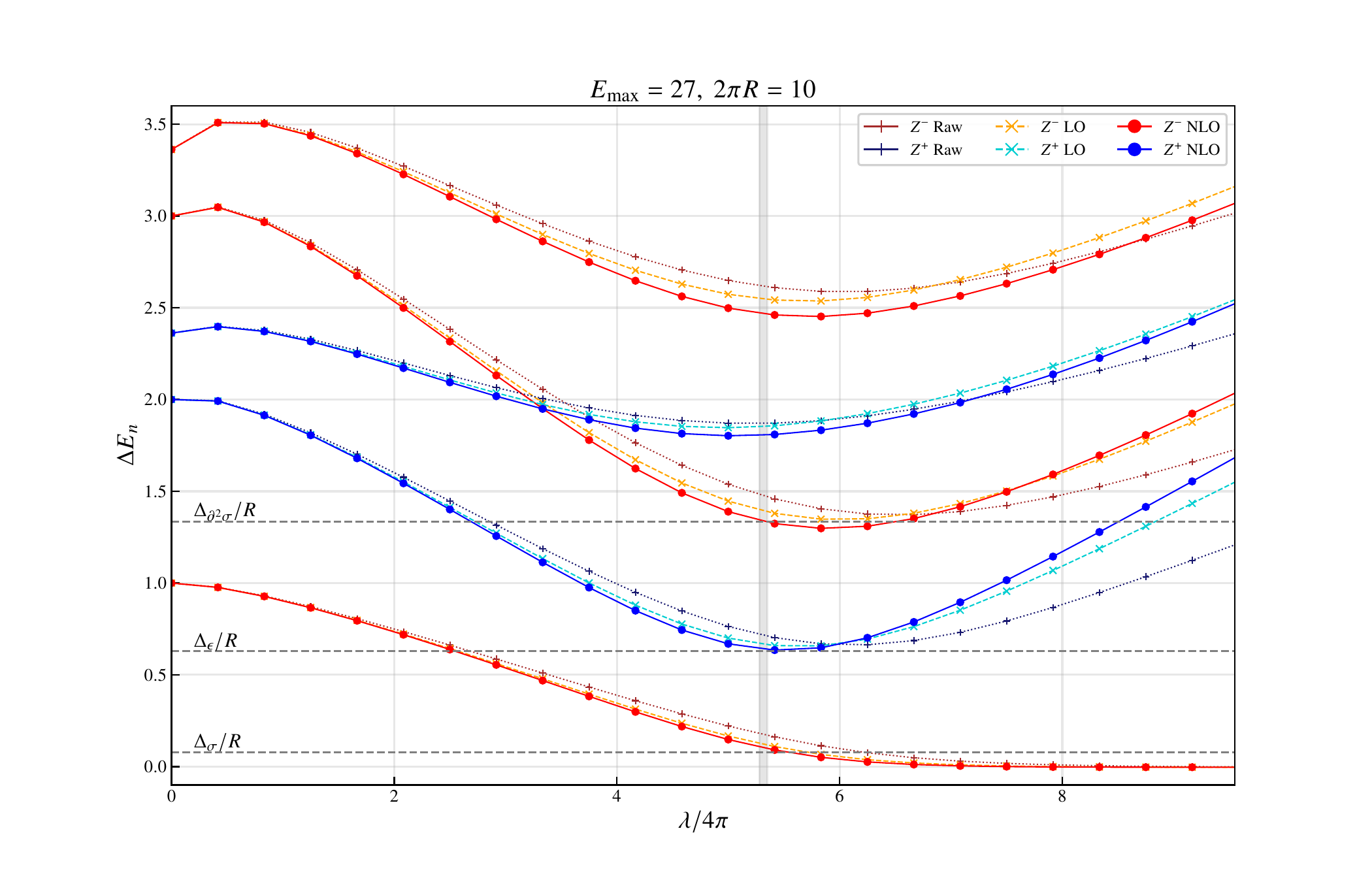}}
	\caption{ Excitation energy spectra plotted against the coupling $\lambda/4\pi$ for $R = 10/2\pi$ (setting $m=1$). States in the $\mathbb{Z}_2$-even and $\mathbb{Z}_2$-odd sectors are distinguished by color. Results from $H_{\rm eff}^{\rm (raw)}$ are shown in dark red (dark blue), from $H_{\rm eff}^{\rm (LO)}$ in yellow (cyan), and from $H_{\rm eff}^{\rm (NLO)}$ in red (blue) for the $\mathbb{Z}_2$-odd (-even) states. Horizontal lines indicate the known theoretical values for the energy levels of the 2D critical Ising model on a cylinder of radius $R$. The vertical gray band marks the predicted critical coupling.}
	\label{fig:E_g_r10}
	\end{minipage}
\end{figure}

\begin{figure}[h!]
	\centering
	\begin{minipage}{0.9\textwidth}
	\centering

	\raisebox{-.02cm}{\includegraphics[trim= 2.5cm	1cm	 2.5cm	0.05cm, clip=true, width=0.96\textwidth]{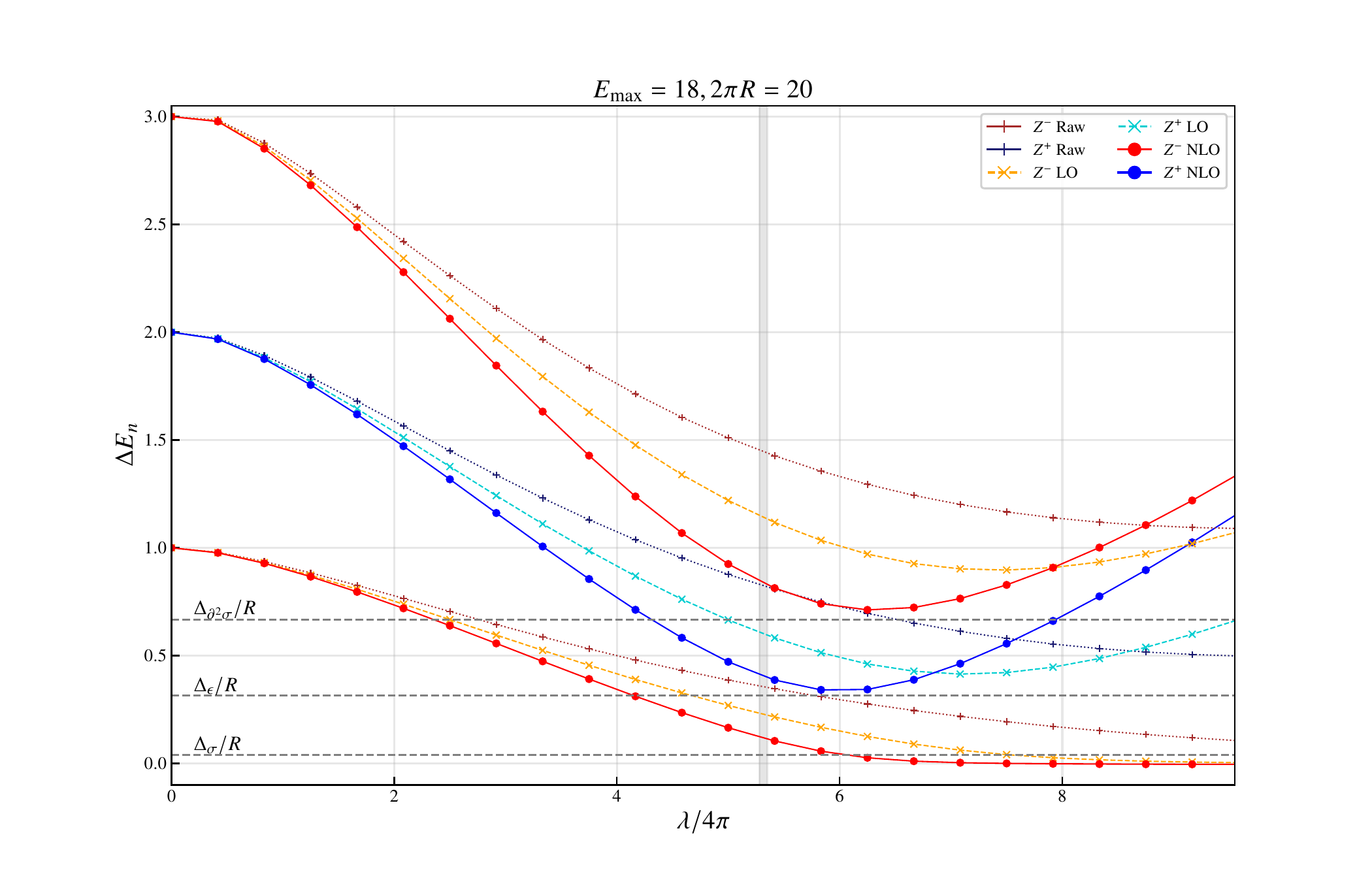}}
	\caption{ Excitation energy spectra plotted against the coupling $\lambda/4\pi$ for $R = 20/2\pi$. States in the $\mathbb{Z}_2$-even and $\mathbb{Z}_2$-odd sectors are distinguished by color. Results from $H_{\rm eff}^{\rm raw}$ are shown in dark red (dark blue), from $H_{\rm eff}^{\rm LO}$ in yellow (cyan), and from $H_{\rm eff}^{\rm NLO}$ in red (blue) for the $\mathbb Z_2$-odd (-even) states. Horizontal lines indicate the known theoretical values for the operator dimensions of 2D Ising model, divided by the radius $R$. The vertical gray band marks the predicted critical coupling. 
	}
	\label{fig:E_g_r20}
	\end{minipage}
\end{figure}

We conclude our numerical analysis by using the low-lying excitation spectrum of $H_{\rm eff}^{\rm (NLO)}$ to probe the location of the critical coupling. 
It is well known that in infinite volume the $1+1$D $\lambda \phi^4$ theory at strong coupling exhibits a second-order phase transition that lies in the Ising universality class. This gives two theoretical expectations at criticality:
	\begin{itemize}
	\item The $\mathbb Z_2$ symmetry should be spontaneously broken, which for the finite volume theory will appear as the emergence of a degeneracy between $\mathbb{Z}_2$ even and odd states
	\item The low-lying excitation energies of the finite volume spectrum should match the scaling dimensions of local operators in the 2D Ising conformal field theory (CFT).
	\end{itemize}
We test these expectations by first examining how the low-lying spectrum evolves with coupling in Figs.~\ref{fig:E_g_r10} and~\ref{fig:E_g_r20}, corresponding to the lower- and higher-volume benchmarks, respectively. 
In both figures, we plot the first five excitation energies $\Delta E_n$ as a function of the coupling $\lambda/4\pi$, scanning the range $[0, 10]$. Each excitation level is shown for $H_{\rm eff}^{\rm (raw)}$ (dark colors), $H_{\rm eff}^{\rm (LO)}$ (light colors), and $H_{\rm eff}^{\rm (NLO)}$ (bright colors). States originating from the $\mathbb Z_2$-even and -odd sectors are distinguished by hue, as indicated in the plot's legend.

We see indication of the $\mathbb Z_2$ spontaneous symmetry breaking in that the first excited energy gap approaches zero for couplings $\lambda/4\pi \gtrsim 6.5$. This signals that the states of the two $\mathbb Z_2$ sectors of the theory are becoming degenerate. This behavior does not appear to be reproduced for the higher excited states, possibly due to a lack of overlap between the low-lying states of the free theory $H_0$ and the low-lying states of $H_{\rm eff}$ at very strong couplings. 
We next compare the energy levels of $H_{\rm eff}$ to those of the radially quantized 2D Ising CFT on a cylinder of radius $R$, which are related to the scaling dimensions of the CFT operators via the relation $E_{\rm Ising} = \Delta_{\rm Ising}/R$. The lowest dimension operators in the 2D Ising CFT are the spin operator $\sigma$ with $\Delta_\sigma= 1/8 $, the energy operator $\epsilon$ with $\Delta_\epsilon = 1$, and the level-2 descendant $\partial^2 \sigma$ with $\Delta_{\partial^2\sigma} = 2 +\Delta_\sigma =  17/8$. These three $\Delta/R$ values are drawn as horizontal dashed black lines in Figs.~\ref{fig:E_g_r10} and~\ref{fig:E_g_r20}.\footnote{Note that the $\Delta/R$ values are smaller for the higher volume benchmark shown in Fig.~\ref{fig:E_g_r20}.} We expect the first few excited energy levels should match these values at the critical coupling.  

Some recent previous estimates of the critical coupling are shown on the Figs.~\ref{fig:E_g_r10} and~\ref{fig:E_g_r20} as a vertical gray band. These estimates were made using Hamiltonian truncation methods~\cite{Elias-Miro:2017tup, Chen:2021pgx, Lajer:2023unt}, as well as other approaches such as lattice Monte Carlo (MC) and resummation techniques~\cite{ Milsted:2013rxa, Bronzin:2018tqz, Pelissetto:2015yha, Serone:2018gjo, Serone:2019szm, Heymans:2021rqo, Vanhecke:2021noi, Elias-Miro:2017tup, Lajer:2023unt}.

One can see by eye in Fig.~\ref{fig:E_g_r10} that the excitation energy levels approach the scaling dimensions of the low-lying operators of the 2D Ising CFT near where previous estimates place the critical coupling. Moreover, the NLO result clearly improves upon the previous LO result, showing a systematic improvement in our estimation of the critical coupling.

In Fig.~\ref{fig:E_g_r20}, the agreement between the energy spectra and the scaling dimensions of the 2D Ising model at the estimated critical coupling is notably worse than the lower volume benchmark (Fig.~\ref{fig:E_g_r10}). This is likely due to the reduced energy cutoff $E_{\rm max}=18$, compared to $E_{\rm max}=27$ used in the lower volume case. Nonetheless, for the higher volume benchmark, the $H_{\rm eff}^{\rm (NLO)}$ prediction shows a much clearer improvement over the $H_{\rm eff}^{\rm (LO)}$ result. This again illustrates the systematic nature of our improvement program and highlights its ability to enhance accuracy when computational resources cap $E_{\rm max}$.

\begin{figure}[h!]
	\centering
	\begin{minipage}{.9\textwidth}
	\centering
	\raisebox{0cm}{\includegraphics[trim= .2cm 0cm 7.0cm .3cm, clip=true, width=.75\textwidth]{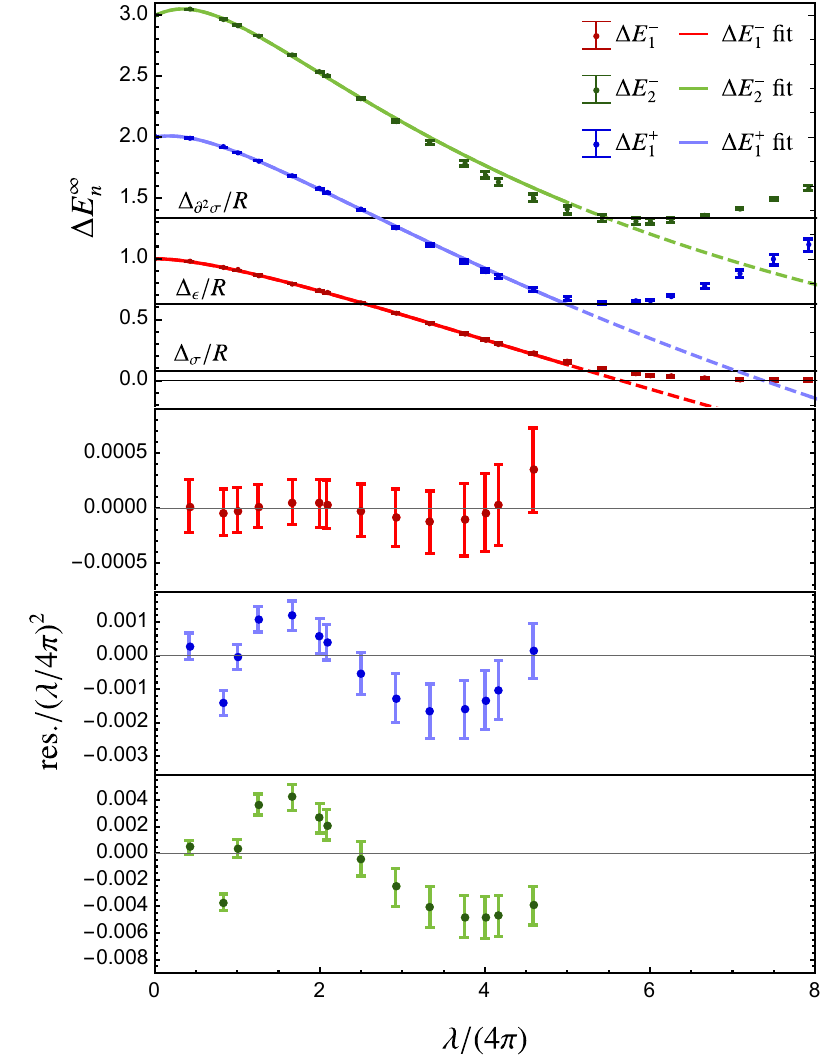}}
	\caption{
	Extrapolated energy gaps  $\Delta E_n^\infty$ plotted against the coupling $\lambda/4\pi$. The top panel shows the $E_{\rm max}\!\to\!\infty$ extrapolations of the first few excitation energies $\Delta E_1^-$ (red), $\Delta E_1^+$ (blue), and $\Delta E_2^-$ (green), with error bars determined as described in the text. Each energy gap is fitted independently using the function $f(\lambda)$ from~\eqref{eq:fit}. The solid portion of each curve indicates the fit region, and the dashed portion shows its extrapolation beyond that domain. Residuals from the fits to $\Delta E_1^-$, $\Delta E_1^+$, and $\Delta E_2^-$, are shown in the second, third, and bottom panels, respectively.	}
	\label{fig:cc-fit}
	\end{minipage}
\end{figure}

Having established that $H_{\rm eff}^{\rm (NLO)}$ yields improved qualitative predictions of the excitation spectrum near the critical coupling, we now use it to extract a more precise quantitative estimate of the critical coupling. Fig.~\ref{fig:cc-fit} illustrates this analysis for the $R=10/2\pi$ benchmark.\footnote{
 The reduced accuracy of the $R=20/2\pi$ benchmark appears to result from limitations imposed by the truncation at $E_{\rm max} = 18$.} We vary the coupling and plot the first two $\mathbb Z_2$-odd excited states and the first $\mathbb Z_2$-even state, which correspond to the lowest three excited states at the critical coupling.

For each excited state, we extract the infinite $E_{\rm max}$ limit by fitting the data to~\eqref{eq:fitp} with $p=4$, consistent with the expected scaling of the NLO-corrected $H_{\rm eff}$. This fit is performed over the range $E_{\rm max} \in [10, 27]$, yielding $\Delta E_{n}^\infty$ values for each coupling. To estimate the uncertainty in the extrapolated values, we use the strategy of~\cite{Elias-Miro:2017tup}: we repeat the fit multiple times, each time excluding a subset of lower $E_{\rm max}$-data points from the range $E_{\rm max} = [10, 14]$. We begin by removing one point a time (yielding five fits), then removing all permutations of two points, and so on, up to a fit with all five points removed. 
The resulting spread in predictions defines the error bars in Fig.~\ref{fig:cc-fit}.

Once the $\Delta E_n^\infty$ data points are obtained, we perform a final fit to extract the critical coupling, shown as solid curves in Fig.~\ref{fig:cc-fit}, with fit residuals shown in the lower panels.  We perform this fit using a rational function similar to what was used in \cite{Elias-Miro:2017tup} but modified for the finite volume case,\footnote{
 Our results are computed at a fixed volume, which avoids any potential uncontrolled effects in the extrapolation to infinite volume.}
\begin{align}
f(\lambda) &=E(0) \frac{c + \left(1-c + \lambda\left(\frac{1}{g_1} + \frac{1}{g_2} + \frac{1}{g_3} + \frac{(1-c)}{\lambda_c}\right) + a \lambda^2\right)\left( 1 - \frac{\lambda}{\lambda_c}\right)^\nu}{\left( 1 + \frac{\lambda}{g_1}\right)\left( 1 + \frac{\lambda}{g_2}\right)\left( 1 + \frac{\lambda}{g_3}\right)} \,,
\label{eq:fit}
\\[8pt]
\textnormal{with } c &= \frac{1}{E(0)} \frac{\Delta_{\rm Ising}}{R} \left( 1 + \frac{\lambda_c}{g_1}\right)\left( 1 + \frac{\lambda_c}{g_2}\right)\left( 1 + \frac{\lambda_c}{g_3}\right) \,,
\label{eq:cdef}
\end{align}
where $E(0)$ are the energy eigenvalues of the free Hamiltonian $H_0$ at $\lambda = 0$ (note the vacuum energy is set to zero in the free Hamiltonian). This function reproduces expected properties of the coupling dependence of the spectrum. Near $\lambda=0$, we have $f(\lambda) = E(0) + \mathcal{O}(\lambda^2)$, consistent with perturbation theory. Near the critical coupling, $f(\lambda\approx \lambda_c) = \Delta_{\rm Ising}/R  +  \# (\lambda- \lambda_c)^\nu$, as expected from an EFT expansion around the 2D Ising model in finite volume, see e.g. \cite{Reinicke:1987zg, Reinicke:1986jq, Lajer:2023unt, Lauchli:2025fii}. We fix the critical exponent  to $\nu = (2-\Delta_\epsilon)^{-1} = 1$, in which case this function becomes a two point Pad\'e approximant. In the infinite volume limit, $c\to0$, and for the first excited energy gap we recover the fit function to $m_{\rm phys}$ in ~(4.7) of~\cite{Elias-Miro:2017tup}.

\begin{table}[b]
\centering
\begin{minipage}{.9\textwidth}
\centering
\begin{tabular}{ |cl|c| }
\Xhline{2\arrayrulewidth}
{\bf Method} & \bf{Year} & {\bf $g_c$} \\
\Xhline{2\arrayrulewidth}
Lattice MC   	& 2009 \cite{Schaich:2009jk}& $2.70^{+0.025}_{-0.013}$\\
 Lattice MC  	& 2015 \cite{Bosetti:2015lsa}& 2.788(15)(8)\\
 Lattice MC   	& 2019 \cite{Bronzin:2018tqz}& 2.7638(35)\\
 \hline
  Matrix Product States 
   			& 2013  \cite{ Milsted:2013rxa}& 2.7690(20),\ 2.7625(8)\\
 Boundary Matrix Product States 
 			& 2021 \cite{Vanhecke:2021noi}& 2.774250(78)\\
\hline
  Lattice MC + Resummation
  			&2015  \cite{Pelissetto:2015yha}& 2.750(10)\\
 \hline
 Resummed PT 
 			& 2018 \cite{Serone:2018gjo} & 2.807(34)\\
 Resummed PT, $\cancel{\mathbb{Z}_2}$
 			& 2019 \cite{Serone:2019szm}	& 2.64(11)\\
Optimized PT  	& 2021 \cite{ Heymans:2021rqo}& 2.779(25)\\
\hline
  Tensor Network RG
  			& 2019   \cite{Kadoh:2018tis}& 2.728(14)\\
 Tensor Network RG 
 			& 2020 \cite{Delcamp:2020hzo}&  2.7715(23)\\
\hline
 Krylov Space truncation method
 			& 2024 \cite{Lajer:2023unt}&2.788(15)\\
\hline
Hamiltonian truncation, LO RG
			& 2015  \cite{Rychkov:2014eea}& 2.97(14) \\
Hamiltonian truncation, raw, $\cancel{\mathbb{Z}_2}$
 			& 2016  \cite{Bajnok:2015bgw} & 2.78(6)\\
Hamiltonian truncation, NLO RG 
 			& 2017 \cite{Elias-Miro:2017tup} & 2.76(3)\\
\hline
 \bf This work	& &	2.752(5)	\\
 \hline
\end{tabular}
\caption{  Comparison of critical coupling estimates using various methods. Results with $\cancel{\mathbb{Z}_2}$ indicate the prediction was made in the $\mathbb{Z}_2$ broken sector, and the result extrapolated to the unbroken sector using Chang duality. The numbers for $g_c$ here are quoted in units $g_c = \lambda_c/4!$ where $\lambda$ is the coupling normalized as in \eqref{eq:Lagrangian}. 
}
\label{tab:cc}
\end{minipage}
\end{table}

Using this fit function, we extract a prediction for the critical coupling.  We first performed a $\chi^2$ fit of $f(\lambda)$ to the data, minimizing 
	$
	\chi^2 =\sum_i  (\Delta E_n^\infty (\lambda_i) - f(\lambda_i))^2/ err_i^2
	$
over the undetermined parameters $g_1, g_2, g_3, a$ and $\lambda_c$ in \eqref{eq:fit}. To estimate uncertainties, we follow the method used in Ref.~\cite{Elias-Miro:2017tup} for consistency. Specifically, we repeat the $\chi^2$ minimization while holding $\lambda_c$ fixed to the value obtained in our initial fit. We denote the resulting minimum value as $\bar\chi^2$, and then define a threshold at $3\sqrt{\bar\chi^2/N}$, where $N$ is the number of fitted points. The allowed range of $\lambda_c$ is then determined by the values that yield $\sqrt{\chi^2/N} < 3\sqrt{\bar\chi^2/N}$, which we use the define our error bars. This method was developed for earlier work, and we use it for ease of comparison, but we note that the uncertainties it yields may not fully capture all errors at our level of precision. The first excited energy level $\Delta E_1^-$ gives the most precise prediction using this method, yielding our prediction for the critical coupling given in Table \ref{tab:cc}. Note that the critical coupling we extract from this fit function will be the coupling in our finite volume theory. The analysis of~\cite{Rychkov:2014eea} estimates the size of the shift in $m^2$ from the finite volume to infinite volume theory, see their Eq.~(2.18). This induces a small shift in our estimate of $g_c$ of $\sim 10^{-6}$ in our units, well below the threshold of our study's precision.

Our result compares favorably with a wide range of non-perturbative techniques, including lattice MC simulations, matrix product state calculations, and tensor-network-based renormalization group (RG) methods. We are also consistent with various resummed perturbation theory (PT) predictions. That being said, caution is warranted when interpreting the uncertainties achieved here, as the estimation method was developed for an earlier study and may not adequately capture systematic effects given the precision of our results.  Still, using this method, our approach yields a more precise estimate than previous Hamiltonian truncation methods, while remaining consistent with the results from a broad range of independent techniques.

\section{Conclusions\label{sec:conclusions}}

In this work we have demonstrated that HTET, which applies EFT techniques to Hamiltonian truncation, improves numerical convergence systematically in powers of the inverse cutoff beyond leading order. We implemented this method in 1+1D $\lambda \phi^4$ theory with a cutoff on total energy $E_{\rm max}$, which was studied previously using HTET at leading order in \cite{Cohen:2021erm}. There it was shown that numerical convergence of eigenvalues as a function of $E_{\rm max}$ agrees with the predictions of EFT: with no correction the raw truncation scales like $1/E_{\rm max}^2$, and including the leading order (local) corrections the convergence scales like $1/E_{\rm max}^3$. In this paper we extended this to include the next order corrections, which includes highly nontrivial, nonlocal terms, and showed that adding these corrections causes the eigenvalues to converge like $1/E_{\rm max}^4$ as predicted by EFT power counting. We showed in Appendix \ref{sos} that separation of scales follows straightforwardly from the leading order calculation done in \cite{Cohen:2021erm}, again as expected from EFT. We also performed a prediction for the critical coupling for this theory,  which is compatible with previous results. This demonstrates that HTET is a robust formalism for systematically adding in corrections to the low energy Hamiltonian which can theoretically be extended to arbitrary order. 

At this order in the EFT expansion, we encountered several interesting new features. First is the necessary introduction of nonlocal corrections in our effective Hamiltonian due to the nonlocal nature of the total energy cutoff $E_{\rm max}$. The nonlocal terms at this order are of the form $H_0\, \mathcal{O}$ or $\mathcal{O}\, H_0$, where $\mathcal{O}$ is itself a local operator (e.g. $\mathcal{O} = \int\, R\, d\theta\, \phi^2$). Based on the EFT expansion used in HTET, we expect this type of behavior to continue at higher orders and that all nonlocal terms will be of the form of either $(H_0)^n\, \mathcal{O}$ or $\mathcal{O}\, (H_0)^n$, which by power counting means they will be suppressed by $(E_{\rm max})^{n}$ \cite{Rutter:2018aog}. 
The second notable feature is that at this order, the truncation error is small enough that the numerical noise, already present at LO in \cite{Cohen:2021erm}, can now obscure the EFT scaling. We  conjecture that this noise is caused in part by finite volume effects \cite{Luscher:1985dn, Luscher:1986pf}, and  showed that by increasing the volume these effects are suppressed and we get the correct $1/E_{\rm max}^4$ scaling behavior. The higher volume benchmark did not yield a better estimate for the critical coupling, however, likely because we did not reach a high enough $E_{\rm max}$. Questions concerning the reliability of larger volume benchmarks and the challenges of extrapolating to infinite volume in light of the eventual orthogonality catastrophe~\cite{Anderson:1967zze} are compelling, but we leave them for future work.

This demonstration that HTET holds to higher orders paves the way to apply this method to more complex systems. In particular it would be interesting to study systems with more complicated matter content, including chiral fermions and gauge bosons, as well as systems in higher dimensions. Complementary to the lattice, chiral fermions are straightforward to formulate in Hamiltonian truncation, while gauge bosons require more care due to the nonlocal nature of the cutoff. We expect HTET to extend straightforwardly to systems with non-trivial UV divergences \cite{Delouche:2023wsl}, and a clear next target to test this is 2+1D $\lambda \phi^4$ theory, for which initial calculations were done in \cite{Cohen:2021erm} and which has been studied using Hamiltonian truncation methods previously in \cite{Elias-Miro:2020qwz, Anand:2020qnp}. Another interesting future direction is to use the higher order calculations done in this paper to extract precision information about the critical point. This could include a more precise estimation of the critical coupling and estimations of the critical exponents using subleading terms in finite volume conformal perturbation theory as was done in \cite{Delouche:2023wsl}. Finally, we might hope to take advantage of the Lorentzian nature of this formalism to construct real-time observables. In particular, the matching operator used in this work \eqref{eq:Sigma} could provide an avenue towards constructing asymptotic states to be used for scattering in finite volume, see e.g. \cite{Molinari_2007, Ingoldby:2025bdb}. The results presented here establish HTET as a precise and extendable approach, and provide a foundation for further progress in analyzing strongly coupled systems.

\acknowledgments
The authors would like to thank Tim Cohen, Olivier Delouche, Joan Elias Mir\'o, James Ingoldby, Anatoly Konechny, Markus Luty, and Matt Walters for useful discussions. RH and ED were supported by the Institute for Fundamental Theory at the University of Florida. The work of KF is supported by the Swiss National Science Foundation under grant no. 200021-205016. RH also thanks the CERN Theoretical Physics Department for hospitality while this research was being carried out.

\clearpage
\appendix


\section{Diagrammatic rules\label{sec:rules}}
\label{app:rules}

For the diagrammatic calculations, we first separate $\phi(t,x)$ into its positive and negative frequency components
\begin{align}
\phi(t,x) 
&= \frac{1}{\sqrt{2\pi R}} \sum_k  \left[ \phi_k^{(-)} e^{ i k\cdot x}+\phi_k^{(+)}e^{-ik \cdot x}\right] \,,
\end{align}
with 
\begin{align}
\phi_k^{(-)} = \frac{1}{\sqrt{2\omega_k}}\, a_k^\dagger\quad{} \textnormal{and}\quad{} \phi_k^{(+)} = \frac{1}{\sqrt{2\omega_k}}\, a_k \,.
\end{align}
The Wick contraction of these fields is then 
\begin{align}
\phi_{p_1} \phi_{p_2} =\ :\!\phi_{p_1} \phi_{p_2}\!: +\ \bcontraction{}{\phi}{_{p_1}}{\phi}\phi_{p_1} \phi_{p_2} \,,
\end{align}
with 
\begin{align}
\bcontraction{}{\phi}{_{p_1}}{\phi}\phi_{p_1} \phi_{p_2} = \frac{\delta_{p_1,p_2}}{2\omega_{p_1}} \,.
\end{align}
Because we are only working with normal-ordered operators, diagrams with lines beginning and ending at the same vertex are forbidden. The diagrammatic rules at $\mathcal{O}(V^n)$ were derived in \cite{Cohen:2021erm}, and we reproduce the final list here:\footnote{Note that our choice of normal-ordered scheme  is equivalent to setting $m_V^2 = 0$ in \cite{Cohen:2021erm}.} 
\begin{itemize}
\item Draw all possible diagrams, including disconnected diagrams, with $n$ ordered vertices. 
\item Assign an independent momentum variable to each internal and external line. Sum over the internal momenta.
\item For each vertex, include the factor 
\begin{align}
\includegraphics[valign=c,scale=0.75]{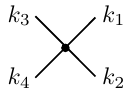} 
= \frac{\lambda}{2\pi R} \delta_{k_1 + \cdots + k_4} \,,
\end{align}
where all the momenta are taken to flow into the vertex.
\item For each internal line with momentum $k$, include the factor 
\begin{align}
\includegraphics[valign=c,scale=0.75]{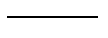}  = \frac{1}{2\omega_k} \,.
\end{align}
\item For a diagram with $i$ lines connected to the initial state
and $j$ lines connected to the final state, include the factor 
\begin{align}
\label{eq:matrixelem}
\langle f| \phi^{(-)}_{k_{i+j}} \cdots \phi^{(-)}_{k_{i+1}}
\phi^{(+)}_{k_i} \cdots \phi^{(+)}_{k_1} |i \rangle \,.
\end{align}
\item
Each vertex also contributes an energy denominator:
\begin{align}
\frac{1}{E_{f\alpha} + i\epsilon} = \frac{1}{E_f - E_\alpha + i \epsilon} \,,
\end{align}
where $E_f$ is the energy of the final state,
and $E_\alpha$ is the energy of the state to the right of the vertex.  
The final energy denominator factor of $1/E_{fi}$ is omitted.
\item
Include an overall symmetry factor
\begin{align}
S = \left( \frac{1}{4!} \right)^{\!\! n} \! C \,,
\end{align}
where $n$ is the number of vertices, and $C$ is the number of Wick contractions that give the same diagram. To determine $C$, take initial state particles to be identical to each other and final state particles to be identical to each other, but distinguish initial state particles from final state particles. The coefficient $C$ is identical to that which appears
in front of the matrix element of the form in 
\eqref{eq:matrixelem} 
when using Wick's theorem. \hspace{-3pt}

\end{itemize}

In the main text these rules are used to calculate the corrections to our effective Hamiltonian at  $\mathcal{O}(V)$ and $\mathcal{O}(V^2)$:
\begin{align}
\langle f|H_1|i \rangle &=\langle f| V|i\rangle\,,\\ 
\langle f| H_2|i\rangle &=\sum^>_\alpha \frac{\langle f|V|\alpha\rangle \langle \alpha|V|i \rangle}{E_f - E_\alpha} \,.
\end{align}
We then expand these expressions in $1/E_{\rm max}$ to find the leading order and next-to-leading-order corrections. 

\newpage

\section{Separation of scales\label{sos}}

Separation of scales at the order computed in this work follows directly from the calculation of separation of scales shown in \cite{Cohen:2021erm}. Here we reproduce this calculation at the next order in our $1/E_{\rm max}$ expansion, demonstrating that this feature of EFT is still manifest at this order. In particular we show that, with the choice of renormalization scale $\mu\sim E_{\rm max}$, terms with UV/IR mixing in the coefficients of non-normal-ordered operators vanish.  

The choice to use normal-ordered operators, employed in the bulk of the paper, is equivalent to adding a mass counterterm to our Langrangian of the form
\begin{align}
\delta m^2 = \frac{\lambda}{8\pi R} \sum_{|k| \leq \Lambda R} \frac{1}{\omega_k} \,,
\end{align}
which completely removes all the UV divergences from the theory. This is a scheme-dependent choice, and we could have just as easily chosen the counterterm
\begin{align}
\delta m^2(\mu) = \frac{\lambda}{8\pi R} \sum_{\mu R \leq |k| \leq \Lambda R} \frac{1}{\omega_k} \,,
\end{align}
which still removes the UV divergences, but differs from the normal-ordered scheme by a finite shift in the mass. We then separate the complete mass term into two pieces
\begin{align}
m^2+\delta m^2 = m_Q^2 + m_V^2 \,,
\end{align}
using one to quantize the theory (the so-called ``quantization mass'' $m_Q^2$) and treating the other as a new interaction term in $V$ with
\begin{align}
\label{eq:renorm}
m_V^2 = m_R^2(\mu) - m_Q^2 + \frac{\lambda}{8\pi R} \sum_{|k|\leq \mu R} \frac{1}{\omega_k} \,.
\end{align}
Here $m_R(\mu)^2$ is now a renormalized mass parameter that depends on the renormalization scale $\mu$ (see \cite{Cohen:2021erm} for a more detailed explanation). In the main text, the choice to use normal-ordered operators is equivalent to choosing $\mu = 0$ and $m_V^2 = 0$. 

In the more general renormalization scheme with $\mu \neq 0$, our diagrammatic expansion now contains another interaction vertex of the form
\begin{align}
\includegraphics[valign=c]{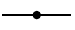} = m_V^2
\end{align}
that must be included in our calculations. At $\mathcal{O}(V^2)$, this contributes the new diagrams
\begin{subequations}
\begin{gather}
\includegraphics[valign=c,scale=0.65]{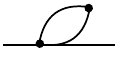}\quad{} \includegraphics[valign=c,scale=0.65]{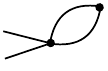}\quad{} \includegraphics[valign=c,scale=0.65]{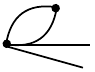} \quad{}\reflectbox{\includegraphics[valign=c,scale=0.65]{figs/snail_f}}\quad{} \reflectbox{\includegraphics[valign=c,scale=0.65]{figs/snail2_f}}\quad{}\reflectbox{\includegraphics[valign=c,scale=0.65]{figs/snail3_f}}\label{eq:mv2diags1}\\
 \includegraphics[valign=c,scale=0.65]{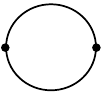} \,,
 \label{eq:mv2diags2}
\end{gather}
\end{subequations}
which must be considered in addition to the diagrams  \eqref{eq:v2diagrams1}. We organize our check of separation of scales by the number of external legs (excluding spectator particles). At each order we consider the diagrammatic contribution using the general renormalization scheme \eqref{eq:renorm}, as well as the contribution from un-normal-ordering operators using the equations
\begin{align}
\label{eq:un-normal}
\begin{split}
:\!\phi^2\!: &=  \phi^2- Z \,, \\
:\!\phi^4\!:&=  \phi^4 - 6Z:\!\phi^2\!: - 3Z^2  \\
&=  \phi^4 - 6Z \phi^2 + 3Z^2 \,,\\
\end{split}
\end{align}
with
\begin{align}
Z = \frac{1}{4\pi R} \sum_k \frac{1}{\omega_k} \,.
\end{align}
The effective Hamiltonian we calculated at $\mathcal{O}(V^2)$ in Section \ref{sec:H2} of the main text can be written as
\begin{align}
H_{2} = \mathbb{C}_0\int R\,d\theta\,  \mathbb{1} +\frac{1}{2} \mathbb{C}_2 \int R\,d\theta\,  :\!\phi^2\!:+\frac{1}{4!} \mathbb{C}_4\int R\,d\theta\,  :\!\phi^4\!: \,,
\end{align}
where the $\mathbb{C}$'s can depend on $E_i,\ E_f$ and thus contain nonlocal information. The additional diagrams coming from adding a renormalization scale $\mu$ will contribute new terms to the coefficients of $:\!\phi^2\!:$ \eqref{eq:mv2diags1} and $\mathbb{1}$ \eqref{eq:mv2diags2}, which we denote by $\mathbb{C}_{2,\mu}$ and $\mathbb{C}_{0,\mu}$ respectively:
\begin{align}
H_{2}(\mu) = \left(\mathbb{C}_0+\mathbb{C}_{0,\mu}\right)\int R\,d\theta\,  \mathbb{1} +\frac{1}{2}\left( \mathbb{C}_2+ \mathbb{C}_{2,\mu}\right) \int R\,d\theta\,  :\!\phi^2\!:+\frac{1}{4!} \mathbb{C}_4\int R\,d\theta\,  :\!\phi^4\!:\,.
\end{align}

Finally rewriting this in terms of non-normal-ordered operators gives
\begin{align}
\begin{split}
H_{2}(\mu) =&\ \left( \mathbb{C}_0+\mathbb{C}_{0,\mu}-\frac{1}{2}Z \mathbb{C}_2 -\frac{1}{2}Z \mathbb{C}_{2,\mu} +\frac{1}{8}Z^2\mathbb{C}_4\right) \int R\,d\theta\,  \mathbb{1} \\
&\qquad{}+\frac{1}{2} \left(  \mathbb{C}_2+\mathbb{C}_{2,\mu}- \frac{1}{2} Z \mathbb{C}_4\right) \int R\,d\theta\, \phi^2+\frac{1}{4!} \mathbb{C}_4\int R\,d\theta\, \phi^4 \,.
\end{split}
\label{eq:unno}
\end{align}
It is in the coefficients of these operators, with the choice $\mu\sim E_{\rm max}$, that we expect any UV/IR overlapping sums to vanish.

\subsection{4 legs}
For 4 external legs, there are no additional diagrams at this order in the more general renormalization scheme, so the coefficient of the non-normal-ordered operator $\phi^4$ is the same as the normal-ordered operator $:\!\phi^4\!:$\, \eqref{eq:unno}
\begin{align}
H_{2}(\mu) \supset  \frac{1}{4!}  \mathbb{C}_4\int R\,d\theta\, \phi^4 \,.
\end{align}
 This was calculated in the main text \eqref{eq:c4main} to be 
\begin{align}
\label{eq:c4}
\mathbb{C}_4 &=  \frac{3\lambda^2}{ 8\pi R} \sum_{k}   \bigg[\frac{\Theta(E_f + 2\omega_k-E_{\rm max})}{\omega_k^2(-2\omega_k)}-\frac{1}{2}E_{fi}\frac{\delta(2\omega_k-E_{\rm max})}{\omega_k^2(-2\omega_k)}-\frac{1}{2} E_{fi}\frac{\Theta( 2\omega_k-E_{\rm max})}{\omega_k^2(-2\omega_k)^2}\bigg] \,.
\end{align}
In the main text this expression was expanded further and written in terms of $H_0$ rather than $E_f$ and $E_i$. The expression in \eqref{eq:c4} contains no sums with overlapping UV/IR dependence (the step function ensures we sum only over momenta $k\gtrsim E_{\rm max} R$), so separation of scales is trivially manifest here.

\subsection{2 legs}
For 2 external legs, to find the coefficient of the non-normal-ordered operator $\phi^2$, we must now include three separate contributions,
\begin{align}
H_2(\mu) \supset  \frac{1}{2} \left( \mathbb{C}_2+\mathbb{C}_{2,\mu} - \frac{1}{2} Z \mathbb{C}_4\right) \int R\, d\theta\, \phi^2 \,.
\end{align}
 The term $\mathbb{C}_{2,\mu}$ comes from calculating the new diagrams in \eqref{eq:mv2diags1} from the more general renormalization scheme. The first and third terms come from un-normal-ordering the operators $\phi^2$ and $\phi^4$ \eqref{eq:unno}, whose coefficients were calculated in the main text.

The diagrammatic contribution to $:\!\phi^2\!:$ in $H_{\rm eff}$ was calculated in the main text to be (see \eqref{eq:c2main})
\begin{align}
\begin{split}
\mathbb{C}_2 =&\  \frac{\lambda^2}{(2\pi R)^2} \frac{1}{24}  \sum_{k, k'} \bigg[\frac{\Theta( E_f + \omega_{k}+ \omega_{k'}+ \omega_{k+k'} -E_{\rm max})}{\omega_{k}\omega_{k'}\omega_{k+k'}( - \omega_{k}- \omega_{k'}- \omega_{k+k'})} \\
&-  \frac{1}{2}E_{fi} \frac{\delta(\omega_{k}+ \omega_{k'}+ \omega_{k+k'} -E_{\rm max})}{\omega_{k}\omega_{k'}\omega_{k+k'}(- \omega_{k}- \omega_{k'}- \omega_{k+k'})} \\
& -   \frac{1}{2}E_{fi}  \frac{\Theta( \omega_{k}+ \omega_{k'}+ \omega_{k+k'} -E_{\rm max})}{\omega_{k}\omega_{k'}\omega_{k+k'}(- \omega_{k}- \omega_{k'}- \omega_{k+k'})^2}\bigg] \,.
\end{split}
\end{align}
These sums contain only one type of term (with multiplicity 3) with UV/IR mixing, where either $k$, $k'$ or $k+k'$ is much smaller than the other two momenta:
\begin{align}
\begin{split}
\mathbb{C}_{2} \supset&\  \frac{\lambda^2}{(2\pi R)^2} \frac{1}{8} \sum_{k' \leq k_{\rm max}} \frac{1}{\omega_{k'}}\\
& \times \sum_{k} \bigg[\frac{\Theta( E_f + 2\omega_{k}-E_{\rm max})}{\omega_{k}^2( - 2\omega_{k})}-  \frac{1}{2}E_{fi} \frac{\delta(2\omega_{k} -E_{\rm max})}{\omega_{k}^2(- 2\omega_{k})}-   \frac{1}{2}E_{fi}  \frac{\Theta(2 \omega_{k} -E_{\rm max})}{\omega_{k}^2(-2 \omega_{k})^2}\bigg] \,.
\end{split}
\end{align}
The diagrams in \eqref{eq:mv2diags1} contribute to the effective Hamiltonian a term with the coefficient
\begin{align}
\mathbb{C}_{2,\mu} =&\,  \frac{\lambda\, m_V^2}{8\pi R} \sum_q \bigg[ \frac{\Theta(E_f+2\omega_q-E_{max})}{\omega_q^2(-2\omega_q)}-\frac{1}{2}E_{fi}\frac{\delta(2\omega_q-E_{max})}{\omega_q^2( - 2\omega_q)}-\frac{1}{2} E_{fi}\frac{\Theta(2\omega_q-E_{max})}{\omega_q^2( - 2\omega_q)^2}\bigg] \,. 
\end{align}
The terms in brackets are all sums over UV states, so to focus on terms with UV/IR mixing, we keep only the part of $m_V^2$ containing an IR sum:
\begin{align}
\hspace{-.1cm}
\mathbb{C}_{2,\mu} 
	&\supset \frac{\lambda^2}{16(2\pi R)^2}\sum_{|k|\leq \mu R} \frac{1}{\omega_k} \nonumber\\
	& \ \times \!\sum_q \! \bigg[ \frac{\Theta(E_f+2\omega_q-E_{max})}{\omega_q^2(-2\omega_q)}-\frac{1}{2} E_{fi}\frac{\delta(2\omega_q-E_{max})}{\omega_q^2( - 2\omega_q)}-\frac{1}{2}E_{fi}\frac{\Theta(2\omega_q-E_{max})}{\omega_q^2( - 2\omega_q)^2}\bigg] \,.
\end{align}

Finally, the contribution from un-normal-ordering $:\!\phi^4\!:$ gives the coefficient (see  \eqref{eq:unno} and \eqref{eq:c4})
\begin{align}
\hspace{-.2cm}
-\frac{1}{2} Z \mathbb{C}_{4} 
	=\ \frac{-3\lambda^2}{ 16(2\pi R)^2} \sum_{k' \leq k_{\rm max}}  \sum_{k}  
		\frac{1}{\omega_{k'}}  \bigg[ &
			\frac{\Theta(E_f + 2\omega_k-E_{\rm max})}{\omega_k^2(-2\omega_k)}
			-\frac{1}{2}E_{fi}\frac{\delta(2\omega_k-E_{\rm max})}{\omega_k^2(-2\omega_k)}
	\nonumber\\
	& \qquad	
			-\frac{1}{2} E_{fi}\frac{\Theta( 2\omega_k-E_{\rm max})}{\omega_k^2(-2\omega_k)^2}\bigg] \,,
\end{align}
where we have only kept the part of $Z$ containing IR momenta with $k \leq k_{\rm max} \equiv E_{\rm max} R$. The total contribution to the UV/IR mixing of the coefficient of $\phi^2$ is then:
\begin{align}
& \mathbb{C}_2
 	+\mathbb{C}_{2,\mu} - \frac{1}{2} Z \mathbb{C}_4  \nonumber\\
	&\ \ \ =\ \frac{\lambda^2}{8(2\pi R)^2}\left(\frac{1}{2} \sum_{|k|\leq \mu R} +  \sum_{|k|\leq k_{\rm max}} - \frac{3}{2} \sum_{|k|\leq k_{\rm max}}  \right) \frac{1}{\omega_k} 
	\nonumber\\[12pt]
	&\quad \ \,  \times \sum_q \bigg[ \frac{\Theta(E_f+2\omega_q-E_{max})}{\omega_q^2(-2\omega_q)}-\frac{1}{2} E_{fi}\frac{\delta(2\omega_q-E_{max})}{\omega_q^2( - 2\omega_q)}-\frac{1}{2}E_{fi}\frac{\Theta(2\omega_q-E_{max})}{\omega_q^2( - 2\omega_q)^2}\bigg] \,.
\end{align}
So we see, as was the case for the local approximation, for the choice of renormalization scale $\mu \sim E_{\rm max} = k_{\rm max}/R$, these terms cancel and we have separation of scales.

\subsection{0 legs}
Similarly, there are five types of contribution to the identity operator at $\mathcal{O}(V^2)$, 
\begin{align}
H_{\rm eff}(\mu)\big|_{\mathcal{O}(V^2)}  \supset  \left( \mathbb{C}_0+\mathbb{C}_{0,\mu}-\frac{1}{2}Z \mathbb{C}_2 -\frac{1}{2}Z \mathbb{C}_{2,\mu} +\frac{1}{8}Z^2\mathbb{C}_4\right) \int R\, d\theta\, \mathbb{1} \,.
\end{align}
The term $\mathbb{C}_{0,\mu}$ comes from calculating the diagram \eqref{eq:mv2diags2} and the term $\mathbb{C}_{0}$ comes from calculating the other vacuum diagram (\ref{eq:c0-diagram}-\ref{eq:c0main}). The terms proportional to $\mathbb{C}_{2,\mu},\ \mathbb{C}_{2}$ and $\mathbb{C}_{4}$ come from un-normal-ordering the $\phi^2$ and $\phi^4$ operators \eqref{eq:unno}.

The first term was calculated in the main text, where we found the expression \eqref{eq:c0main}
\begin{align}
\mathbb{C}_{0}
=\frac{1}{384}\frac{\lambda^2}{(2\pi R)^3}\sum_{1,\ldots,4}\delta_{1234,0} \frac{\Theta(E_f + \omega_1+\omega_2+\omega_3+\omega_4 - E_\text{max})}{\omega_1\omega_2\omega_3\omega_4(-\omega_1-\omega_2-\omega_3-\omega_4)} \,.
\end{align}
This contains two types of terms with overlapping UV/IR sums: one with multiplicity 6 where two momenta are small and two are large, and one with multiplicity 4 where one momentum is small and three are large:
\begin{align}
\begin{split}
\mathbb{C}_{0}
\supset&\ \frac{1}{64} \frac{\lambda^2}{(2\pi R)^3}\sum_{k,k' \leq k_{\rm max}} \frac{1}{\omega_k \omega_{k'}}\sum_q \frac{\Theta(E_f +2 \omega_q - E_\text{max})}{\omega_q^2(-2\omega_q)}\\
&+\frac{1}{96}\frac{\lambda^2}{(2\pi R)^3}\sum_{k \leq k_{\rm max}} \frac{1}{\omega_k} \sum_{q,q' \geq k_{\rm max}}  \frac{\Theta(E_f + \omega_q+\omega_{q'}+\omega_{q+q'}- E_\text{max})}{\omega_q\omega_{q'}\omega_{q+q'}(-\omega_q-\omega_{q'}-\omega_{q+q'})} \,.
\end{split}
\end{align}

Calculating the second term coming from the diagram \eqref{eq:mv2diags2} gives
\begin{align}
\mathbb{C}_{0,\mu} = \frac{1}{8} \frac{1}{2\pi R}  \left(m_V^2\right)^2 \sum_q \frac{\Theta(E_f + 2\omega_q - E_{\rm max})}{ \omega_q^2 (-2\omega_q)} \,.
\end{align}
Due to the step function, the sum in this expression is over UV modes, so to focus on terms with overlapping UV/IR sums we must include at least one of the IR sums present in $m_V^2$:
\begin{align}
\begin{split}
\mathbb{C}_{0,\mu} \supset&\ \frac{1}{16} \frac{\lambda}{(2\pi R)^2}  \left( m^2_R(\mu) - m_Q^2\right) \sum_{k \leq \mu R} \frac{1}{\omega_k}\sum_q \frac{\Theta(E_f + 2\omega_q - E_{\rm max})}{ \omega_q^2 (-2\omega_q)}\\
&+\frac{1}{128} \frac{\lambda^2}{(2\pi R)^3} \sum_{k,k' \leq \mu R} \frac{1}{\omega_k\omega_{k'}}\sum_q \frac{\Theta(E_f + 2\omega_q - E_{\rm max})}{ \omega_q^2 (-2\omega_q)} \,.
\end{split}
\end{align}

Un-normal-ordering the $\phi^2$ term from the main text we get:
\begin{align}
-\frac{1}{2} Z \mathbb{C}_{2} &=-\frac{1}{96}\frac{\lambda^2}{(2\pi R)^3}  \sum_{k} \frac{1}{\omega_k}  \sum_{q, q'} \bigg[\frac{\Theta( E_f + \omega_{q}+ \omega_{q'}+ \omega_{q+q'} -E_{\rm max})}{\omega_{q}\omega_{q'}\omega_{q+q'}( - \omega_{q}- \omega_{q'}- \omega_{q+q'})}\bigg] \,.
\end{align}
Here we have dropped the terms proportional the $E_{fi}$, since these vanish when multiplying the identity operator. This contains two types of terms with overlapping UV/IR sums: one where $q$, $q'$ and $q+q'$ are all large and the IR sum coming from $Z$, and another term (with multiplicity 3) where two momenta of $q$, $q'$ and $q+q'$ are large and one is small:
\begin{align}
\begin{split}
-\frac{1}{2} Z \mathbb{C}_{2} \supset&\ -\frac{1}{96}\frac{\lambda^2}{(2\pi R)^3}  \sum_{k\leq k_{\rm max}} \frac{1}{\omega_k}  \sum_{q, q' \geq k_{\rm max}} \bigg[\frac{\Theta( E_f + \omega_{q}+ \omega_{q'}+ \omega_{q+q'} -E_{\rm max})}{\omega_{q}\omega_{q'}\omega_{q+q'}( - \omega_{q}- \omega_{q'}- \omega_{q+q'})}\bigg]\\
& -\frac{1}{16}\frac{\lambda^2}{(2\pi R)^2}  Z \sum_{k'\leq k_{\rm max}}\frac{1}{ \omega_{k'}}  \sum_{q} \bigg[\frac{\Theta( E_f +2 \omega_{q} -E_{\rm max})}{\omega_{q}^2( - 2\omega_{q})}\bigg] \,.
\end{split}
\end{align}
In the last line we have kept $Z$ general, so the sum is over both UV and IR modes. 

For the new diagrams using the general renormalization scheme, we also get a contribution from un-normal-ordering $\phi^2$:
\begin{align}
-\frac{1}{2} Z \mathbb{C}_{2,\mu } &=-\frac{1}{8}  \frac{\lambda}{(2\pi R)} m_V^2 Z \sum_q \bigg[ \frac{\Theta(E_f+2\omega_q-E_{max})}{\omega_q^2(-2\omega_q)}\bigg] \,.
\end{align}
This contains two terms with overlapping UV/IR sums: one where $Z$ contains IR modes and one where $m_V^2$ contains IR modes
\begin{align}
\begin{split}
-\frac{1}{2} Z \mathbb{C}_{2,\mu } =&\ -\frac{1}{16}  \frac{\lambda}{(2\pi R)^2}  \left( m^2_R(\mu) - m_Q^2\right)  \sum_{k\leq k_{\rm max}} \frac{1}{\omega_k}\sum_q \bigg[ \frac{\Theta(E_f+2\omega_q-E_{max})}{\omega_q^2(-2\omega_q)}\bigg]\\
&-\frac{1}{32}  \frac{\lambda^2}{(2\pi R)^2} Z \sum_{k'\leq \mu R} \frac{1}{\omega_{k'}}  \sum_q \bigg[ \frac{\Theta(E_f+2\omega_q-E_{max})}{\omega_q^2(-2\omega_q)}\bigg] \,.
\end{split}
\end{align}
The last term comes from un-normal-ordering $\phi^4$:
\begin{align}
\begin{split}
\frac{1}{8}Z^2\mathbb{C}_4 =&\ \frac{3}{32} Z  \frac{\lambda^2}{(2\pi R)^2}\sum_{k\leq k_{\rm max}} \frac{1}{\omega_k}\sum_{q}   \bigg[\frac{\Theta(E_f + 2\omega_q-E_{\rm max})}{\omega_q^2(-2\omega_q)}\bigg]\\
&-\frac{3}{128}  \frac{\lambda^2}{(2\pi R)^3}\sum_{k,k' \leq k_{\rm max}}\frac{1}{\omega_k\omega_{k'}} \sum_{q}   \bigg[\frac{\Theta(E_f + 2\omega_q-E_{\rm max})}{\omega_q^2(-2\omega_q)}\bigg] \,,
\end{split}
\end{align}
where we've kept the IR part of at least one of the $Z$'s.

Finally, looking at the overlapping UV/IR sums in the full coefficient of the identity operator we have
\begin{align}
 \mathbb{C}_0+ &\mathbb{C}_{0,\mu}-\frac{1}{2}Z \mathbb{C}_2 -\frac{1}{2}Z \mathbb{C}_{2,\mu} +\frac{1}{8}Z^2\mathbb{C}_4
 	\nonumber\\
 \supset 
 	&\ \frac{1}{16} \frac{\lambda}{(2\pi R)^2}  \left( m^2_R(\mu) - m_Q^2\right)\left( \sum_{k \leq \mu R} -\sum_{k\leq k_{\rm max}}\right)\frac{1}{\omega_k}\sum_q \frac{\Theta(E_f + 2\omega_q - E_{\rm max})}{ \omega_q^2 (-2\omega_q)}
	\nonumber\\
	&+\frac{1}{128} \frac{\lambda^2}{(2\pi R)^3}\left( \sum_{k,k' \leq \mu R} +2\sum_{k,k' \leq k_{\rm max}} -3\sum_{k,k' \leq k_{\rm max}}\right)\frac{1}{\omega_k\omega_{k'}}\sum_q \frac{\Theta(E_f + 2\omega_q - E_{\rm max})}{ \omega_q^2 (-2\omega_q)}
	\nonumber\\
	&+\frac{1}{32}  \frac{\lambda^2}{(2\pi R)^2} Z\left(- \sum_{k\leq \mu R} -2\sum_{k\leq k_{\rm max}}+3 \sum_{k\leq k_{\rm max}}\right) \frac{1}{\omega_{k}}  \sum_q \bigg[ \frac{\Theta(E_f+2\omega_q-E_{max})}{\omega_q^2(-2\omega_q)}\bigg]
	\nonumber\\
	&+\frac{1}{96}\frac{\lambda^2}{(2\pi R)^3} \!\!
		\left(\sum_{k \leq k_{\rm max}}
			-  \sum_{k\leq k_{\rm max}} \right)  \!
		\frac{1}{\omega_k}  \sum_{q,q' \geq k_{\rm max}}   \!\!\!\!
			\frac{\Theta(E_f + \omega_q+\omega_{q'}+\omega_{q+q'}- E_\text{max})}{\omega_q\omega_{q'}\omega_{q+q'}(-\omega_q-\omega_{q'}-\omega_{q+q'})} \,,
\end{align}
which again vanishes for the choice $\mu \sim E_{\rm max} = k_{\rm max}/R$, demonstrating separation of scales at this order.

\newpage

\clearpage

\normalem
\bibliography{HTET_NLO.bib}
\bibliographystyle{JHEP.bst}

\end{document}